\documentclass[twocolumn]{article} 
\usepackage[utf8]{inputenc}
\usepackage{subcaption}
\usepackage{amsmath}
\usepackage{graphicx}
\usepackage[a4paper, total={6in, 8in}]{geometry}
\usepackage{listings}
\usepackage{float}
\usepackage{placeins}
\usepackage[nottoc,numbib]{tocbibind}
\usepackage{xcolor}
\usepackage{hyperref}
\hypersetup{colorlinks,linkcolor={blue},citecolor={blue},urlcolor={blue}}
\usepackage[]{lineno}
\usepackage{bm}
\usepackage{url}
\usepackage[affil-it]{authblk}

\usepackage[style=phys]{biblatex}
\title{An Electromagnetic Particle-Particle Method for Relativistic Electron Bunch Dynamics from Early Expansion to Long-Range Transport}

\author[1]{Yinjian Zhao}
\author[1]{Yibo Liang}
\author[1,2,3,*]{Yanan Zhang}
\author[2,3]{Xiaochun Ma}
\author[1]{Hui Liu}

\affil[1]{School of Energy Science and Engineering,
Harbin Institute of Technology,
Harbin 150001, People’s Republic of China}
\affil[2]{Harbin Boiler Company Limited,
Harbin 150040, People’s Republic of China}
\affil[3]{State Key Laboratory of Low-carbon Thermal Power Generation Technology and Equipments,
Harbin 150040, People’s Republic of China}
\affil[*]{Corresponding author: Yanan Zhang, zhangyn@hbc.com.cn}

\date{\today}

\begin{document}


\twocolumn[
  \begin{@twocolumnfalse}
    \maketitle
\begin{abstract}
Particle-mesh methods, such as the particle-in-cell (PIC) method, cannot retain exact pairwise interaction at sub-cell scales. For dense nonneutral relativistic electron bunches, this makes it difficult to accurately capture
the inter-particle electromagnetic interaction and the associated bunch divergence. In this work, the previously developed electromagnetic particle-particle (EM-PP) model for relativistic two-particle interaction is extended to many-particle electron bunch transport in the Earth's magnetosphere. The method combines the Li\'enard--Wiechert fields, an improved retarded-time evaluation procedure, and a relativistic particle pusher, and adopts a two-stage strategy to couple the dense early self-field-dominated evolution to the later long-range geomagnetic-field-controlled transport. The method provides a practical mesh-free approach for accurately simulating long-range transport
of relativistic electron bunches when short-range electromagnetic interaction is important.
\end{abstract}
  \end{@twocolumnfalse}
]

\section{Introduction}\label{sec:introduction}

Particle-based methods are widely used in plasma physics and beam dynamics because they can naturally describe charged-particle motion under self-consistent and external fields. Besides Vlasov-type solvers that directly evolve kinetic equations \cite{cui2021grid,cui2024vlasov}, the most commonly used particle methods include the particle-in-cell (PIC) method and the particle-particle (PP) method. In our previous work \cite{zhang2025comparison,zhang2025electromagnetic}, we have elaborated the physical principles, applicable scope and strengths and limitations of PIC and PP methods. In PIC, particles are deposited onto a mesh to obtain charge or current density, the fields are solved on the grid, and then interpolated back to particle positions.\cite{birdsall1991plasma} 
This treatment is computationally efficient for large-scale systems, but it does not retain exact pairwise interaction at the shortest scales. In contrast, the PP method computes the inter-particle forces are evaluated directly between particle pairs, so that short-range Coulomb forces are resolved explicitly, although at a much higher computational cost, typically scaling as $O(N_p^2)$.\cite{hockney1988computer}. Therefore, the two methods are suitable for different physical regimes: PIC is generally more effective for large-scale weakly coupled or quasi-neutral plasmas dominated by smooth collective fields, whereas PP becomes attractive when accurate short-range interaction, non-neutrality, or early dense-stage expansion plays a dominant role\cite{zhao2017investigation,zhao2018three,fan2024electron}.  Our previous comparison between electrostatic PP and PIC for electron bunch expansion in vacuum further showed that, for dense nonneutral bunches, PIC accuracy does not necessarily improve monotonically with decreasing cell size, and large computational domains are required in long-time expansion to avoid nonphysical boundary effects, whereas PP can describe the same process more naturally in an effectively unbounded vacuum\cite{zhang2025comparison}.


However, the above PP studies were restricted to electrostatic interaction. For relativistic charged-particle dynamics, electromagnetic effects and retardation must be included, and the appropriate point-charge description is given by the Li\'enard--Wiechert potentials
\cite{Griffiths,Jackson1999}.
In our recent work\cite{zhang2025electromagnetic}, as an first application of an electromagnetic particle-particle (EM-PP) model, the technique was developed by extending the PP framework to the Li\'enard--Wiechert fields, implicit retarded-time evaluation, and a relativistic particle pusher \cite{HCpusher}. That work demonstrated that EM-PP can accurately resolve relativistic binary collisions in vacuum, and also pointed out several natural next steps, including its extension to more particles, dynamic time stepping, and more efficient treatment of larger systems\cite{zhang2025electromagnetic}.

Here, the EM-PP method is applied to a dense non-neutral relativistic electron bunch injected into the Earth's magnetosphere, which has been proposed as a powerful diagnostic tool for probing the magnetosphere, ionosphere, and upper atmosphere \cite{winckler1980application,Powis,borovsky2022modification,sanchez2019relativistic,borovsky2020mission}. Previous studies have primarily addressed the beam trajectory through simulations or observations, whereas the self-consistent evolution of the bunch envelope has not yet been fully characterized. The dynamics of relativistic electron bunch propagating towards Earth contain two distinct stages: a compact early stage dominated by strong short-range inter-particle electromagnetic interaction, and a later long-range stage governed mainly by the large-scale geomagnetic field. This combination makes a mesh-free particle-particle formulation particularly attractive. For this non-neutral bunch problem, the early evolution is controlled precisely by these strong local interactions. At the same time, carrying the full electromagnetic particle-particle interaction throughout the entire long-distance transport would be unnecessarily expensive. Therefore, the challenge is not simply to use EM-PP, but to use it where it is physically most needed and to couple it efficiently to the later field-dominated transport.

These considerations naturally motivate the present work. Full electromagnetic particle-particle interactions are retained during the initial dense stage, when short-range collective effects are strongest. After the bunch expands, the description is switched to a weakly coupled model, since the subsequent dynamics are mainly controlled by the geomagnetic field. This two-stage treatment preserves the mesh-free accuracy of EM-PP in the regime where it is most needed, while making long-distance magnetospheric transport computationally affordable. Thus, the work extends the particle-particle method from electrostatic bunch expansion and validation in relativistic binary-collision problems to the transport of many-particle relativistic bunches in space.



\section{Method}\label{sec:method}
\subsection{Governing Equations}\label{sec:GE}

The electromagnetic particle-particle model adopted in this work
is based on the Li\'enard--Wiechert fields generated by moving point charges.
Part of the formulation has been introduced in
\cite{zhang2025electromagnetic}
for relativistic two-particle interactions.
Here, the same governing equations are retained
and later extended to a bunch of interacting particles.

\begin{figure}[!ht]
\centering
\includegraphics[width=0.3\textwidth]{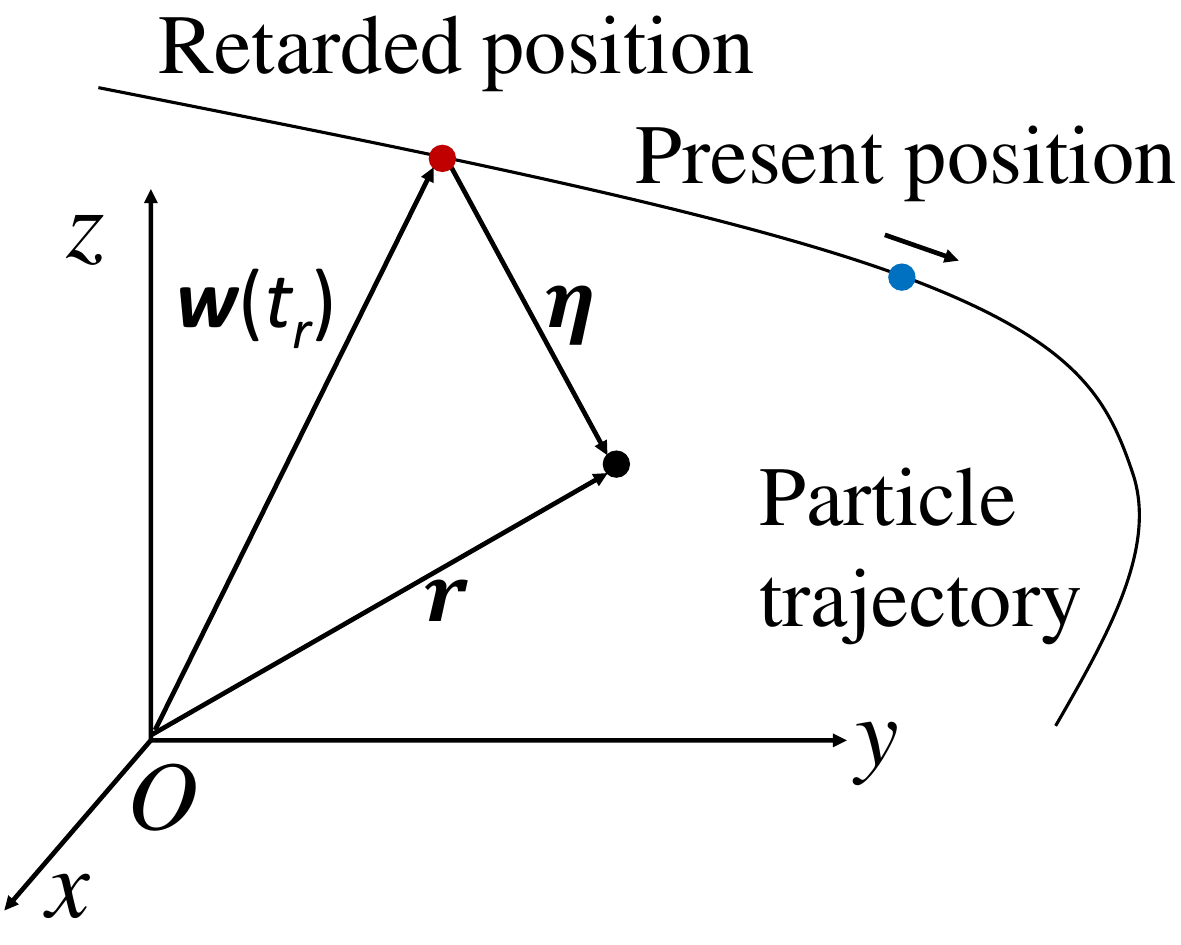}
\caption{
Illustration of the retarded time
and the vector $\bm{\eta}$.
}
\label{fig:EMP}
\end{figure}

Consider a point charge $q$ moving in vacuum.Its trajectory is denoted by $\bm{w}(t)$. Because electromagnetic signals propagate at the finite speed of light $c$,
the electric and magnetic fields observed at position $\bm{r}$ and time $t$ are determined by the source-particle state
at an earlier time, namely the retarded time $t_r$. As illustrated in Fig.~\ref{fig:EMP}, the retarded time satisfies
\begin{equation}
    |\bm{\eta}| \equiv |\bm{r} - \bm{w}(t_r)| = c(t-t_r),
\end{equation}
where
\begin{equation}
    \bm{\eta} \equiv \bm{r} - \bm{w}(t_r)
\end{equation}
is the vector from the retarded source position
to the field point.

The corresponding Li\'enard--Wiechert scalar and vector potentials
are written as \cite{Griffiths}
\begin{equation}
    V(\bm{r},t) =
    \dfrac{1}{4 \pi \varepsilon_0}
    \dfrac{q c}
    {\eta c - \bm{\eta}\cdot\dot{\bm{w}}(t_r)},
\end{equation}
\begin{equation}
    \bm{A}(\bm{r},t) =
    \dfrac{\dot{\bm{w}}(t_r)}{c^2} V(\bm{r},t),
\end{equation}
where $\varepsilon_0$ is the vacuum permittivity,
and $\dot{\bm{w}}(t_r)$ is the source-particle velocity
evaluated at the retarded time.

From these potentials,
the electric and magnetic fields at $(\bm{r},t)$
can be expressed as
\begin{equation}\label{eq:LW_E}
    \bm{E}(\bm{r},t) =
    \dfrac{
    q \eta
    \left[
    (c^2-v^2)\bm{v}'
    + \bm{\eta}\times(\bm{v}'\times\bm{a})
    \right]
    }{
    4\pi\varepsilon_0
    (\bm{\eta}\cdot\bm{v}')^3
    },
\end{equation}
\begin{equation}\label{eq:LW_B}
    \bm{B}(\bm{r},t) =
    \dfrac{1}{c}\,
    \bm{\hat{\eta}}\times\bm{E}(\bm{r},t),
\end{equation}
where
\begin{equation}
    \bm{v}' \equiv c\bm{\hat{\eta}} - \bm{v},
\end{equation}
and the shorthand notations
\begin{equation}
    \bm{v} \equiv \dot{\bm{w}}(t_r),
    \qquad
    \bm{a} \equiv \ddot{\bm{w}}(t_r)
\end{equation}
denote the retarded velocity and acceleration of the source particle,
respectively.

The particle motion is governed by the relativistic equations of motion,
\begin{equation}\label{eq:drdt2}
    \dfrac{d\bm{r}}{dt}
    = \bm{v}
    = \dfrac{\bm{u}}{\gamma},
\end{equation}
\begin{equation}\label{eq:dvdt2}
    \dfrac{d\bm{u}}{dt}
    = \dfrac{q}{m}
    \left(
    \bm{E} + \bm{v}\times\bm{B}
    \right),
\end{equation}
where $\bm{u}$ is the proper velocity,
$m$ is the particle mass, and
\begin{equation}
    \gamma =
    \left(1-\dfrac{v^2}{c^2}\right)^{-1/2}
    =
    \left(1+\dfrac{u^2}{c^2}\right)^{1/2}
\end{equation}
is the Lorentz factor.
The radiation-reaction force is neglected in the present study
\cite{Griffiths}.

For a system of multiple particles,
the total electromagnetic field acting on particle $i$
is obtained by superposing the contributions from all other particles,
together with the prescribed background geomagnetic field.
Therefore, the total fields can be written as
\begin{equation}
    \bm{E}_i^{\mathrm{tot}}
    =
    \sum_{j\neq i}\bm{E}_{ij},
\end{equation}
\begin{equation}
    \bm{B}_i^{\mathrm{tot}}
    =
    \sum_{j\neq i}\bm{B}_{ij}
    + \bm{B}_{\mathrm{geo}}(\bm{r}_i),
\end{equation}
where $\bm{E}_{ij}$ and $\bm{B}_{ij}$ are the Li\'enard--Wiechert fields
generated by particle $j$ and evaluated at the location of particle $i$,
and $\bm{B}_{\mathrm{geo}}$ denotes the background geomagnetic field.
In this way, the present formulation naturally extends
the two-particle EM-PP model to a relativistic electron bunch.
\subsection{Determination of the Retarded Time}\label{sec:tr}

To evaluate the Li\'enard--Wiechert fields in
Eqs.~(\ref{eq:LW_E}) and (\ref{eq:LW_B}),
the retarded time $t_r$ must be determined implicitly
for each source particle and each field evaluation point.
This is one of the key numerical difficulties
in the electromagnetic particle-particle formulation.
In the previous work \cite{zhang2025electromagnetic},
a bracketing procedure based on trajectory history
was introduced.
The same idea is retained here,
and a linear reconstruction is further employed
to improve the accuracy of the retarded quantities.

During the simulation,
the particle trajectory $\bm{w}(t)$
and velocity history $\bm{v}(t)$
are stored at discrete time steps.
For a given field point $\bm{r}$ at time $t$,
the retarded time $t_r$ is first bracketed
within an interval $[t_1,t_2]$ such that
\begin{equation}
    t_1 \leq t_r \leq t_2 .
\end{equation}
The search starts from the maximum available interval $[t_0,t]$,
where $t_0$ is the earliest stored time.
If the field point is sufficiently close to the present particle position,
namely
\begin{equation}
    |\bm{r}-\bm{w}(t-\Delta t)| < c\Delta t,
\end{equation}
where $\Delta t$ is the simulation time step,
the retarded time is directly bracketed by
\begin{equation}
    t_1 = t-\Delta t,
    \qquad
    t_2 = t.
\end{equation}
Otherwise, a bisection procedure is applied.

The bisection search begins from the midpoint
\begin{equation}
    t_m = \frac{t_0+t}{2},
\end{equation}
which may also be written as
\begin{equation}
    t_m = t_0 + \delta t = t - \delta t,
\end{equation}
with
\begin{equation}
    \delta t = \frac{t-t_0}{2}.
\end{equation}
At each iteration,
the quantity $|\bm{r}-\bm{w}(t_m)|$ is compared with $c\delta t$.
If
\begin{equation}
    c\delta t \leq |\bm{r}-\bm{w}(t_m)|,
\end{equation}
then the retarded time must lie in the interval
$[t_0,t_m]$.
Otherwise, if
\begin{equation}
    c\delta t > |\bm{r}-\bm{w}(t_m)|,
\end{equation}
then the retarded time lies in $(t_m,t]$.
The search interval is halved recursively
until the condition
\begin{equation}
    t_2 - t_1 = \Delta t
\end{equation}
is satisfied.

In \cite{zhang2025electromagnetic},
once $t_1$ and $t_2$ are obtained,
the retarded quantities are approximated by midpoint values,
\begin{equation}\label{eq:mid}
\begin{aligned}
t_r &\approx \frac{t_1+t_2}{2}, \\
\bm{w}(t_r) &\approx \frac{\bm{w}(t_1)+\bm{w}(t_2)}{2}, \\
\bm{v}(t_r) &\approx \frac{\bm{v}(t_1)+\bm{v}(t_2)}{2}, \\
\bm{a}(t_r) &\approx \frac{\bm{v}(t_2)-\bm{v}(t_1)}{\Delta t}.
\end{aligned}
\end{equation}
Here, the acceleration history is not stored explicitly,
and is approximated by a finite difference of the velocity history.

In the present work,
the midpoint approximation is improved
by introducing a local linear reconstruction
within the interval $[t_1,t_2]$.
Let
\begin{equation}
    \delta t \equiv t_r - t_1,
\end{equation}
and denote
\begin{equation}
    \bm{w}_1 \equiv \bm{w}(t_1),
    \qquad
    \bm{v}_1 \equiv \bm{v}(t_1),
    \qquad
    \bm{v}_2 \equiv \bm{v}(t_2).
\end{equation}
Assuming that the trajectory is locally linear over one time step,
the retarded position is approximated by
\begin{equation}
    \bm{w}(t_r) \approx \bm{w}_1 + \bm{v}_1 \delta t.
\end{equation}
Substituting this relation into the retardation condition yields
\begin{equation}
    c[t-(t_1+\delta t)]
    =
    \left|
    \bm{r}-(\bm{w}_1+\bm{v}_1\delta t)
    \right|.
\end{equation}
This relation leads to a quadratic equation for $\delta t$,
whose physically relevant solution is
\begin{equation}\label{eq:dt}
    \delta t
    =
    \dfrac{-C_1-\sqrt{C_1^2-4C_2C_0}}{2C_2},
\end{equation}
where the coefficients are given by
\begin{equation}
    C_2 = c^2 - v_1^2,
\end{equation}
\begin{equation}
    C_1 =
    2\left[
    (\bm{r}-\bm{w}_1)\cdot\bm{v}_1
    - c^2(t-t_1)
    \right],
\end{equation}
\begin{equation}
    C_0 =
    c^2(t-t_1)^2 - |\bm{r}-\bm{w}_1|^2.
\end{equation}
The positive-sign root is discarded,
since it gives $\delta t > t-t_1$.
For physically admissible particle velocities,
the discriminant remains positive.

Once $\delta t$ is obtained,
the retarded quantities are reconstructed as
\begin{equation}\label{eq:linear}
\begin{aligned}
t_r &\approx t_1 + \delta t, \\
\bm{w}(t_r) &\approx \bm{w}_1 + \bm{v}_1 \delta t
= \bm{w}_1(1-\delta f)+\bm{w}_2\delta f, \\
\bm{v}(t_r) &\approx \bm{v}_1(1-\delta f)+\bm{v}_2\delta f, \\
\bm{a}(t_r) &\approx \frac{\bm{v}_2-\bm{v}_1}{\Delta t},
\end{aligned}
\end{equation}
where
\begin{equation}
    \delta f = \frac{\delta t}{t_2-t_1},
\end{equation}
and
\begin{equation}
    \bm{w}_2 \equiv \bm{w}(t_2).
\end{equation}
Compared with the midpoint approximation in Eq.~(\ref{eq:mid}),
the above treatment provides a more accurate estimate
of the retarded position and velocity
without introducing additional trajectory storage.

A special case arises when the true retarded time is older than
the earliest stored time,
namely
\begin{equation}
    t_r < t_0 = t_1,
\end{equation}
or equivalently,
\begin{equation}
    c(t-t_0) < |\bm{r}-\bm{w}_0|,
\end{equation}
where $\bm{w}_0=\bm{w}(t_0)$.
In this case, Eq.~(\ref{eq:dt}) yields $\delta t<0$,
which can still be used consistently.
The retarded quantities are then approximated as
\begin{equation}\label{eq:linear2}
\begin{aligned}
t_r &\approx t_1 + \delta t, \\
\bm{w}(t_r) &\approx \bm{w}_1 + \bm{v}_1 \delta t, \\
\bm{v}(t_r) &\approx \bm{v}_1, \\
\bm{a}(t_r) &\approx 0.
\end{aligned}
\end{equation}
Therefore, the trajectory history should be stored over as long a time span
as affordable in practice,
so that the occurrence of $t_r<t_0$ is minimized
and the approximation error remains small.

\subsection{Relativistic Particle Pusher}

The particle trajectories are advanced with the relativistic
structure-preserving pusher proposed by Higuera and Cary \cite{HCpusher}.
Its implementation follows the formulation summarized by
Ripperda et al.~\cite{Ripperda_2018},
which is suitable for relativistic charged-particle motion in prescribed electric and magnetic fields.

For a particle with charge $q$ and mass $m$,
the update from time level $n$ to $n+1$
is carried out in three steps.
First, the proper velocity is accelerated by
half of the electric impulse,
\begin{equation}
    \bm{u}^{-}
    =
    \bm{u}^{n}
    +
    \frac{q\Delta t}{2m}
    \bm{E}(\bm{r}^{n+1/2}),
\end{equation}
where $\Delta t$ is the time step
and $\bm{r}^{n+1/2}$ denotes the particle position
at the half step.

Next, the magnetic rotation is applied.
Introducing the auxiliary quantities
\begin{equation}
    \gamma^{-}
    =
    \left[
    1+\frac{(\bm{u}^{-})^{2}}{c^{2}}
    \right]^{1/2},
\end{equation}
\begin{equation}
    \bm{\tau}
    =
    \bm{B}(\bm{r}^{n+1/2})
    \frac{q\Delta t}{2m},
\end{equation}
\begin{equation}
    u^{*}
    =
    \frac{\bm{u}^{-}\cdot\bm{\tau}}{c},
\end{equation}
\begin{equation}
    \sigma
    =
    (\gamma^{-})^{2}-\tau^{2},
\end{equation}
the updated Lorentz factor after magnetic rotation is
\begin{equation}
    \gamma^{+}
    =
    \sqrt{
    \frac{
    \sigma+\sqrt{\sigma^{2}+4(\tau^{2}+(u^{*})^{2})}
    }{2}
    }.
\end{equation}
Then,
\begin{equation}
    \bm{t}=\frac{\bm{\tau}}{\gamma^{+}},
    \qquad
    s=\frac{1}{1+t^{2}},
\end{equation}
and the rotated proper velocity is written as
\begin{equation}
    \bm{u}^{+}
    =
    s\left[
    \bm{u}^{-}
    +
    (\bm{u}^{-}\cdot\bm{t})\bm{t}
    +
    \bm{u}^{-}\times\bm{t}
    \right].
\end{equation}

Finally, the second half electric acceleration is applied,
\begin{equation}
    \bm{u}^{n+1}
    =
    \bm{u}^{+}
    +
    \frac{q\Delta t}{2m}
    \bm{E}(\bm{r}^{n+1/2})
    +
    \bm{u}^{+}\times\bm{t}.
\end{equation}
As noted in \cite{Ripperda_2018},
the last term should be
$\bm{u}^{+}\times\bm{t}$;
using $\bm{u}^{-}\times\bm{t}$ there would be a typo.

Once $\bm{u}^{n+1}$ is obtained,
the particle velocity follows from
\begin{equation}
    \bm{v}^{n+1}
    =
    \frac{\bm{u}^{n+1}}{\gamma^{n+1}},
\end{equation}
with
\begin{equation}
    \gamma^{n+1}
    =
    \left[
    1+\frac{(\bm{u}^{n+1})^{2}}{c^{2}}
    \right]^{1/2}.
\end{equation}
The particle position is then advanced accordingly.

Compared with the classical Boris scheme,
the Higuera--Cary pusher preserves the correct
relativistic $\bm{E}\times\bm{B}$ drift
while maintaining good long-time numerical stability.
For the present study,
this property is important because the beam particles
undergo relativistic motion over a long propagation distance
under the combined influence of the geomagnetic field
and the inter-particle electromagnetic forces.

\subsection{Adaptive Time-Step Selection}\label{sec:adaptive_dt}

In the early dense stage of the bunch evolution,
the inter-particle electromagnetic forces can vary rapidly
because the particles remain close to each other
and their relative motion changes on a very short time scale.
To resolve this stage accurately while avoiding an unnecessarily small
time step during the later weakly coupled stage,
an adaptive time-step strategy is employed.

At each time step,
the same field-evaluation loop used to compute the total fields
is also used to estimate a characteristic interaction time scale.
For each interacting particle pair $(i,j)$,
the retarded separation vector $\bm{\eta}_{ij}$
and the corresponding retarded source-particle velocity
$\bm{v}_j(t_{r,ij})$ are obtained from the retarded-time solver.
Based on these quantities,
two global measures are introduced:
the minimum effective particle separation,
\begin{equation}
    \eta_{\min}
    =
    \min_{i\neq j,\ |\bm{\eta}_{ij}|\ge \eta_{\mathrm{allow}}}
    |\bm{\eta}_{ij}|,
\end{equation}
and the maximum relative speed,
\begin{equation}
    v_{\max}
    =
    \max_{i\neq j,\ |\bm{\eta}_{ij}|\ge \eta_{\mathrm{allow}}}
    \left|
    \bm{v}_i(t)-\bm{v}_j(t_{r,ij})
    \right|,
\end{equation}
where $\eta_{\mathrm{allow}}$ is a prescribed lower cutoff
introduced to exclude extremely small effective separations
from the time-step estimate.

The time step is then updated according to
\begin{equation}
    \Delta t
    =
    C_{\Delta t}\,
    \frac{\eta_{\min}}{v_{\max}},
\end{equation}
where $C_{\Delta t}$ is a safety factor.
In the present work,
\begin{equation}
    C_{\Delta t} = 0.05.
\end{equation}
This expression corresponds to taking the time step
as a small fraction of the shortest resolved interaction time scale
among the active particle pairs.

To prevent the time step from becoming either too small
or unnecessarily large,
it is further constrained by
\begin{equation}
    \Delta t
    \leftarrow
    \min\!\left(
    \Delta t_{\max},
    \max(\Delta t_{\min},\Delta t)
    \right),
\end{equation}
with
\begin{equation}
    \Delta t_{\min}=10^{-15}\ \mathrm{s},
    \qquad
    \Delta t_{\max}=10^{-12}\ \mathrm{s}.
\end{equation}
Therefore, the adaptive rule automatically reduces the time step
when the bunch is compact and strongly interacting,
and increases it as the bunch expands
and the inter-particle coupling weakens.

This adaptive time-step strategy is used in the fully coupled stage
of the simulation and provides the numerical basis
for efficiently resolving the early self-field-driven expansion.
Its behavior for different particle numbers
and its relation to the switching criterion
are examined later in Sec.~\ref{sec:timestep}.

\subsection{Beam Initialization and Setup}\label{sec:setup}

Following the configuration considered by
Powis et al.~\cite{Powis},
the injected relativistic electron bunch
is initialized by a Gaussian phase-space distribution.
In the beam frame, the distribution function is written as
\begin{equation}
\begin{aligned}
f(\bm{r},\bm{v})
&=
f_0
\exp\left[
-\frac{x^2+y^2}{2r_{b,\perp}^2}
-\frac{z^2}{2r_{b,\parallel}^2}
\right]
\\
&\quad\times
\exp\left[
-\frac{v_x^2+v_y^2}{2\langle v_\perp\rangle^2}
-\frac{(v_z-v_0)^2}{2\langle v_\parallel\rangle^2}
\right],
\end{aligned}
\label{eq:initial-distribution}
\end{equation}
where $f_0$ is a normalization constant,
$r_{b,\perp}$ is the initial transverse beam radius,
and $r_{b,\parallel}$ is the axial beam radius.

The mean beam velocity is defined by
\begin{equation}
    v_0 = \beta c,
    \quad
    \gamma_0 = 1 + \frac{E_0}{m_e c^2},
    \quad
    \beta = \sqrt{1-\frac{1}{\gamma_0^2}},
\end{equation}
where $E_0$ is the kinetic beam energy,
$m_e$ is the electron mass,
and $c$ is the speed of light.
The transverse and longitudinal RMS velocity spreads are taken as
\begin{equation}
    \langle v_\perp\rangle
    =
    \frac{\varepsilon_r v_0}{r_{b,\perp}},
    \qquad
    \langle v_\parallel\rangle
    \approx
    \sqrt{\left\langle (v_z-v_0)^2 \right\rangle},
\end{equation}
where $\varepsilon_r$ is the unnormalized radial emittance.

To characterize the bunch evolution,
the transverse and longitudinal RMS beam sizes are defined as
\begin{equation}
    \sigma_x
    =
    \sqrt{
    \frac{\sum_{i=1}^{N} w_i (x_i-\bar{x})^2}
         {\sum_{i=1}^{N} w_i}
    },
\end{equation}
\begin{equation}
    \sigma_y
    =
    \sqrt{
    \frac{\sum_{i=1}^{N} w_i (y_i-\bar{y})^2}
         {\sum_{i=1}^{N} w_i}
    },
\end{equation}
\begin{equation}
    \sigma_{xy}
    =
    \frac{\sigma_x+\sigma_y}{2},
\end{equation}
\begin{equation}
    \sigma_z
    =
    \sqrt{
    \frac{\sum_{i=1}^{N} w_i (z_i-\bar{z})^2}
         {\sum_{i=1}^{N} w_i}
    }.
\end{equation}

The baseline beam parameters are summarized in
Table~\ref{tab:beam-params}.

\begin{table}[!ht]
\centering
\caption{
Beam parameters used to define the initial bunch distribution in Eq.~\eqref{eq:initial-distribution}.
}
\begin{tabular}{lll}
\hline
Quantity    & Symbol  & Value     \\
\hline
Beam kinetic energy          & $E_0$           & $1.0\,\mathrm{MeV}$          \\
Lorentz factor               & $\gamma_0$      & $2.96$                       \\
Speed ratio                  & $\beta$         & $0.941$                      \\
Relative energy spread       & $\delta_E$      & $0.01$                       \\
Radial emittance             & $\varepsilon_r$ & $1.0\,\mathrm{mm\,mrad}$     \\
Transverse beam radius       & $r_{b,\perp}$   & $2\,\mathrm{mm}$             \\
Axial beam radius            & $r_{b,\parallel}$ & $2\,\mathrm{mm}$           \\
\hline
\end{tabular}
\label{tab:beam-params}
\end{table}

In the original setup of Powis et al.~\cite{Powis}, the axial beam size is related to the mini-pulse length by
\begin{equation}
    r_{b,\parallel} = \frac{L_{\mathrm{mp}}}{2}.
\end{equation}
In the present study, to emphasize the role of self-fields in the early dense stage,
the axial bunch size is reduced to be comparable to the transverse size.
Therefore, the baseline calculations adopt
\begin{equation}
    r_{b,\parallel} = r_{b,\perp}.
\end{equation}

The macroparticle positions are sampled from Gaussian distributions,
\begin{subequations}
\label{eq:space-sampling}
\begin{align}
x &\sim \mathcal{N}(0,\sigma_r^2),
&
y &\sim \mathcal{N}(0,\sigma_r^2),
\\
z &\sim \mathcal{N}(0,\sigma_z^2),
\end{align}
\end{subequations}
with
\begin{equation}
    \sigma_r = r_{b,\perp},
    \qquad
    \sigma_z = r_{b,\parallel}.
\end{equation}

The transverse velocities and particle energy are sampled as
\begin{subequations}
\label{eq:vel-sampling}
\begin{align}
v_x &\sim \mathcal{N}(0,\sigma_\perp^2),
&
v_y &\sim \mathcal{N}(0,\sigma_\perp^2),
\\
E &\sim \mathcal{N}(E_0,\sigma_E^2),
\end{align}
\end{subequations}
where
\begin{equation}
    \sigma_\perp = \langle v_\perp\rangle
    = \frac{\varepsilon_r v_0}{r_{b,\perp}},
    \qquad
    \sigma_E = \delta_E E_0.
\end{equation}

For each sampled energy $E$,
the corresponding relativistic longitudinal velocity is obtained from
\begin{subequations}
\label{eq:rel-vz}
\begin{align}
\gamma(E) &= 1 + \frac{E}{m_e c^2},
\\
v_z(E) &= c\sqrt{1-\frac{1}{\gamma(E)^2}}.
\end{align}
\end{subequations}
In the numerical implementation, the sample mean of $v_z$ is shifted back to $v_0$ to ensure the prescribed average injection speed,
\begin{equation}
    v_z
    \leftarrow
    v_z - \left(\langle v_z\rangle - v_0\right).
\end{equation}

If the bunch is injected with a finite angle relative to the local geomagnetic field, the particle coordinates and velocities are rotated
from the beam frame to the laboratory frame according to
\begin{equation}
\begin{aligned}
\bm{x}_i^{(\mathrm{lab})}
&=
R(\theta,\phi)\,
\bm{x}_i^{(\mathrm{beam})},
\\
\bm{v}_i^{(\mathrm{lab})}
&=
R(\theta,\phi)\,
\bm{v}_i^{(\mathrm{beam})},
\end{aligned}
\end{equation}
where
\begin{equation}
    R(\theta,\phi) = R_z(\phi)R_y(\theta),
\end{equation}
with
\begin{equation}
    R_y(\theta)
    =
    \begin{pmatrix}
    \cos\theta & 0 & \sin\theta \\
    0 & 1 & 0 \\
    -\sin\theta & 0 & \cos\theta
    \end{pmatrix},
\end{equation}
\begin{equation}
    R_z(\phi)
    =
    \begin{pmatrix}
    \cos\phi & -\sin\phi & 0 \\
    \sin\phi & \cos\phi & 0 \\
    0 & 0 & 1
    \end{pmatrix}.
\end{equation}
Here $\theta$ is the polar angle between the beam axis and the laboratory $z$ axis, and $\phi$ is the azimuthal angle in the $x$--$y$ plane.

The effective volume of the Gaussian bunch is approximated by
\begin{equation}
    V = (2\pi)^{3/2} r_{b,\perp}^2 r_{b,\parallel}.
\end{equation}
If $N_p$ macroparticles are used to represent
a bunch with peak density $n_0$, the macroparticle weight is given by
\begin{equation}
    w_p = \frac{n_0 V}{N_p}.
\end{equation}

\begin{figure}[!ht]
\centering
\includegraphics[width=0.5\textwidth]{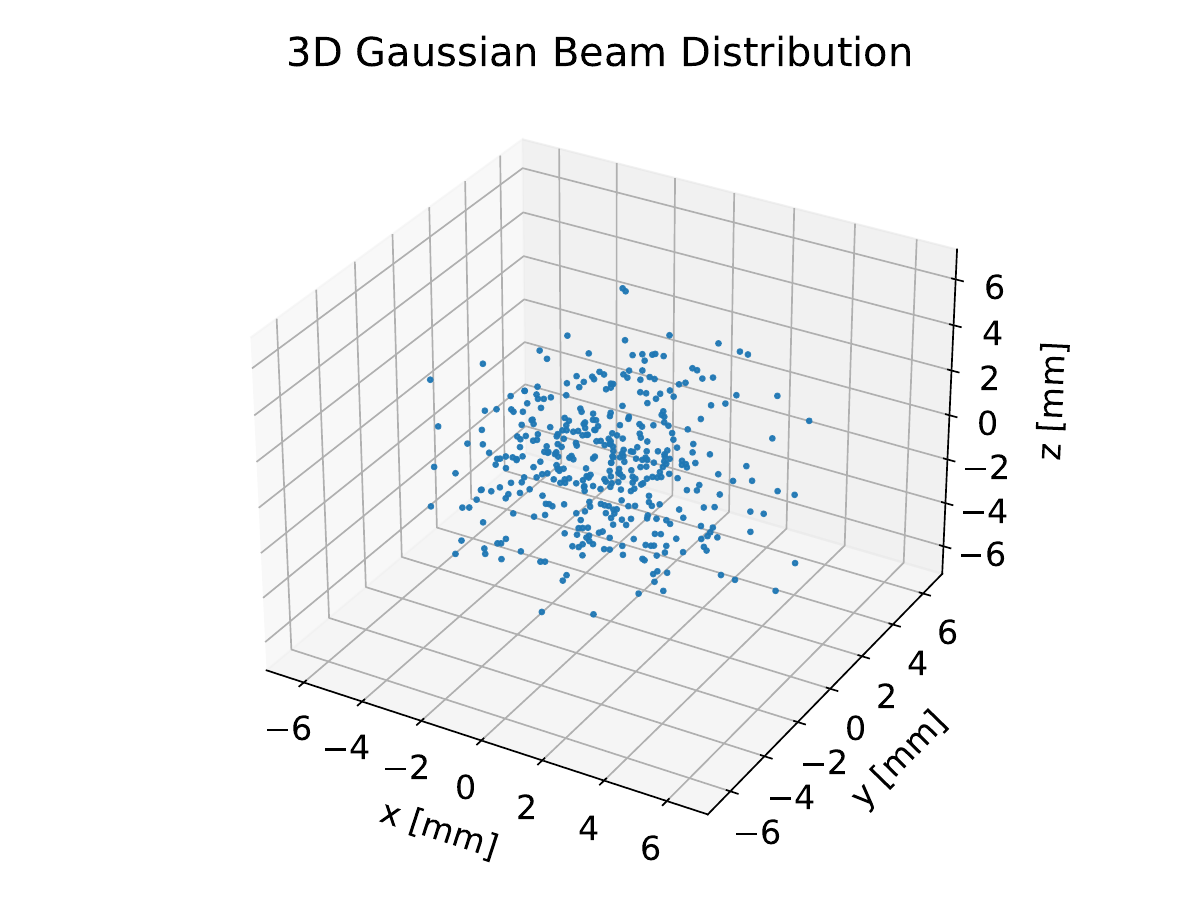}
\caption{
Particle spatial sampling.
}
\label{fig:fs}
\end{figure}

\begin{figure}[!ht]
\centering
\includegraphics[width=0.4\textwidth]{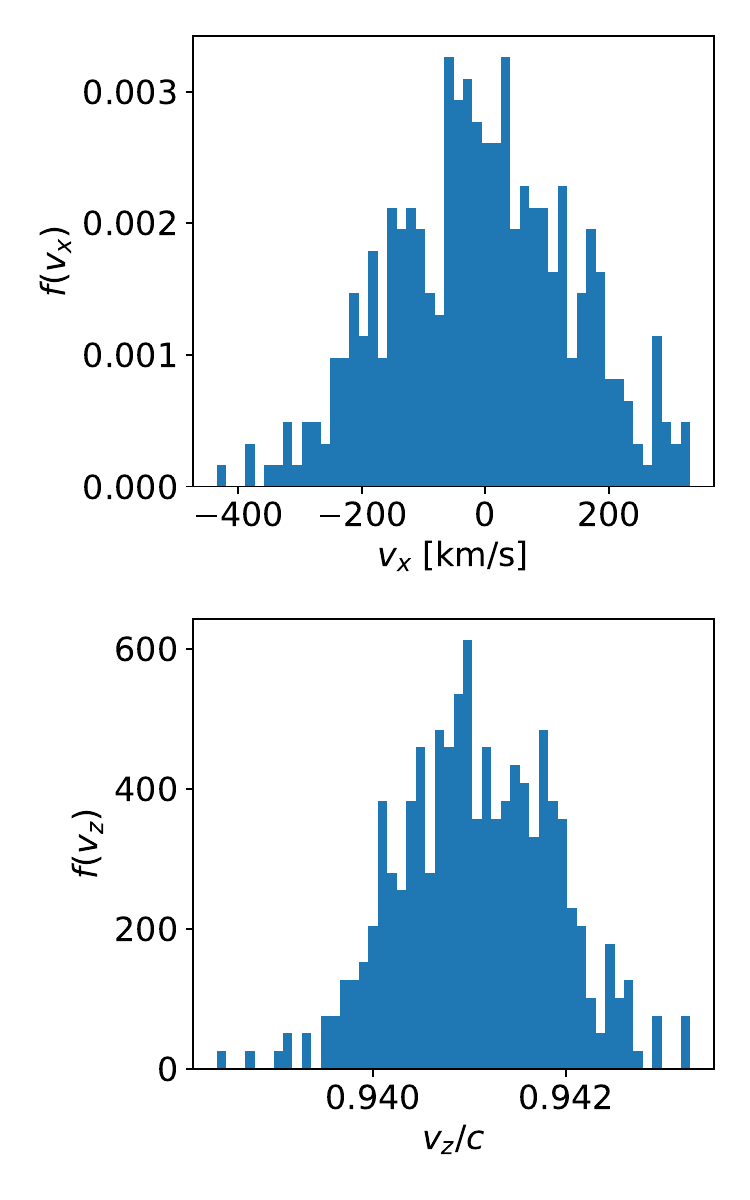}
\caption{
Particle velocity sampling.
}
\label{fig:fv}
\end{figure}

Unless otherwise stated,
the bunch is initialized at
\begin{equation}
    \bm{r}_0 = (-10R_{\mathrm{E}},\,0,\,0),
\end{equation}
where $R_{\mathrm{E}}$ is the Earth radius.
For the reference case with $N_p=400$ and $\theta=\phi=0$, the generated spatial and velocity distributions are shown in Fig.~\ref{fig:fs} and Fig.~\ref{fig:fv},
respectively.

\subsection{Geomagnetic Field Model and Injection Geometry}\label{sec:geomagnetic}

The background magnetic field is prescribed
and modeled as a static centered dipole field.
The origin of the laboratory frame is placed at the center of the Earth, and the dipole moment is aligned with the laboratory $z$ axis,
\begin{equation}
    \bm{M} = M_0 \bm{e}_z,
\end{equation}
with
\begin{equation}
    M_0 = 8.6\times10^{22}\ \mathrm{A\,m^2}.
\end{equation}
For numerical convenience, the geomagnetic field is evaluated in the Cartesian form
given in \cite{Jackson1999},
\begin{equation}
    \bm{B}_{\mathrm{geo}}(\bm{r})
    =
    \frac{\mu_0}{4\pi}
    \frac{3(\bm{M}\cdot\hat{\bm{r}})\hat{\bm{r}}-\bm{M}}{r^3},
\end{equation}
where
\begin{equation}
    r = |\bm{r}|,
    \qquad
    \hat{\bm{r}} = \frac{\bm{r}}{r},
\end{equation}
and $\mu_0$ is the vacuum permeability.
This idealized dipole model is also consistent with the near-Earth magnetic-field configuration adopted in the beam-transport study of Powis et al.~\cite{Powis}.

The relativistic electron bunch is injected from
\begin{equation}
    \bm{r}_0 = (-10R_{\mathrm{E}},\,0,\,0),
\end{equation}
where $R_{\mathrm{E}}$ is the Earth radius.
This injection geometry follows the long-range transport scenario considered by Powis et al.~\cite{Powis}, in which a compact accelerator mounted on a satellite launches a relativistic electron beam into the magnetosphere from a geocentric distance of order $10R_{\mathrm{E}}$.

At the injection point, the beam is initialized approximately along the local geomagnetic field line. Possible deviations from perfect field-aligned injection
are described by the rotation angles $\theta$ and $\phi$ introduced in Sec.~\ref{sec:setup}.
Here $\theta$ denotes the polar angle
between the beam axis and the laboratory $z$ axis, while $\phi$ is the azimuthal angle in the $x$--$y$ plane.
Unless otherwise stated,
the baseline cases adopt $\theta=\phi=0$.

Under this prescribed dipole field,
the particles undergo gyromotion around magnetic field lines
during their transport toward the Earth.
As the geocentric distance decreases,
the magnetic field strength increases,
so that the beam dynamics are increasingly influenced by
magnetic mirroring, curvature drift, and $\nabla B$ drift.
Particles inside the loss cone may continue to precipitate,
whereas particles outside the loss cone can be reflected
before reaching the atmosphere.
The present dipole model therefore provides the large-scale
magnetic transport environment,
on top of which the early-stage self-field-driven bunch expansion
is resolved by the electromagnetic particle-particle model.

\subsection{Two-stage Strategy}\label{sec:twostage}

For the relativistic electron bunch considered in this work, the collective dynamics exhibit two clearly separated stages. Immediately after injection,
the bunch is highly compact and the inter-particle electromagnetic interactions are strong. In this stage, the bunch undergoes rapid transverse expansion and longitudinal stretching driven by its self-fields. As the bunch expands, the particle separation increases rapidly,
and the inter-particle forces become progressively weaker. At later times, the transport is dominated primarily by the prescribed geomagnetic field, while the self-field contribution becomes negligible. 

In principle, the full electromagnetic particle-particle model can be applied throughout the entire transport process. However, such a treatment is computationally very expensive for long-range propagation from the injection point to the near-Earth region, because the Li\'enard--Wiechert interaction between all particle pairs
must be evaluated repeatedly together with the retarded-time search. To reduce the computational cost while retaining the essential physics, the two-stage framework separates the simulation into an early strongly coupled stage and a later weakly coupled transport stage.

In Stage I, the bunch evolution is resolved using the full EM-PP model. Both the inter-particle electric and magnetic fields and the prescribed geomagnetic field are included, and a small adaptive time step is employed to resolve the rapid early expansion. This stage is designed to capture the dominant self-field effects, including the conversion of the initial electrostatic potential energy into particle kinetic energy and the associated redistribution of the bunch envelope.

To determine when the inter-particle interaction can be neglected, the mean kinetic energy of the bunch is monitored,
\begin{equation}
    \bar{E}(t)
    =
    \frac{\sum_{i=1}^{N_p} w_i E_i(t)}
         {\sum_{i=1}^{N_p} w_i},
\end{equation}
where $E_i(t)$ is the kinetic energy of the $i$th macroparticle and $w_i$ is its weight. The switching time $t_{\mathrm{sw}}$ is defined by the condition that the growth rate of the mean energy falls below a prescribed threshold,
\begin{equation}
    \left|
    \frac{d\bar{E}}{dt}
    \right|
    < \varepsilon_E,
\end{equation}
where $\varepsilon_E$ is taken as $30~\mathrm{keV/s}$ or less in the present work. This criterion indicates that the initial strong inter-particle interaction has largely decayed, and that the remaining evolution is no longer sensitive to the self-field contribution.

Once $t_{\mathrm{sw}}$ is obtained from the Stage-I calculation, the long-range propagation is recomputed from the initial state. During this second run, the full EM-PP model is retained only for
$0\le t \le t_{\mathrm{sw}}$,
so that the early self-field-driven expansion is reproduced. For
\begin{equation}
    t > t_{\mathrm{sw}},
\end{equation}
the inter-particle electromagnetic interaction is turned off, and the particles are advanced only under the prescribed geomagnetic field,
\begin{equation}
    \bm{E}_i^{\mathrm{tot}} = \bm{0},
    \qquad
    \bm{B}_i^{\mathrm{tot}}
    =
    \bm{B}_{\mathrm{geo}}(\bm{r}_i).
\end{equation}
At the same time, the time step can be increased substantially, since the later transport is much smoother and is no longer constrained by the short-range self-field dynamics.

The above procedure avoids the prohibitive cost of carrying
the pairwise retarded electromagnetic interaction through the entire long-distance simulation, while preserving the essential collective physics of the dense early stage. Therefore, the two-stage strategy combines high-fidelity mesh-free particle-particle interactions when they are physically important with an efficient long-range propagation model when the bunch dynamics become geomagnetic-field dominated.

For clarity, the two-stage procedure may be summarized as follows:
\begin{enumerate}
    \item Initialize the bunch at $\bm{r}_0$
    with the prescribed phase-space distribution.
    \item Perform a short full EM-PP simulation
    with adaptive small time steps.
    \item Monitor $\bar{E}(t)$ and determine the switching time
    $t_{\mathrm{sw}}$ from the criterion
    $|d\bar{E}/dt|<\varepsilon_E$.
    \item Restart the simulation from the same initial state.
    \item Retain the full EM-PP interaction for
    $0\le t\le t_{\mathrm{sw}}$.
    \item For $t>t_{\mathrm{sw}}$,
    neglect inter-particle interaction,
    increase the time step,
    and continue the propagation under the geomagnetic field
    until the particles either reach the near-Earth region
    or undergo magnetic mirroring.
\end{enumerate}

\section{Numerical Setup and Algorithm Validation}\label{sec:numerical}
\subsection{Baseline Simulation Parameters}\label{sec:baseline}



To examine the effect of macroparticle resolution,
three particle numbers are considered,
\begin{equation}
    N_p = 400,\ 800,\ 1600.
\end{equation}
The case with $N_p=400$ is used as the baseline configuration
for assessing the overall beam evolution and computational efficiency,
while the cases with $N_p=800$ and $N_p=1600$
are used to evaluate the sensitivity of the results
to the macroparticle number.
In addition to the reference energy $E_0=1~\mathrm{MeV}$,
higher-energy cases at
\begin{equation}
    E_0 = 10~\mathrm{MeV}
    \qquad \text{and} \qquad
    E_0 = 100~\mathrm{MeV}
\end{equation}
are also simulated to investigate the energy dependence
of the long-range bunch transport.

For the early dense stage,
the full electromagnetic particle-particle interaction is solved
with an adaptive time step.
The lower and upper bounds of the time step are set to
\begin{equation}
    \Delta t_{\min} = 10^{-15}~\mathrm{s},
    \qquad
    \Delta t_{\max} = 10^{-12}~\mathrm{s}.
\end{equation}
These limits are chosen to resolve the rapid initial bunch expansion
while maintaining acceptable computational cost.
In the two-stage framework,
the transition from the strongly coupled stage
to the weakly coupled long-range transport stage
is determined from the evolution of the mean bunch energy.
Unless otherwise stated,
the switching criterion is taken as
\begin{equation}
    \left|
    \frac{d\bar{E}}{dt}
    \right|
    < 30~\mathrm{keV/s},
\end{equation}
after which the inter-particle interaction is neglected
and the subsequent propagation is advanced
under the geomagnetic field alone.


\subsection{Single-Particle Benchmark Against Powis}\label{sec:SingleParticleTest}

Before investigating the collective dynamics of the relativistic electron bunch,
a single-particle benchmark is carried out to verify
the accuracy of the prescribed geomagnetic-field model
and the relativistic particle pusher.
The reference trajectory is compared with the result reported by
Powis et al.~\cite{Powis},
who studied the long-range transport of relativistic electron beams
in the Earth's magnetosphere.

In this benchmark,
the inter-particle electromagnetic interaction is completely turned off.
Although the numerical implementation still uses
the bunch framework introduced in Sec.~\ref{sec:setup},
all particles are assigned the same initial velocity,
with
\begin{equation}
    v_x = 0,
    \qquad
    v_y = 0,
    \qquad
    v_z = v_0,
\end{equation}
so that the bunch reduces effectively to a set of identical
non-interacting test particles.
In this way, the beam centroid follows the trajectory
of a single relativistic electron injected into the prescribed dipole field.

A constant time step is adopted in this test,
\begin{equation}
    \Delta t = 0.01~\mathrm{ms},
\end{equation}
and the total simulation time is set to
\begin{equation}
    t_{\mathrm{end}} = 289~\mathrm{ms},
\end{equation}
which is sufficient for the particle to propagate
from the injection point at $10R_{\mathrm{E}}$
to the near-Earth region.
The particle trajectory is obtained from the weighted mean position
of the macroparticles,
\begin{equation}
    \bar{\bm{r}}(t)
    =
    \frac{\sum_{i=1}^{N_p} w_i \bm{r}_i(t)}
         {\sum_{i=1}^{N_p} w_i}.
\end{equation}
Since all particles are initialized identically
and no inter-particle forces are present,
the centroid trajectory is equivalent to the trajectory
of the corresponding single test particle.

\begin{figure}[!ht]
\centering
\includegraphics[width=0.45\textwidth]{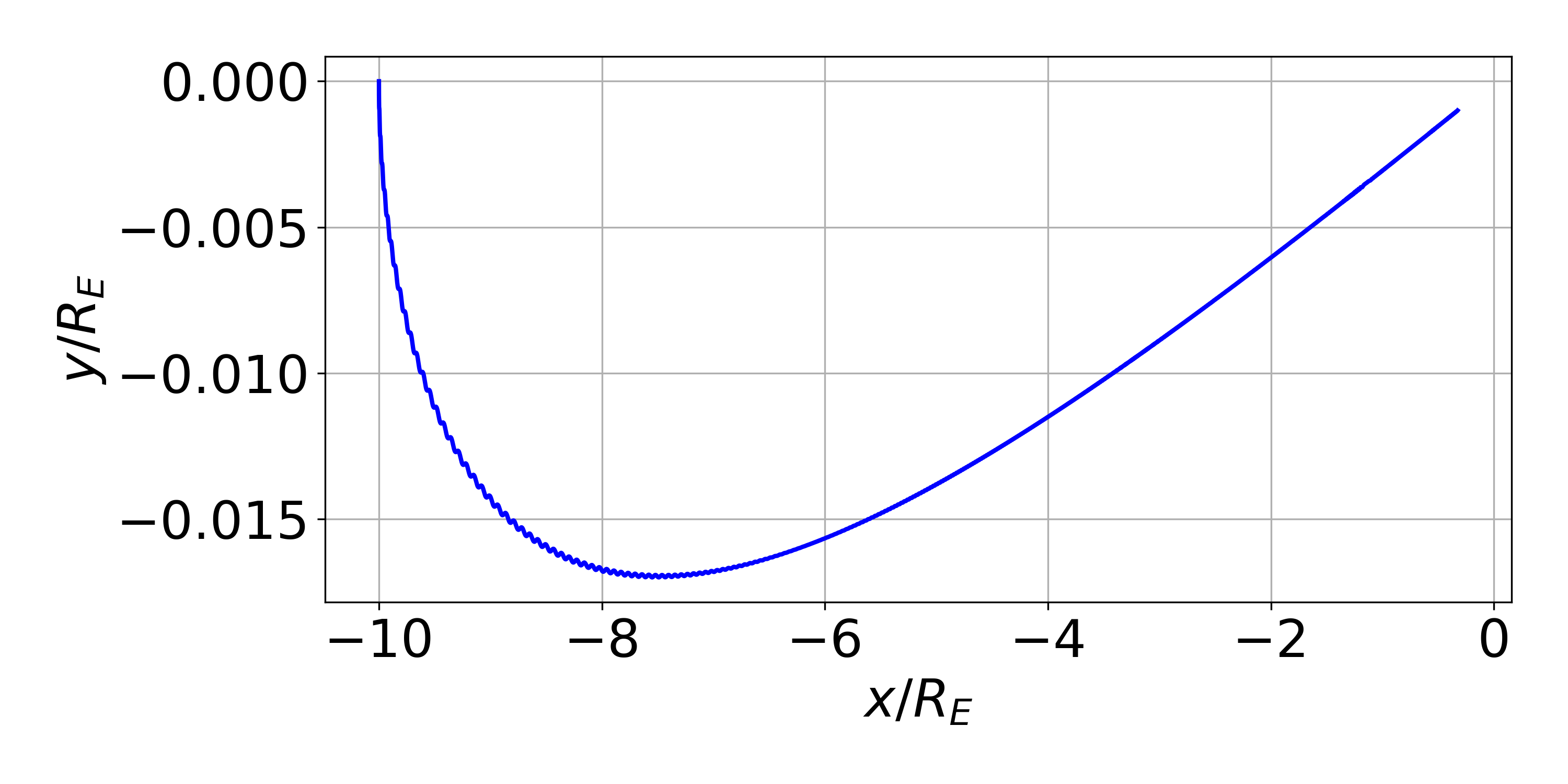}
\includegraphics[width=0.45\textwidth]{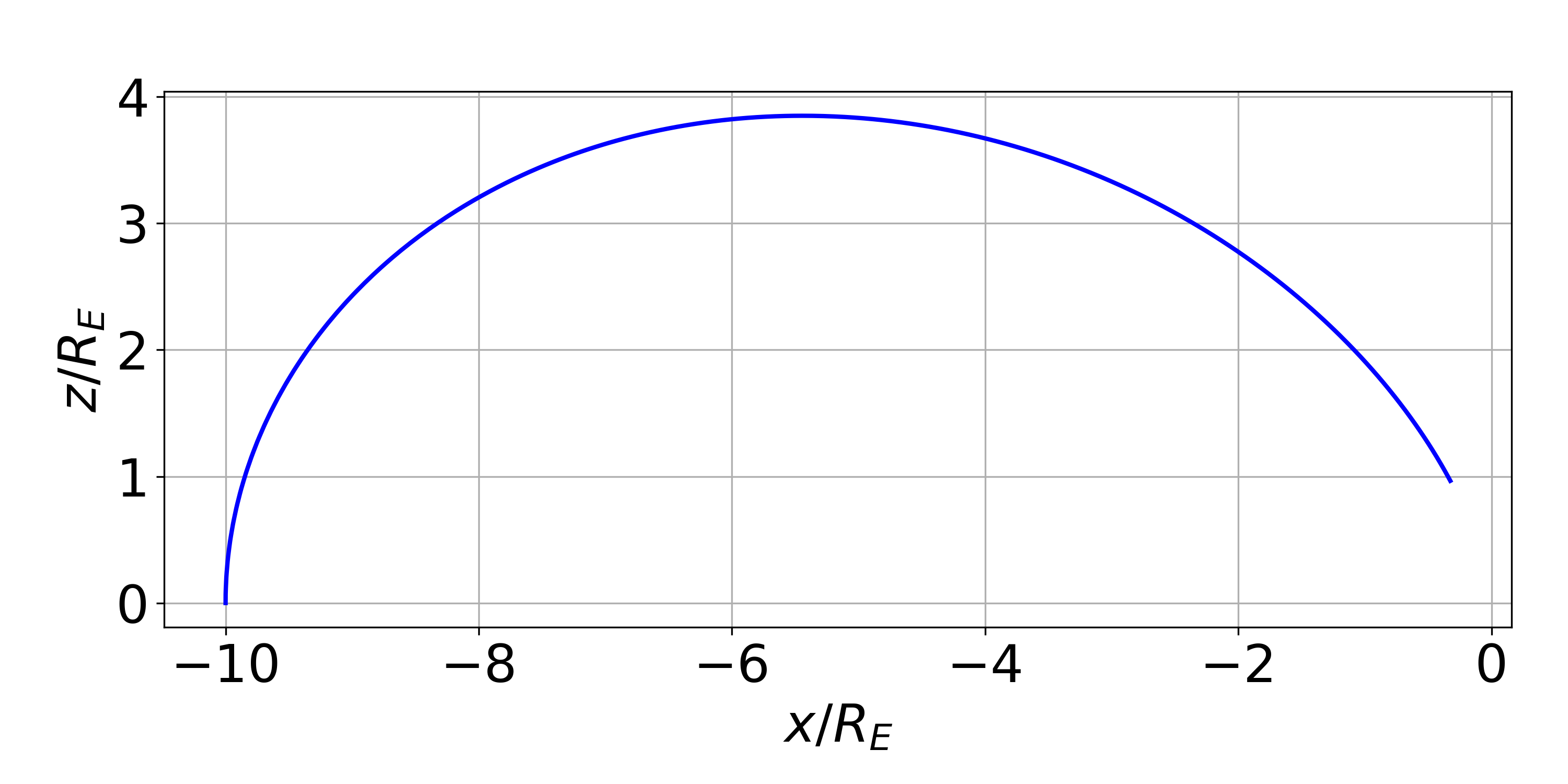}
\includegraphics[width=0.45\textwidth]{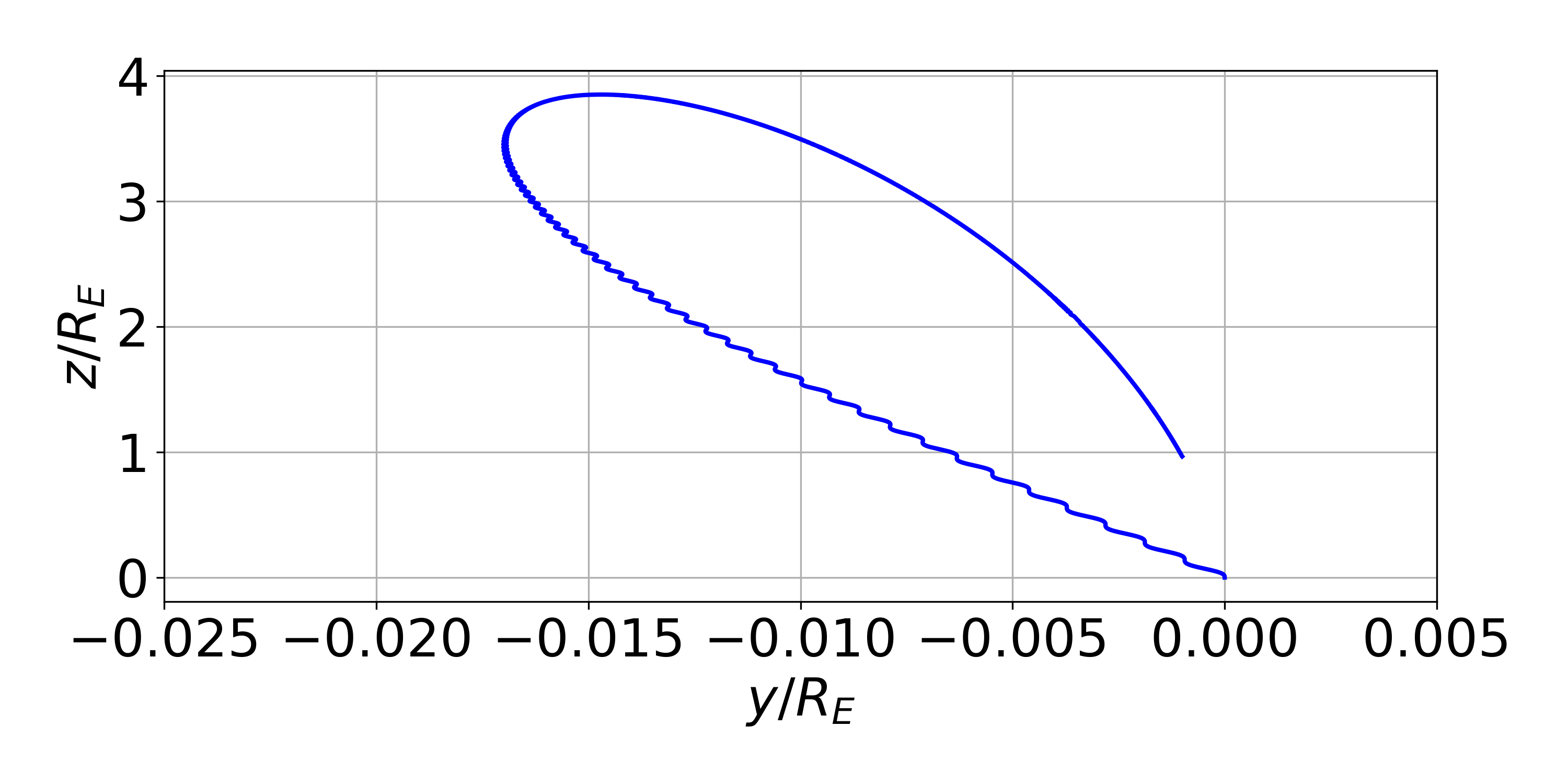}
\caption{
Trajectory projections of the single-particle benchmark
in the prescribed dipole geomagnetic field:
$x$--$y$ (top), $x$--$z$ (middle), and $y$--$z$ (bottom).
}
\label{fig:traj}
\end{figure}

\begin{figure}[!ht]
\centering
\includegraphics[width=0.45\textwidth]{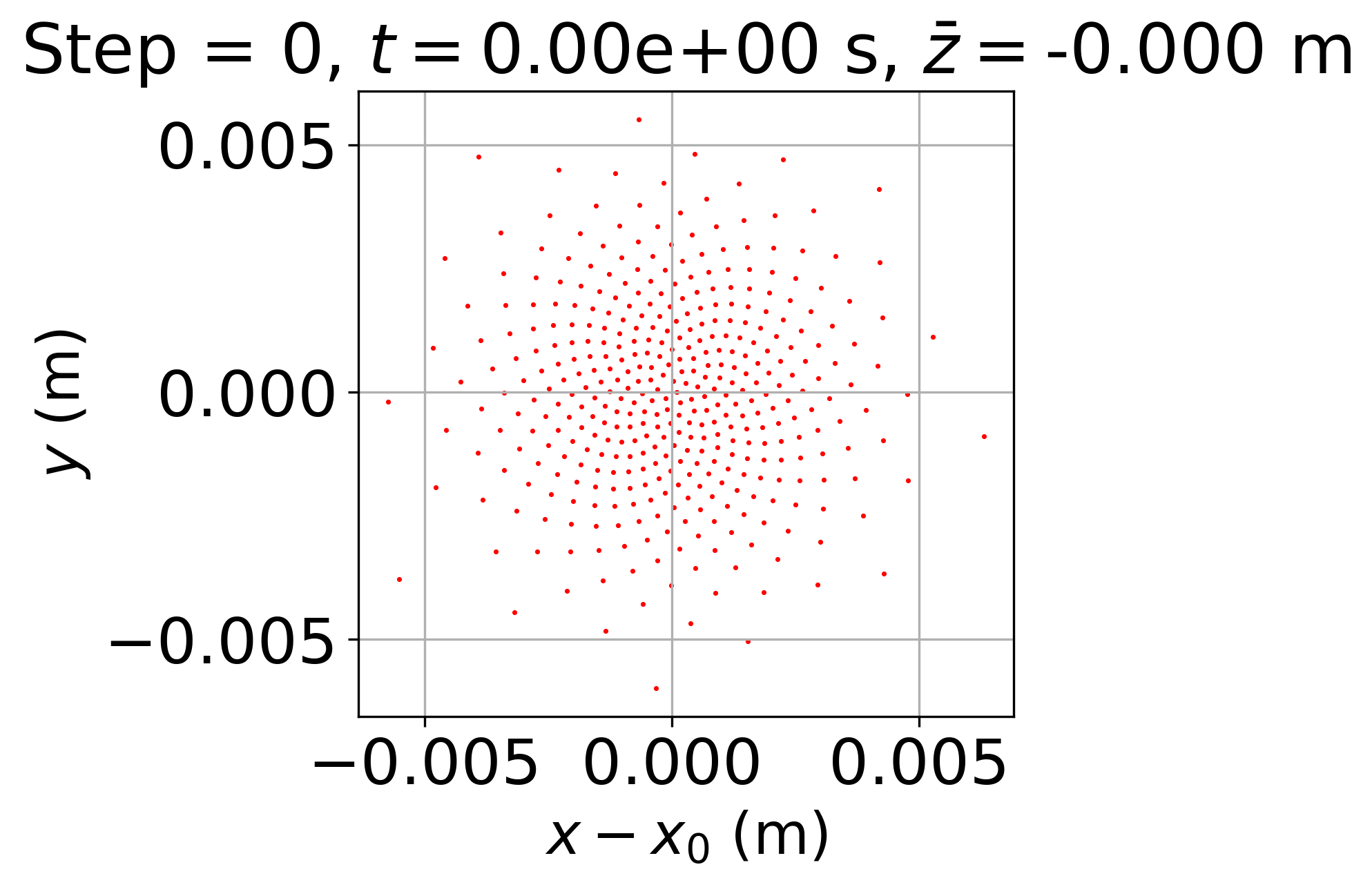}
\includegraphics[width=0.45\textwidth]{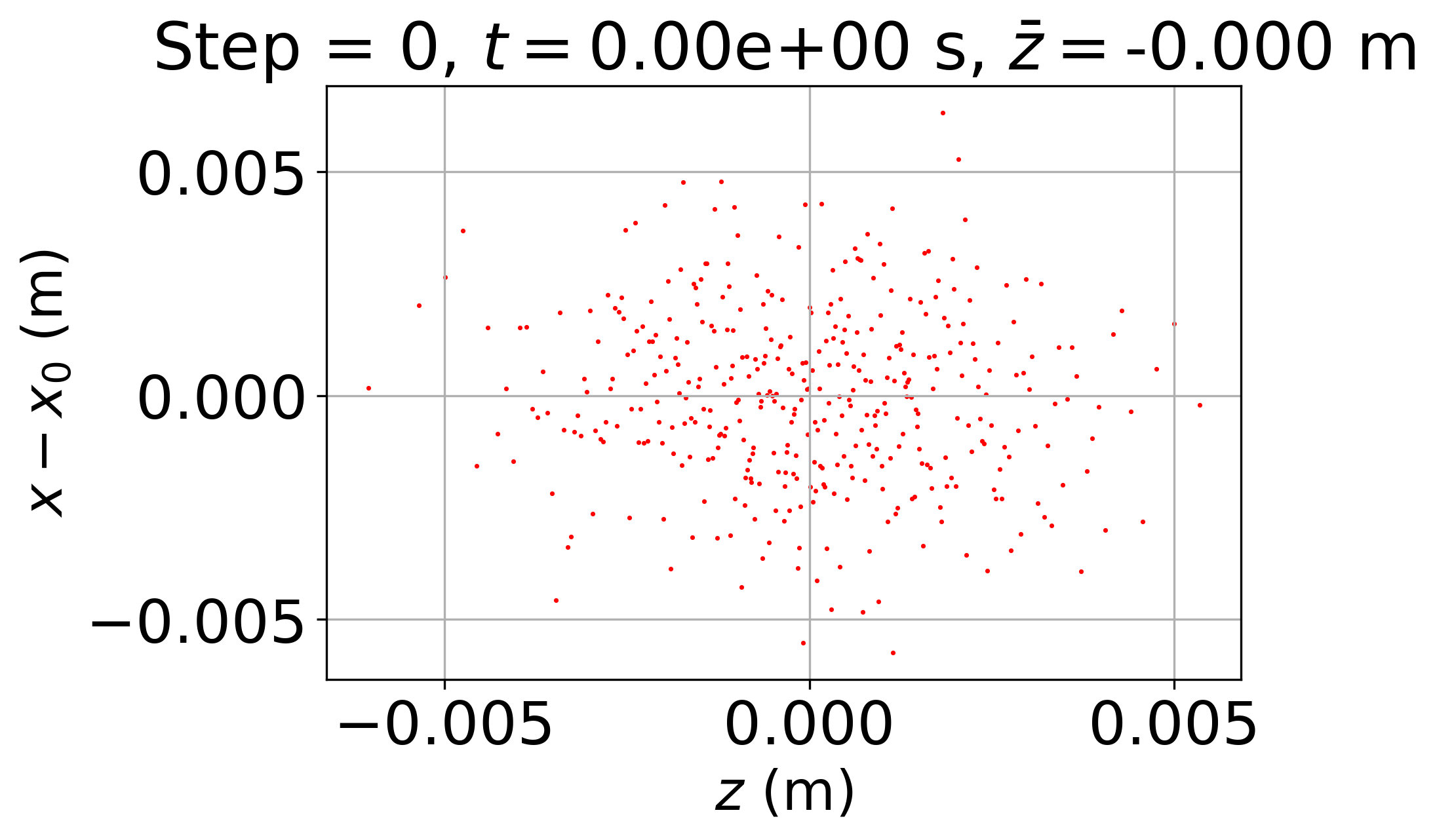}
\caption{
Initial particle distribution of the relativistic electron bunch
for the case with $N_p=400$.
The upper panel shows the transverse distribution in the
$(x-x_0)$--$y$ plane,
and the lower panel shows the longitudinal distribution in the
$z$--$(x-x_0)$ plane.
}
\label{fig:initdis}
\end{figure}

\begin{figure}[!ht]
\centering
\includegraphics[width=0.45\textwidth]{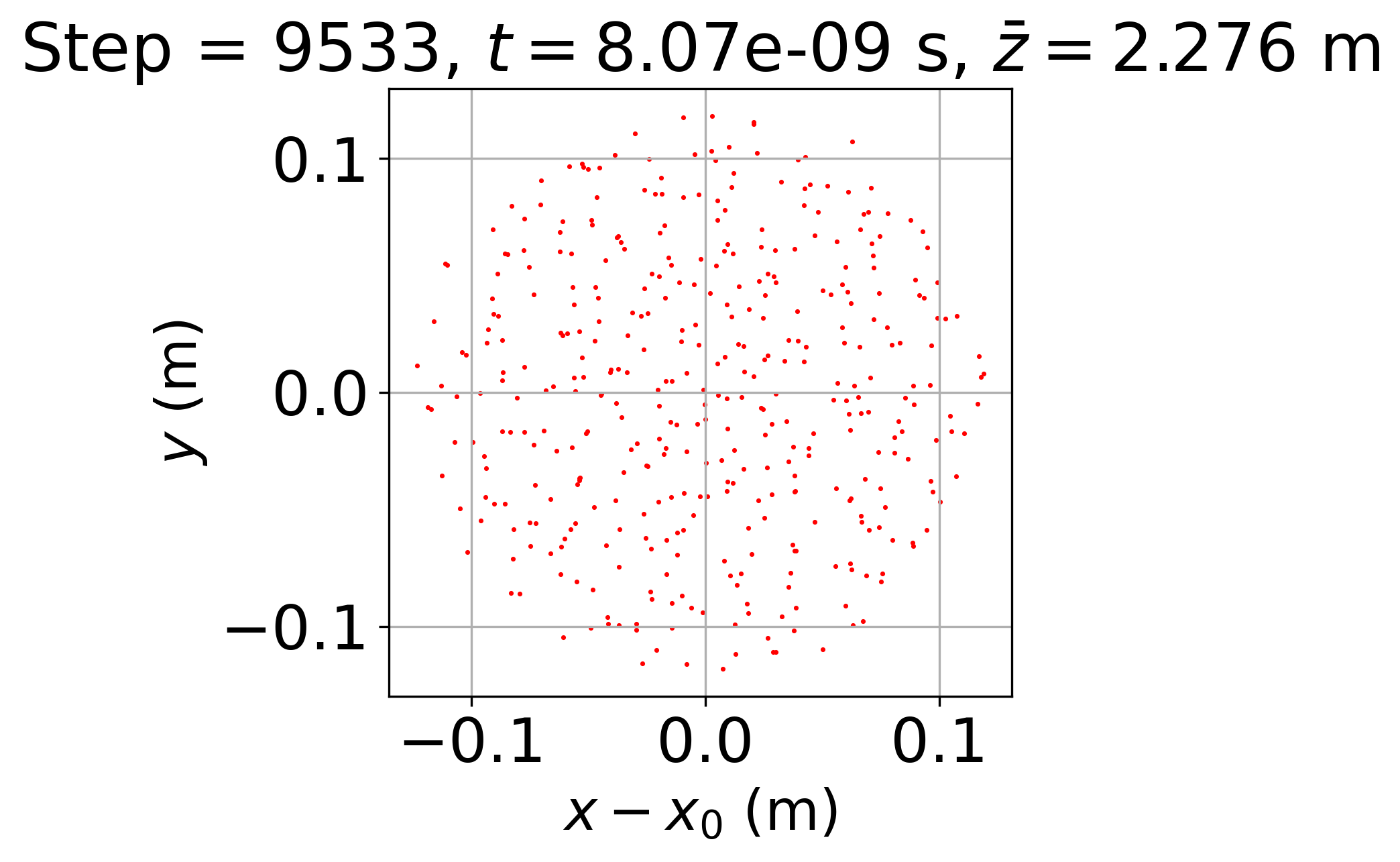}
\includegraphics[width=0.45\textwidth]{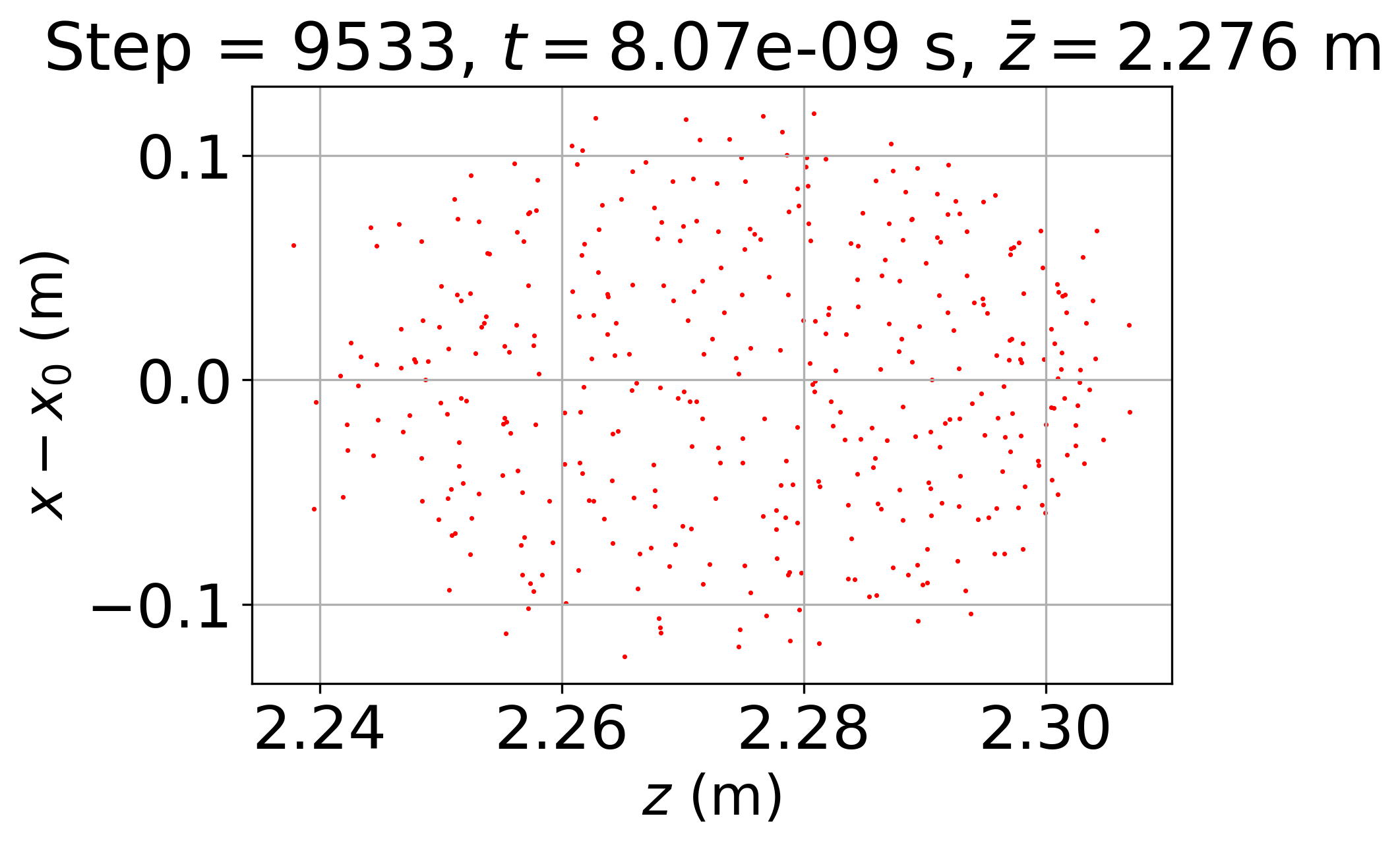}
\caption{
Particle distribution of the relativistic electron bunch
at step 9533 for the case with $N_p=400$
($t=8.07\times10^{-9}\,\mathrm{s}$, $\bar{z}=2.276\,\mathrm{m}$).
The upper panel shows the transverse distribution in the
$(x-x_0)$--$y$ plane,
and the lower panel shows the longitudinal distribution in the
$z$--$(x-x_0)$ plane.
}
\label{fig:dis9533400}
\end{figure}

\begin{figure}[!ht]
\centering
\includegraphics[width=0.45\textwidth]{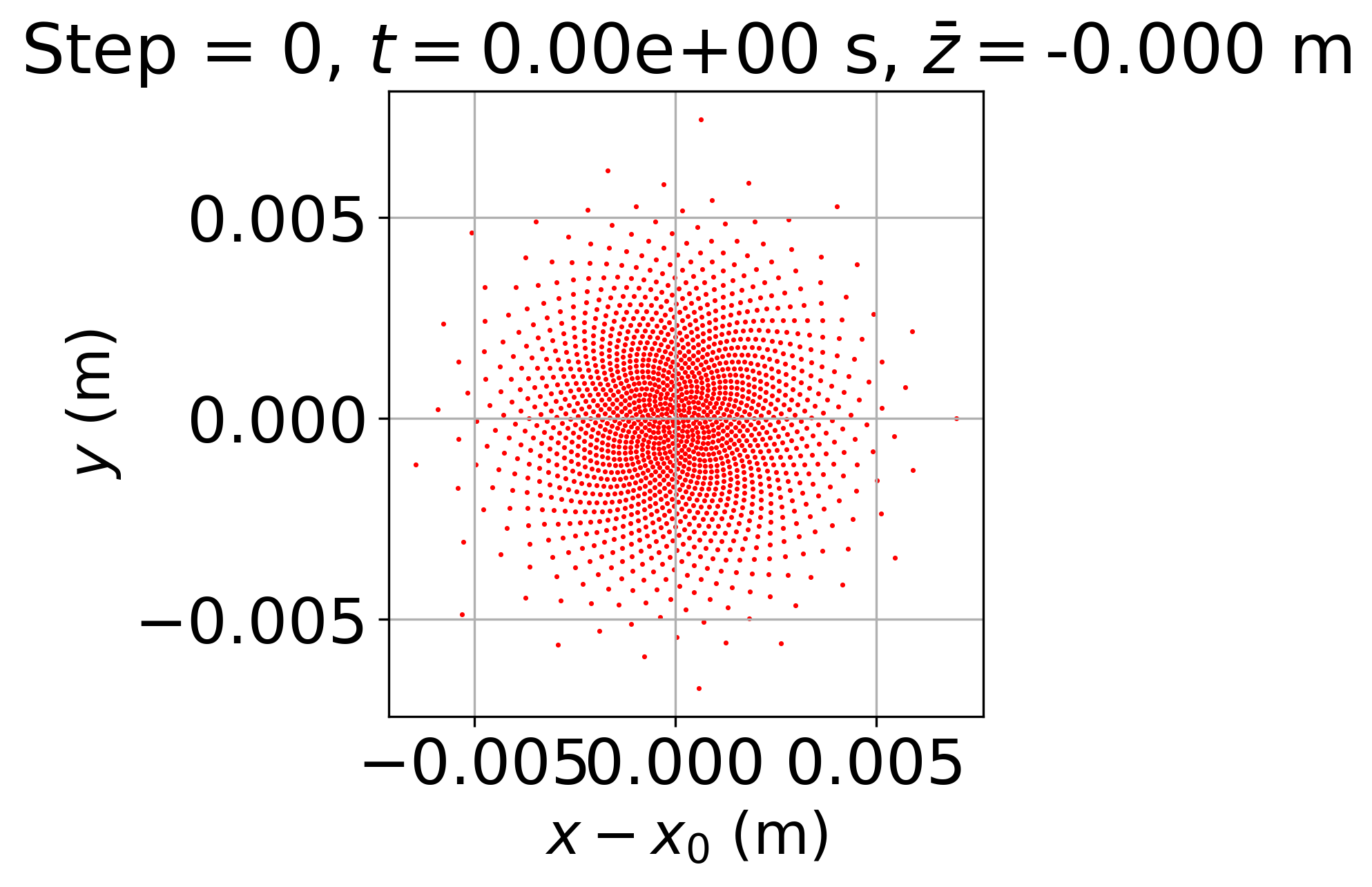}
\includegraphics[width=0.45\textwidth]{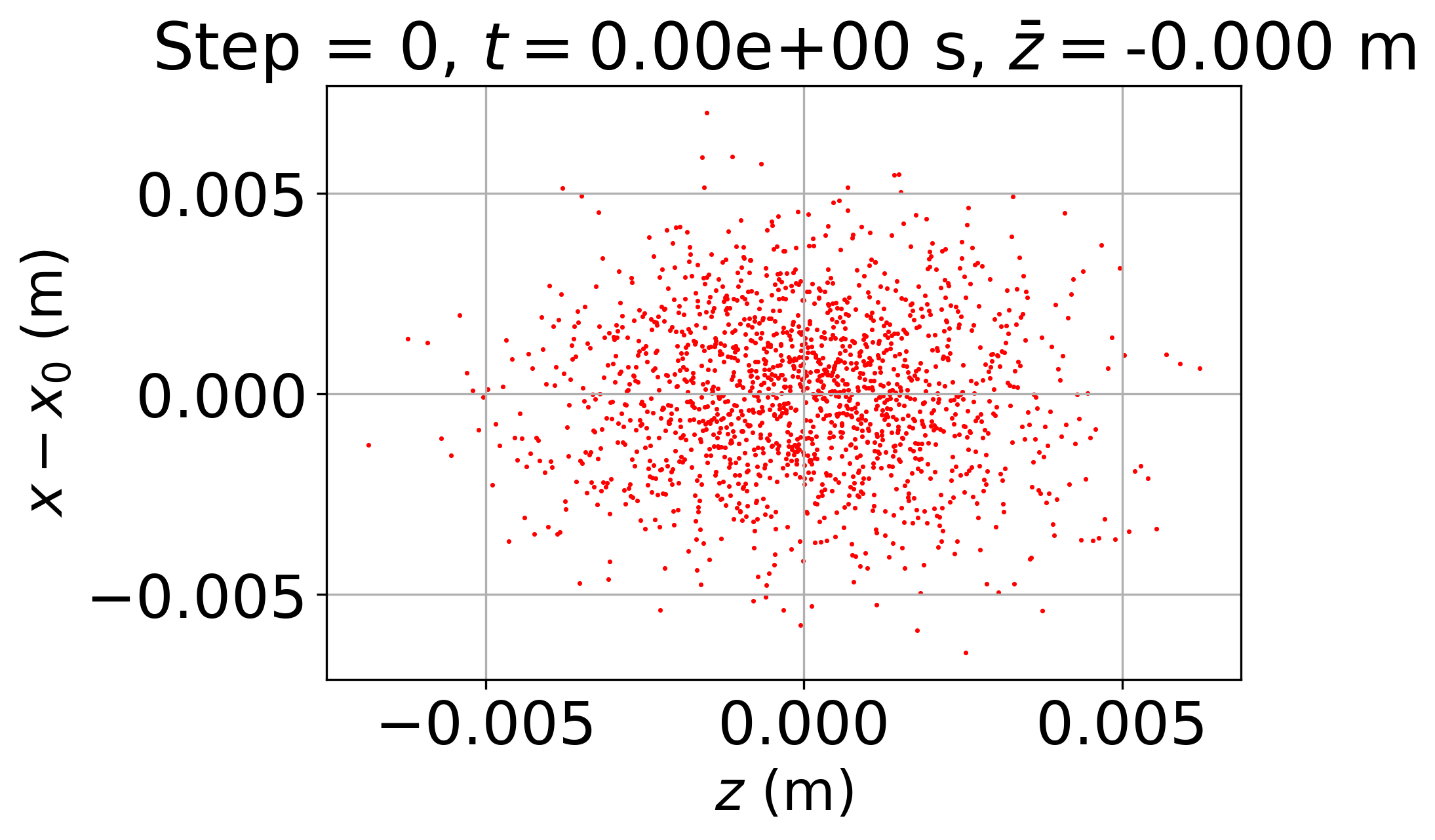}
\caption{
Initial particle distribution of the relativistic electron bunch
for the case with $N_p=1600$.
The upper panel shows the transverse distribution in the
$(x-x_0)$--$y$ plane,
and the lower panel shows the longitudinal distribution in the
$z$--$(x-x_0)$ plane.
}
\label{fig:initdis_1600}
\end{figure}

\begin{figure}[!ht]
\centering
\includegraphics[width=0.45\textwidth]{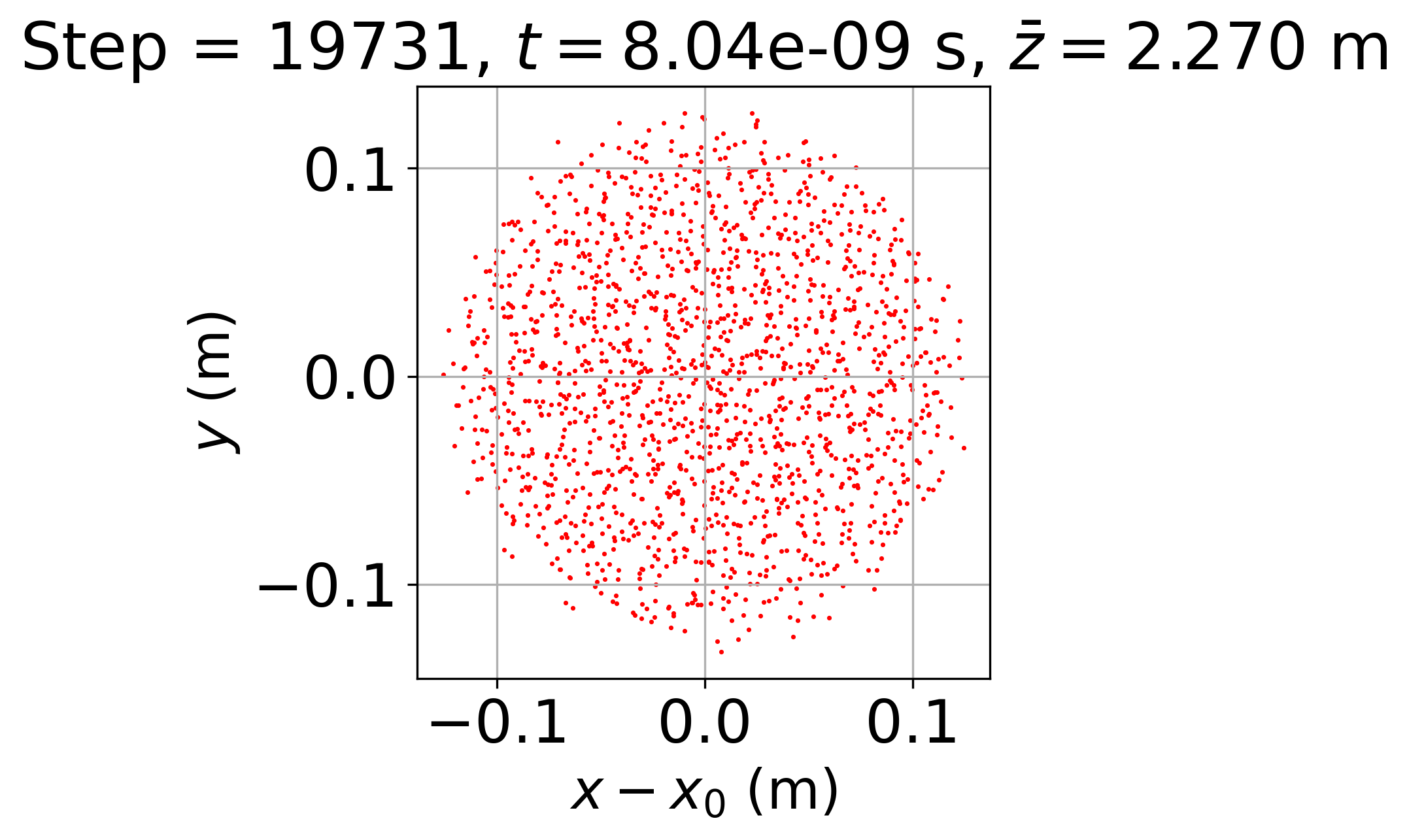}
\includegraphics[width=0.45\textwidth]{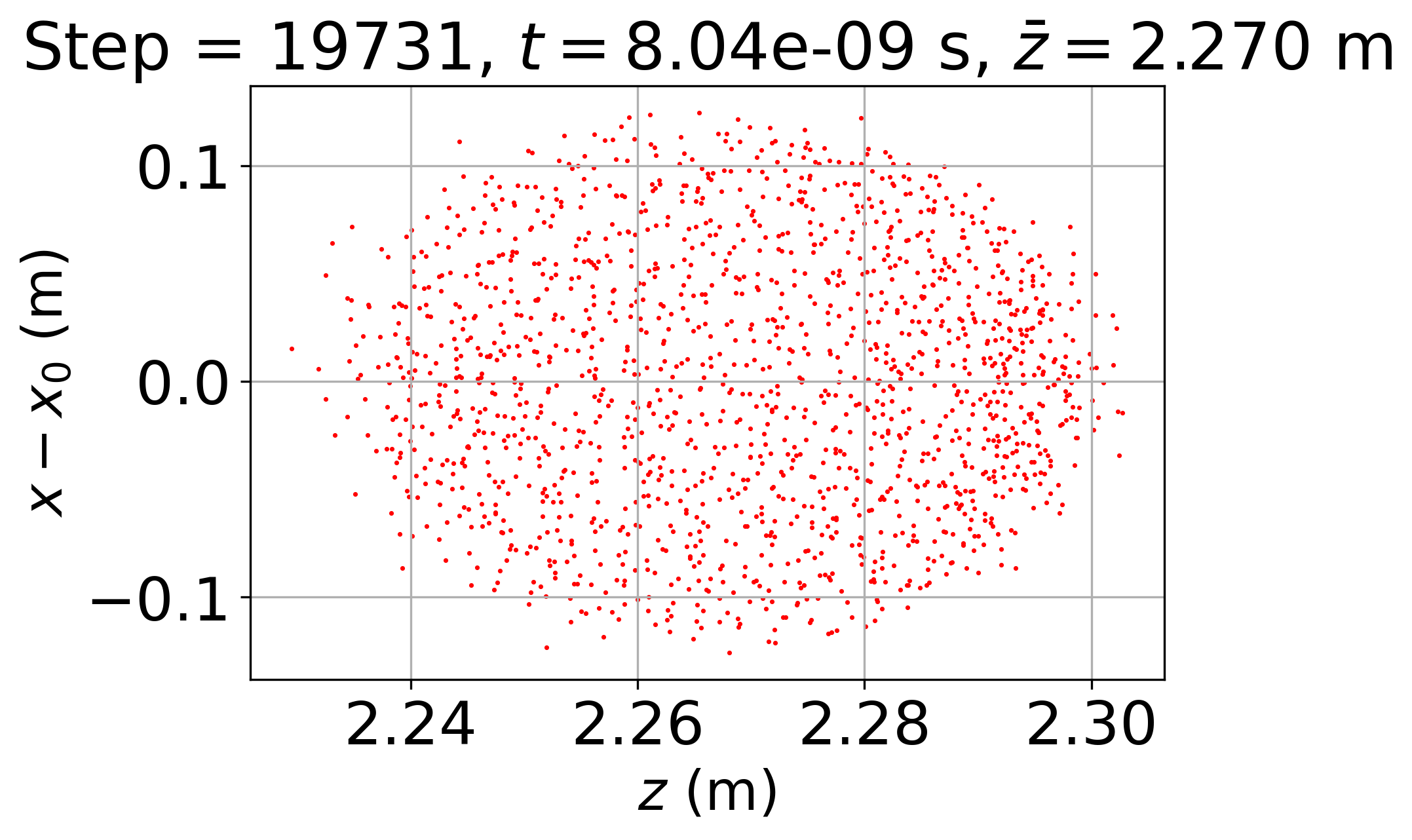}
\caption{
Particle distribution of the relativistic electron bunch
at step 19731 for the case with $N_p=1600$
($t=8.04\times10^{-9}\,\mathrm{s}$, $\bar{z}=2.270\,\mathrm{m}$).
The upper panel shows the transverse distribution in the
$(x-x_0)$--$y$ plane,
and the lower panel shows the longitudinal distribution in the
$z$--$(x-x_0)$ plane.
}
\label{fig:dis19731}
\end{figure}

\begin{figure}[!ht]
\centering
\includegraphics[width=0.5\textwidth]{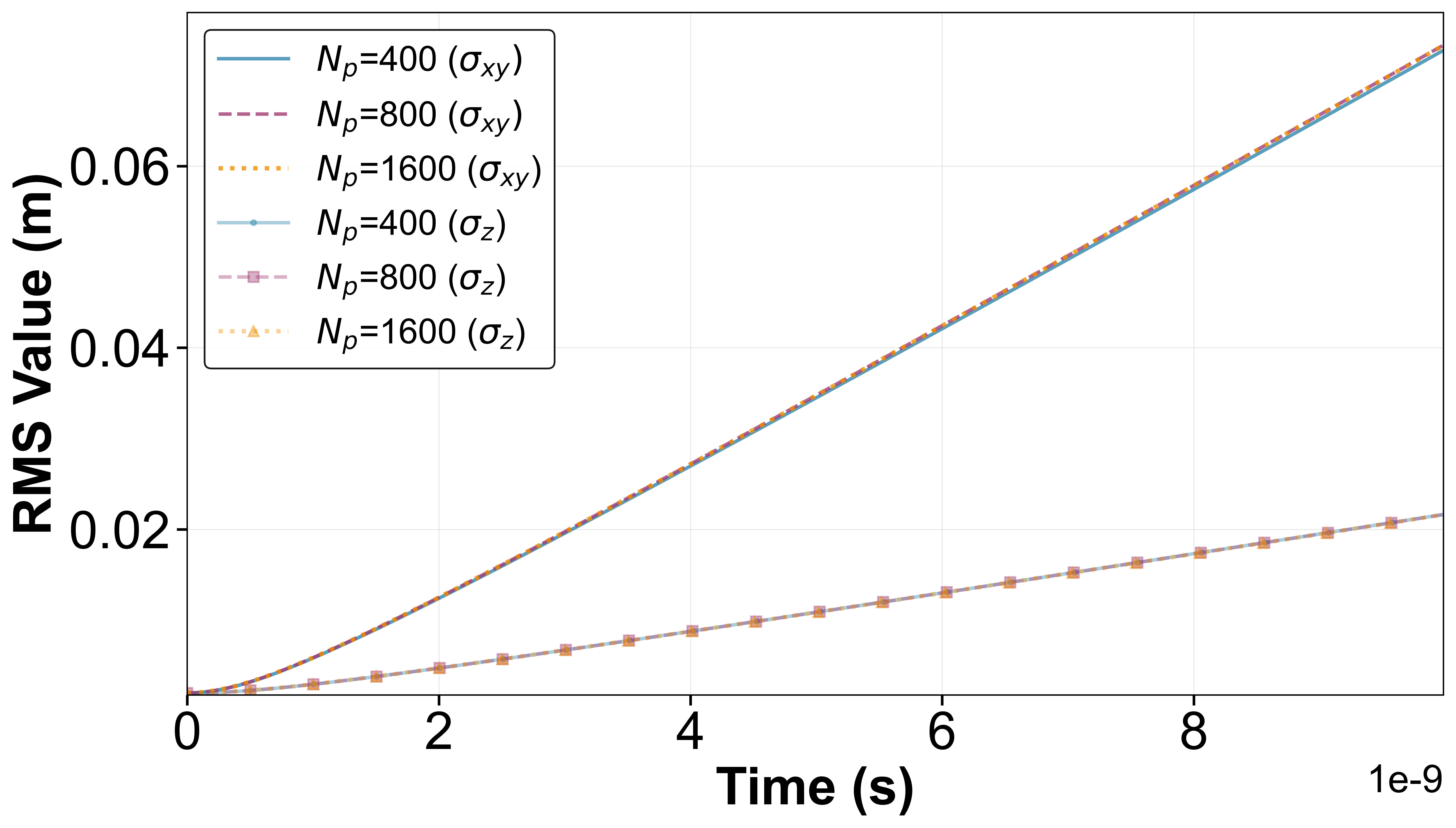}
\caption{
Time evolution of the transverse and longitudinal RMS bunch sizes
for the cases with $N_p=400$, $800$, and $1600$.
Here $\sigma_{xy}$ denotes the transverse RMS size
and $\sigma_z$ denotes the longitudinal RMS size.
}
\label{fig:RMS}
\end{figure}

\begin{figure}[ht]
  \centering
  \includegraphics[width=0.5\textwidth]{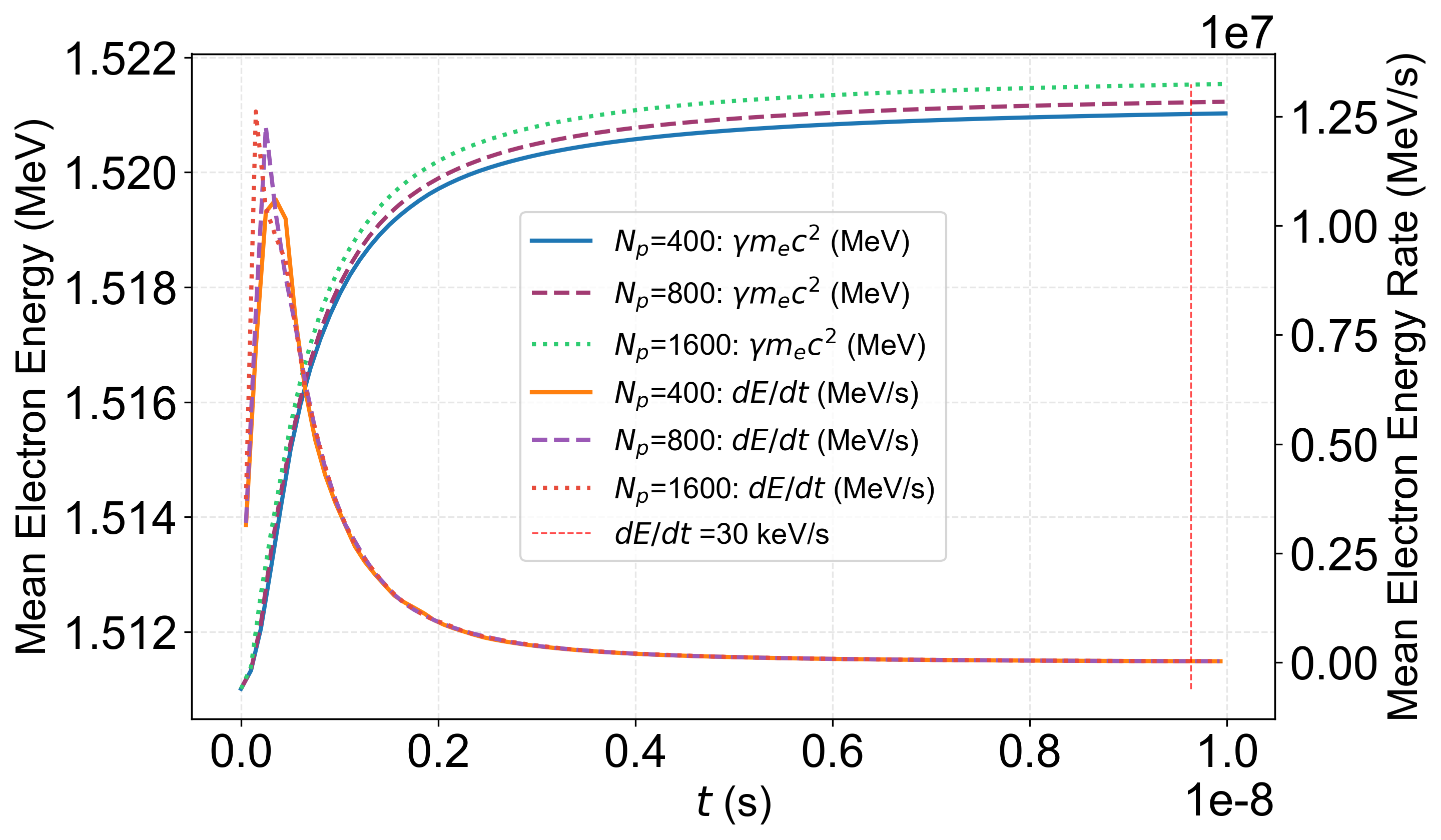}
 \caption{
Mean electron energy and its time derivative
for the cases with $N_p=400$, $800$, and $1600$.
The threshold
$\left|d\bar{E}/dt\right|<30~\mathrm{keV/s}$
is marked using a vertical line.
}
  \label{fig:energy}
\end{figure}

The resulting trajectory projections in the $x$--$y$,
$x$--$z$, and $y$--$z$ planes are shown in Fig.~\ref{fig:traj}.
The simulated path agrees well with the trajectory reported by
Powis et al.~\cite{Powis},
indicating that the present implementation correctly reproduces
the large-scale guiding-center motion in the prescribed geomagnetic field.
This benchmark therefore validates the background-field model
and the relativistic particle pusher independently of the collective
self-field effects considered in the subsequent bunch calculations.

Having verified the single-particle transport,
the simulations are next extended to the full relativistic electron bunch,
for which inter-particle electromagnetic interactions
lead to rapid early-stage expansion and modify the later long-range propagation.
\subsection{Macroparticle-Number Sensitivity}\label{sec:Np}

To evaluate the influence of macroparticle resolution
on the predicted bunch dynamics,
simulations are carried out with three different particle numbers,
namely
\begin{equation}
    N_p = 400,\ 800,\ 1600.
\end{equation}
All other parameters remain identical to the baseline case
introduced in Sec.~\ref{sec:baseline}.
The purpose of this comparison is to assess
whether the early self-field-driven bunch expansion
and the associated statistical quantities
are sensitive to the number of macroparticles used
to represent the relativistic electron bunch.

For the reference case with $N_p=400$,
the initial particle distribution is shown in
Fig.~\ref{fig:initdis}.
The particles are tightly confined within a narrow spatial range,
consistent with the Gaussian phase-space sampling described in
Sec.~\ref{sec:setup}.
At this stage, the bunch number density is high,
and the inter-particle electromagnetic interaction
is expected to be strongest.
As the simulation proceeds,
the bunch undergoes rapid transverse expansion
together with a noticeable longitudinal stretching.
An example of the evolved particle distribution
for the $N_p=400$ case is shown in Fig.~\ref{fig:dis9533400},
where the transverse extent has increased substantially
relative to the initial state.

To examine the effect of increasing particle number,
the same calculation is repeated for
$N_p=800$ and $N_p=1600$.
A representative initial distribution for the
$N_p=1600$ case is shown in Fig.~\ref{fig:initdis_1600},
and an evolved distribution is shown in Fig.~\ref{fig:dis19731}.
As expected, larger $N_p$ values provide a denser sampling
of the prescribed Gaussian distribution,
so the particle cloud appears smoother
and less affected by sampling noise.
However, despite these visual differences,
the overall expansion pattern remains similar for all three cases,
indicating that the dominant physical mechanism
is not altered by the choice of $N_p$.

A more quantitative comparison is made using the transverse
and longitudinal RMS beam sizes,
$\sigma_{xy}$ and $\sigma_z$,
defined in Sec.~\ref{sec:setup}.
Their time evolution is shown in Fig.~\ref{fig:RMS}.
For all three particle numbers,
the bunch exhibits the same qualitative behavior:
both RMS sizes increase with time,
while the transverse radius grows much faster
than the longitudinal one.
This trend reflects the strong early transverse repulsion
within the dense non-neutral bunch.
Most importantly, the curves corresponding to
$N_p=400$, $800$, and $1600$
remain close to one another throughout the simulation,
which demonstrates that the predicted bunch expansion
has largely converged with respect to the macroparticle number.

Further evidence is provided by the evolution of the mean particle energy,
shown in Fig.~\ref{fig:energy}.
In all three cases,
the mean energy first increases rapidly
and then approaches a much slower growth stage.
This behavior indicates that the initial inter-particle potential energy
is progressively converted into kinetic energy during the early expansion,
after which the self-field effect becomes weak.
The close agreement among the three cases confirms
that the energy transfer process is also only weakly dependent
on the macroparticle number.

Based on these comparisons,
it is concluded that the case with $N_p=400$
already captures the essential collective dynamics
with acceptable accuracy,
while larger particle numbers mainly improve
the smoothness of the sampled distribution
and reduce statistical fluctuations.
Therefore, $N_p=400$ provides a reasonable compromise
between numerical cost and accuracy for the baseline calculations,
whereas $N_p=1600$ is retained in selected cases to provide additional confidence in the long-range transport results.

\subsection{Time-Step Adaptation and Switching Criterion}\label{sec:timestep}

The adaptive time-step rule introduced in Sec.~\ref{sec:adaptive_dt}
is examined here for the fully coupled early-stage calculations.
At injection, the bunch is still dense and the inter-particle electromagnetic forces change rapidly.
The time step is therefore limited to
\begin{equation}
    \Delta t_{\min} = 10^{-15}~\mathrm{s},
    \qquad
    \Delta t_{\max} = 10^{-12}~\mathrm{s}.
\end{equation}
The lower bound provides sufficient temporal resolution during the strongest self-field-driven expansion,
whereas the upper bound avoids unnecessary loss of accuracy once the dynamics become smoother.

\begin{figure}[!ht]
\centering
\includegraphics[width=0.5\textwidth]{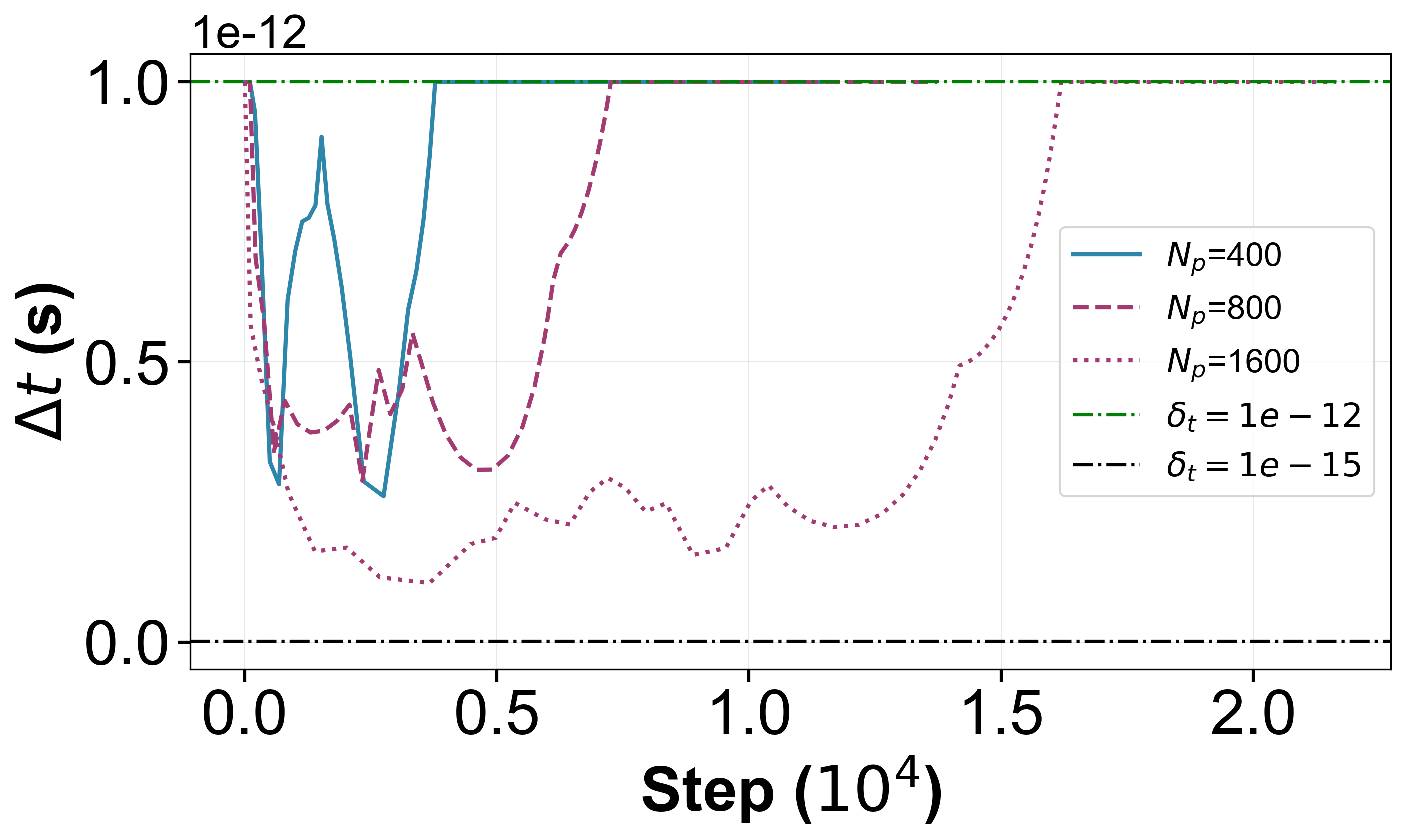}
\caption{
Evolution of the adaptive time step for the cases with
$N_p=400$, $800$, and $1600$ during the early fully coupled calculation.
The time step remains within the prescribed bounds
$\Delta t_{\min}=10^{-15}\,\mathrm{s}$ and
$\Delta t_{\max}=10^{-12}\,\mathrm{s}$,
shown by the dash-dotted lines.
}
\label{fig:Time}
\end{figure}

The resulting time-step histories are shown in Fig.~\ref{fig:Time}
for different macroparticle numbers.
In all cases, the selected time step stays within the prescribed bounds.
At early times, it remains close to $\Delta t_{\min}$ in order to resolve the rapid expansion and strong pairwise interaction.
As the bunch expands and the inter-particle forces weaken, larger time steps become possible.
The similar time-step histories obtained for different $N_p$ indicate that the adaptive strategy is robust and mainly reflects the intrinsic bunch dynamics rather than sampling noise.

To determine when inter-particle interaction can be neglected in the long-range stage,
we monitor the mean kinetic energy of the bunch.
The mean energy increases rapidly at early times and then approaches a weak-growth regime,
indicating that the initial self-field energy has largely been converted into particle kinetic energy.
Once the bunch has expanded sufficiently, the particles are well separated and the remaining inter-particle interaction becomes much less important for the subsequent transport.

The switching criterion is based on the time derivative of the mean energy.
In the present work, we use
\begin{equation}
    \left|
    \frac{d\bar{E}}{dt}
    \right|
    < 30~\mathrm{keV/s}
\end{equation}
as a reference condition.
This threshold is not taken as the exact switching point.
Instead, it marks the onset of a weakly relaxing regime.
The actual switching time is chosen later so that the residual self-field effect has decayed further and the subsequent transport is dominated by the prescribed geomagnetic field.

\begin{figure}[!ht]
\centering
\includegraphics[width=0.45\textwidth]{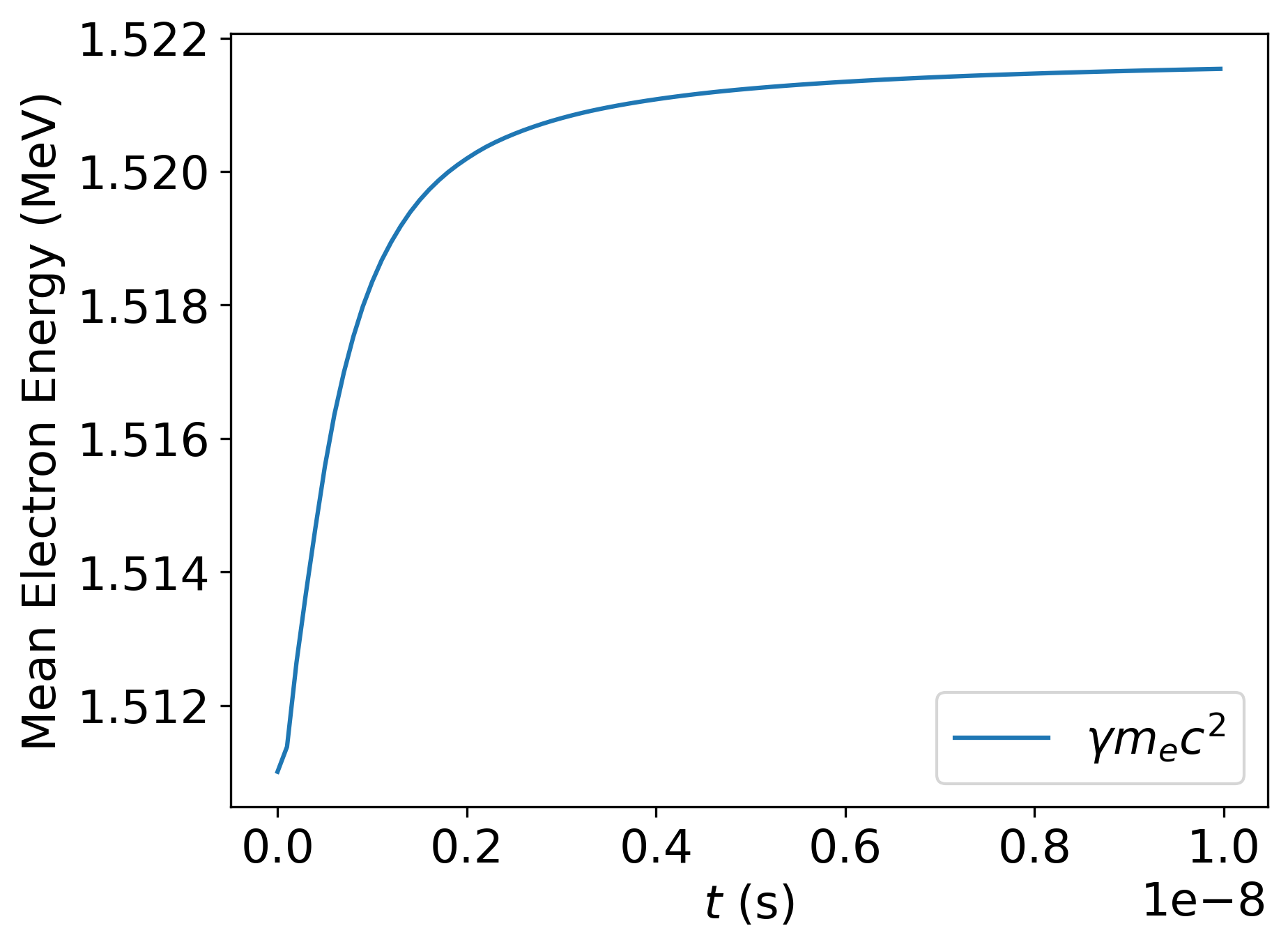}
\includegraphics[width=0.45\textwidth]{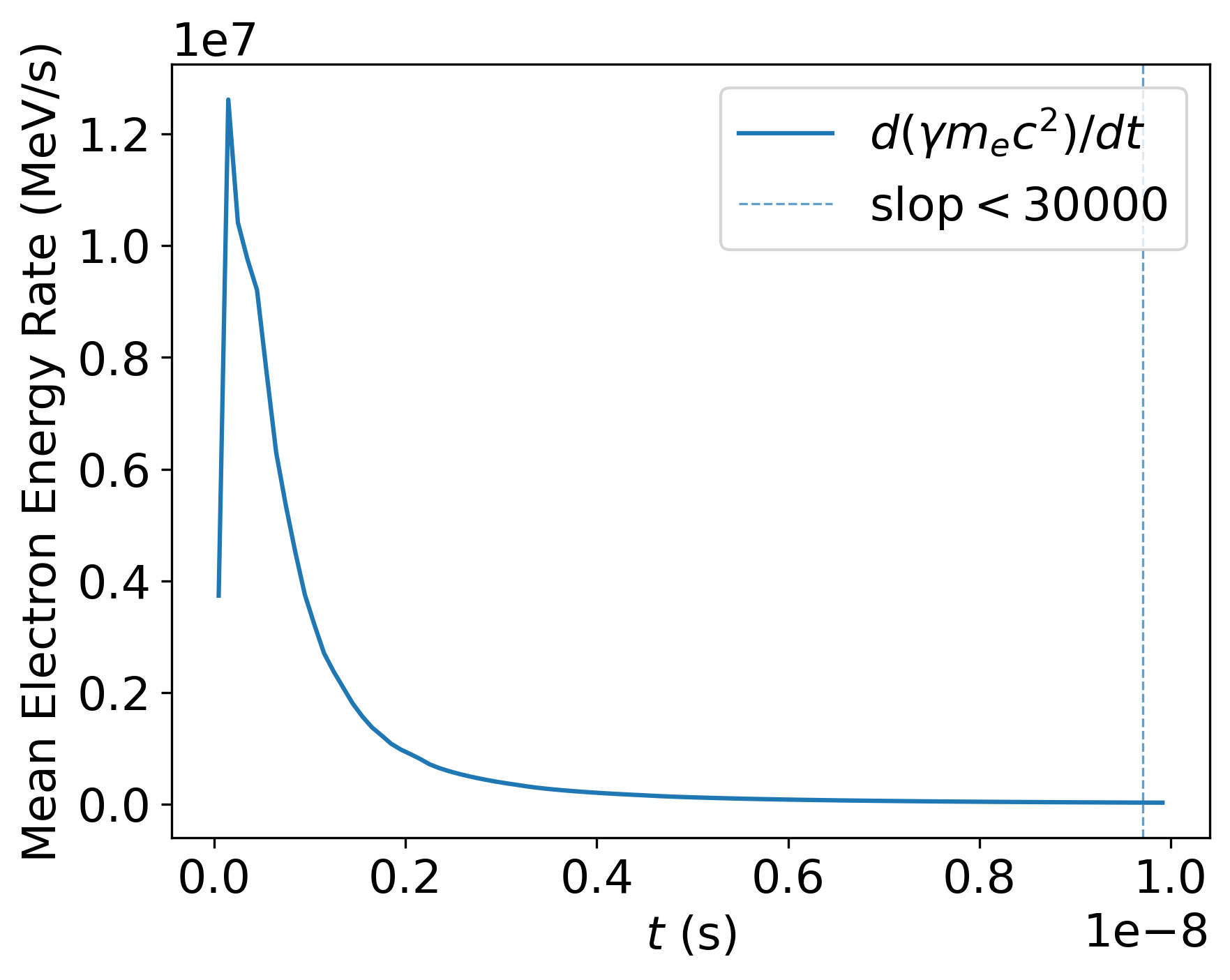}
\caption{
Mean electron energy and its time derivative
for the case with $N_p=1600$ and $E_0=1~\mathrm{MeV}$.
The upper panel shows the evolution of the mean energy,
while the lower panel shows the corresponding energy growth rate.
The dashed vertical line marks 
$\left|d\bar{E}/dt\right|=30~\mathrm{keV/s}$.
}
\label{fig:energy_1600_1Mev}
\end{figure}

\begin{figure}[!ht]
\centering
\includegraphics[width=0.45\textwidth]{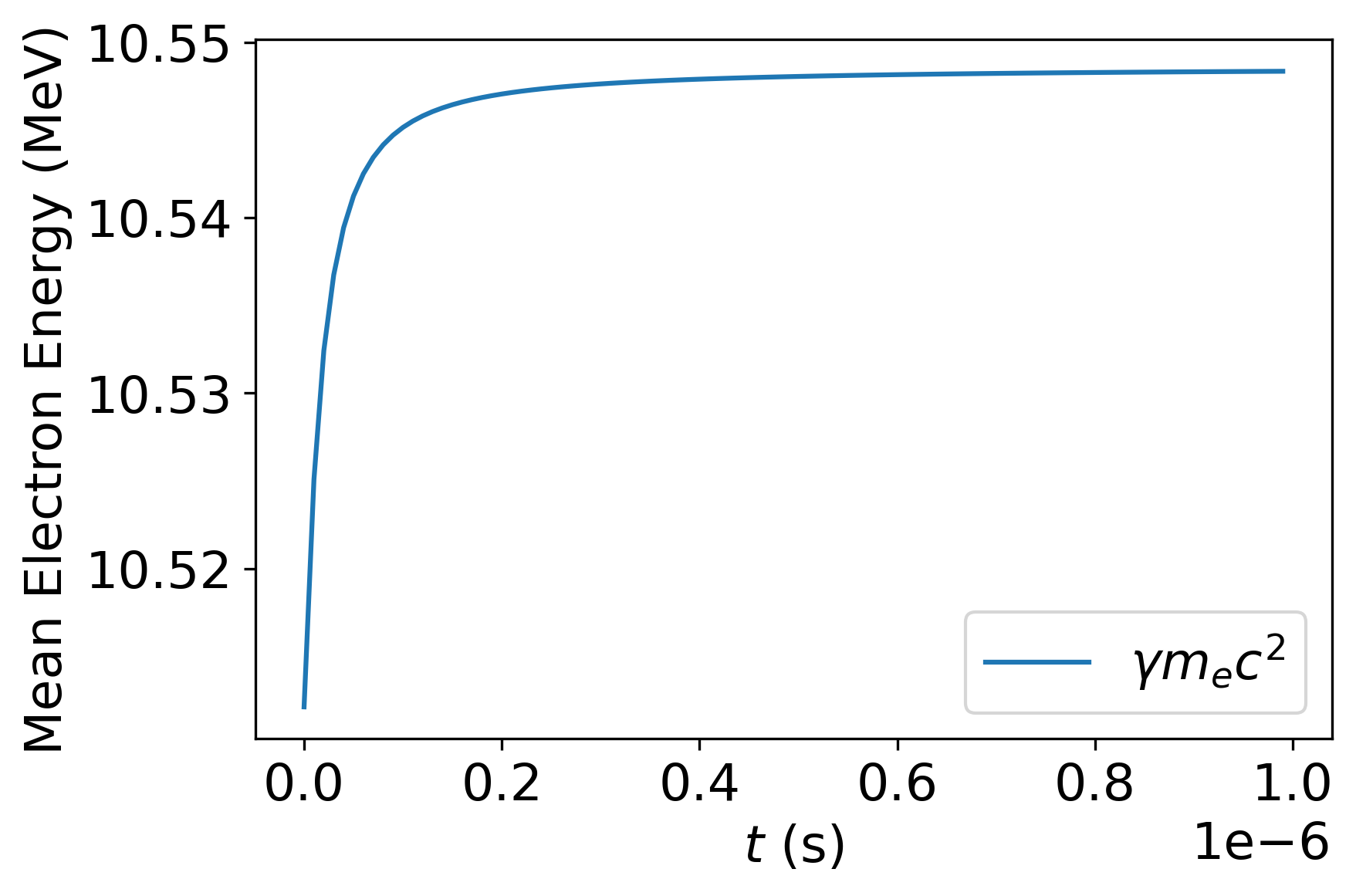}
\includegraphics[width=0.45\textwidth]{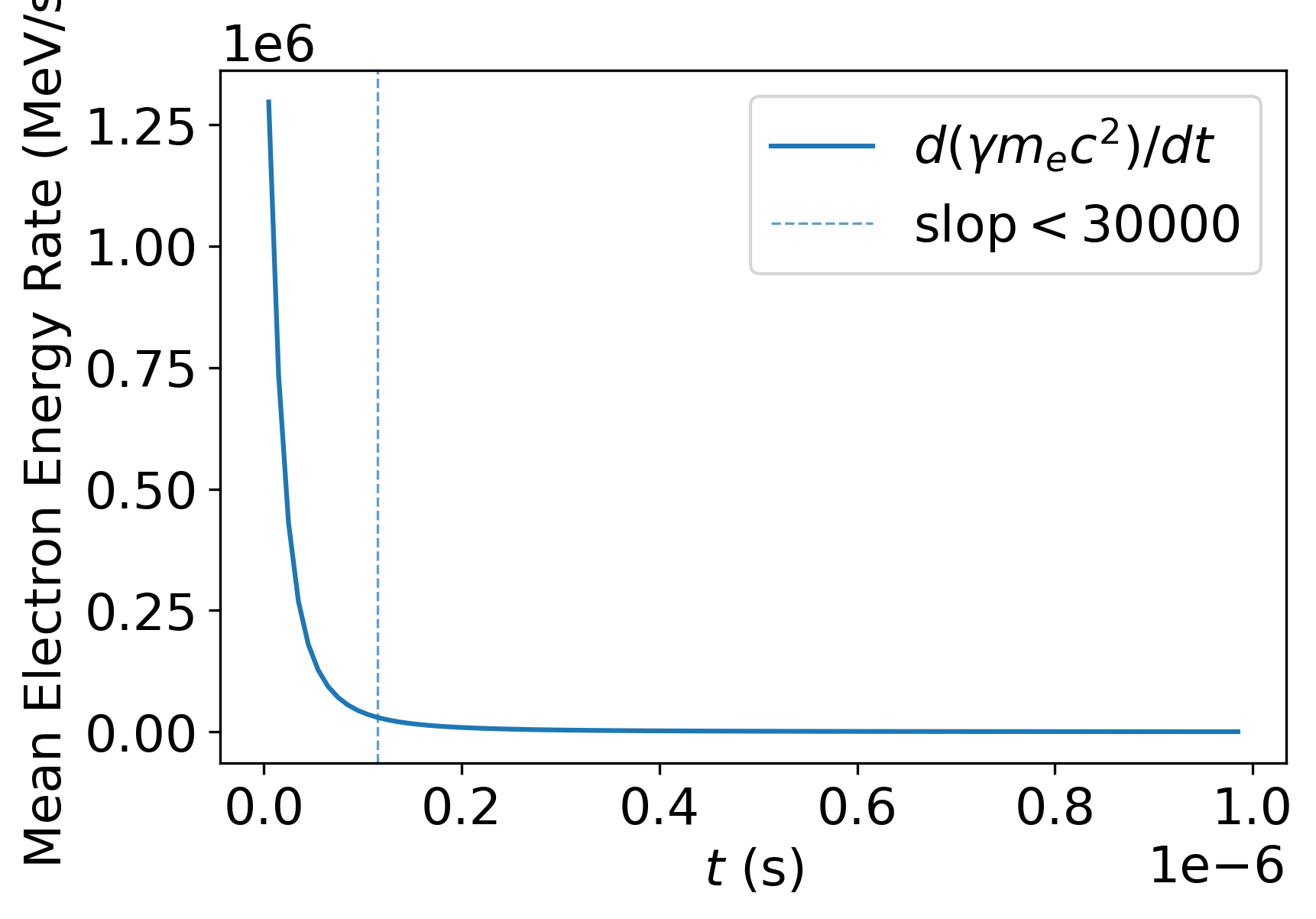}
\caption{
Mean electron energy and its time derivative
for the case with $N_p=1600$ and $E_0=10~\mathrm{MeV}$.
The upper panel shows the evolution of the mean energy,
while the lower panel shows the corresponding energy growth rate.
The dashed vertical line marks 
$\left|d\bar{E}/dt\right|=30~\mathrm{keV/s}$.
}
\label{fig:energy_1600_10Mev}
\end{figure}

\begin{figure}[!ht]
\centering
\includegraphics[width=0.45\textwidth]{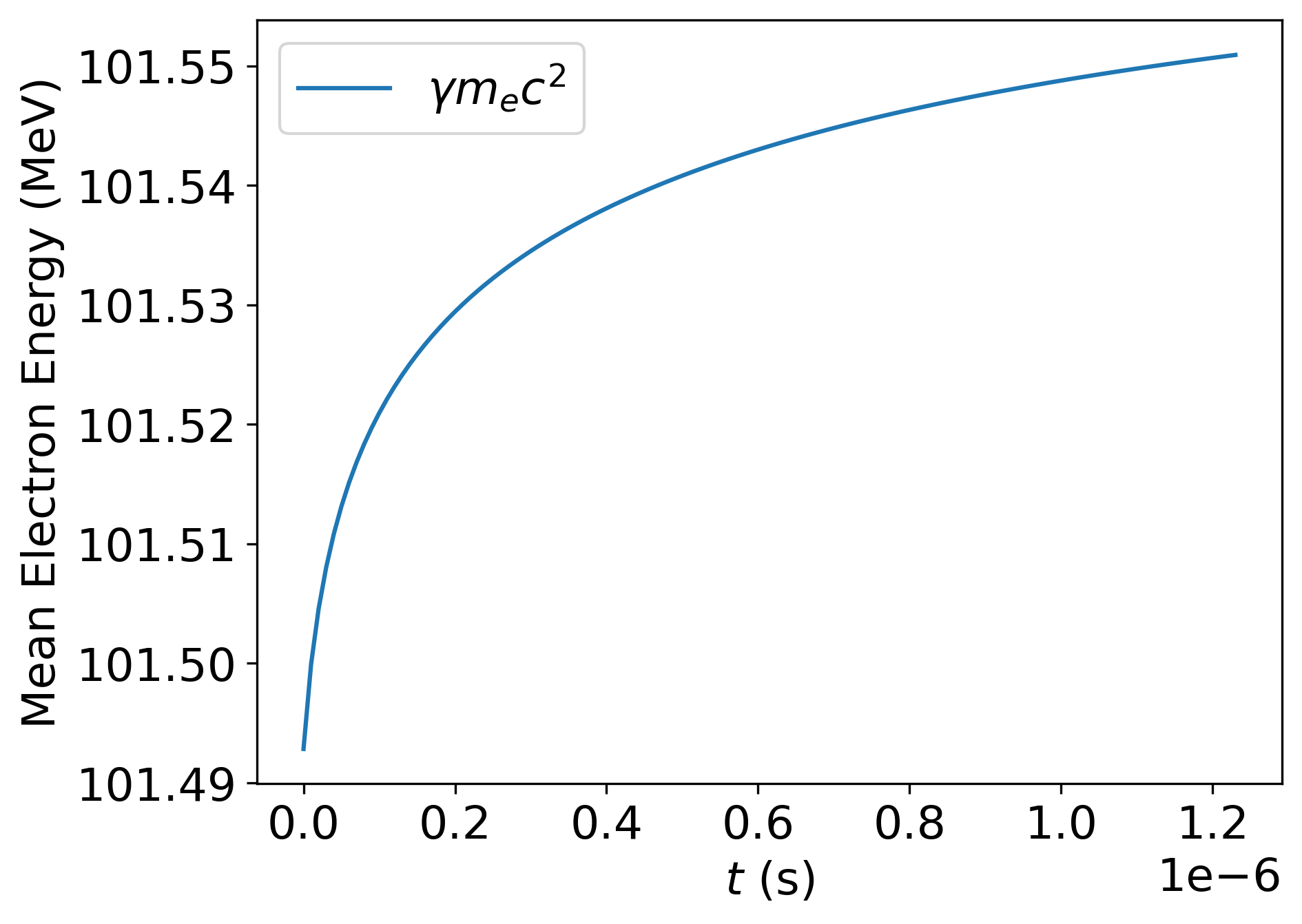}
\includegraphics[width=0.45\textwidth]{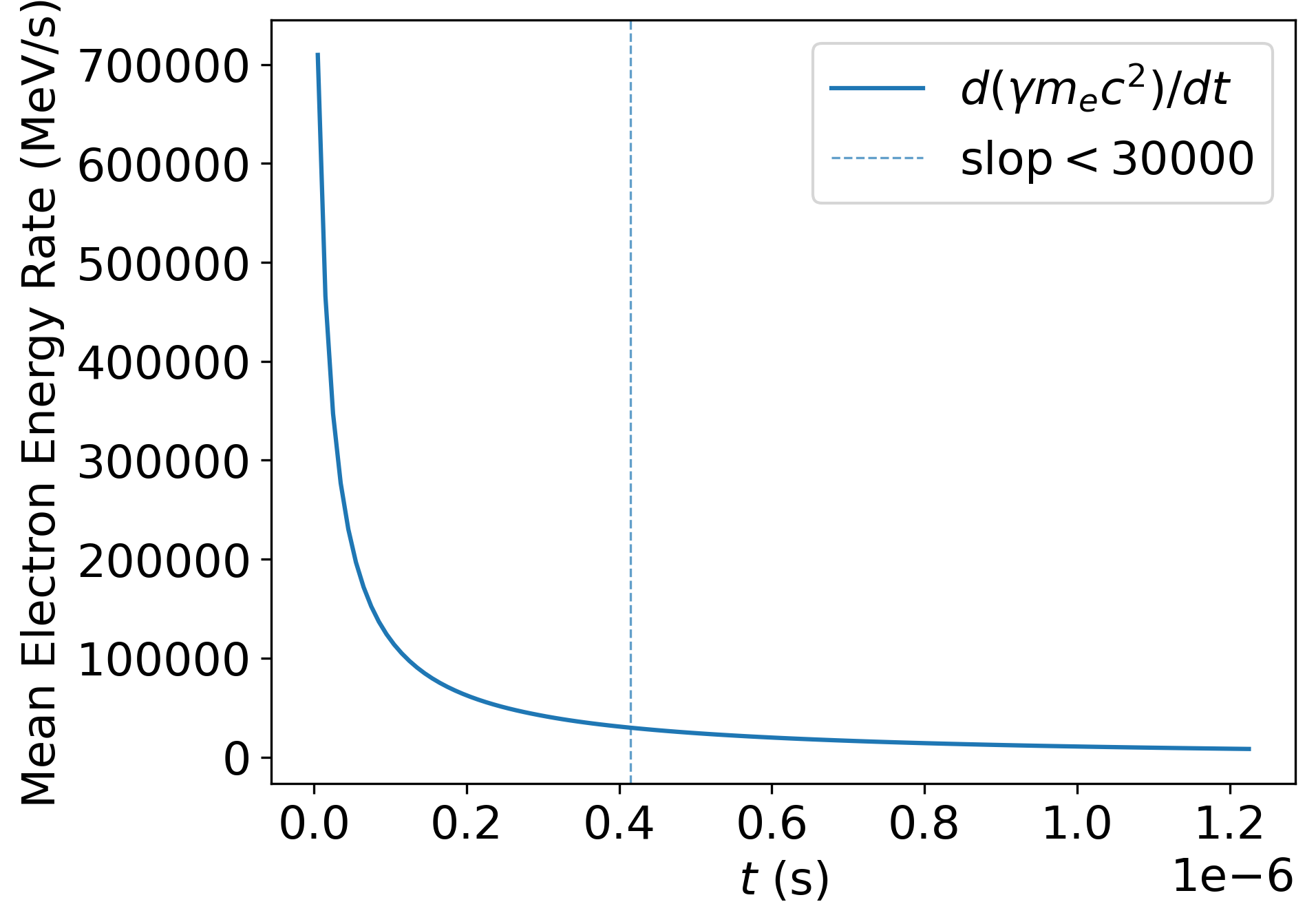}
\caption{
Mean electron energy and its time derivative
for the case with $N_p=1600$ and $E_0=100~\mathrm{MeV}$.
The upper panel shows the evolution of the mean energy,
while the lower panel shows the corresponding energy growth rate.
The dashed vertical line marks 
$\left|d\bar{E}/dt\right|=30~\mathrm{keV/s}$.
}
\label{fig:energy_1600_100Mev}
\end{figure}

Figures~\ref{fig:energy_1600_1Mev}--\ref{fig:energy_1600_100Mev}
show the mean electron energy and its time derivative for the cases with
$N_p=1600$ and
$E_0=1~\mathrm{MeV}$,
$10~\mathrm{MeV}$,
and $100~\mathrm{MeV}$.
In all three cases, the mean energy rises rapidly at first and then tends toward a plateau,
while the corresponding energy growth rate decreases quickly with time.

We define
\begin{equation}
    t_{\mathrm{th}}
    =
    \min \left\{
    t:\left|
    \frac{d\bar{E}}{dt}
    \right| < 30~\mathrm{keV/s}
    \right\}
\end{equation}
as the first time at which the threshold is satisfied.
The dashed vertical lines in
Fig.~\ref{fig:energy_1600_1Mev}--Fig.~\ref{fig:energy_1600_100Mev}
mark this reference time.
In the two-stage calculations, the final switching time $t_{\mathrm{sw}}$ is chosen such that
\begin{equation}
    t_{\mathrm{sw}} > t_{\mathrm{th}},
\end{equation}
so that the fully coupled calculation continues for some time after the threshold is first crossed.

For the $1~\mathrm{MeV}$ case, the threshold is reached only near the end of the short fully coupled run, and the adopted switching time is
\begin{equation}
    t_{\mathrm{sw}} \approx 1.0\times10^{-8}\ \mathrm{s}.
\end{equation}
For the $10~\mathrm{MeV}$ case, the threshold is crossed earlier, whereas the final switching time is taken as
\begin{equation}
    t_{\mathrm{sw}} \approx 2.0\times10^{-7}\ \mathrm{s}.
\end{equation}
For the $100~\mathrm{MeV}$ case, the threshold-crossing time is again earlier than the adopted switching time, and the final value is
\begin{equation}
    t_{\mathrm{sw}} \approx 1.2\times10^{-6}\ \mathrm{s}.
\end{equation}
These results show that higher beam energies require a longer physical time to pass through the early relaxation stage before entering the weakly coupled transport regime.
The switching time should therefore be chosen in an energy-dependent manner rather than as a single universal value.

\section{Results and Discussion}\label{sec:results}
\subsection{Early-Stage Self-Field-Driven Beam Expansion}\label{sec:early}

Sections~3.3 and 3.4 show that the relativistic electron bunch undergoes rapid self-field-driven expansion immediately after injection. Owing to the initially compact and non-neutral beam configuration, inter-particle electromagnetic interactions are strongest at this stage, leading to pronounced transverse broadening and weaker longitudinal stretching. The same trend is observed for $N_p=400$, $800$, and $1600$, which suggests that the behavior is physical rather than numerical.

At the same time, the mean particle energy rises rapidly, indicating conversion of self-field energy into particle kinetic energy. As the bunch expands, the average particle separation increases and the inter-particle interaction becomes weaker. For this reason, the early dense stage must be resolved with the full EM-PP model, while the later long-range transport can be described within the weakly coupled two-stage framework.
\begin{figure*}[!ht]
\centering
\includegraphics[width=0.18\textwidth]{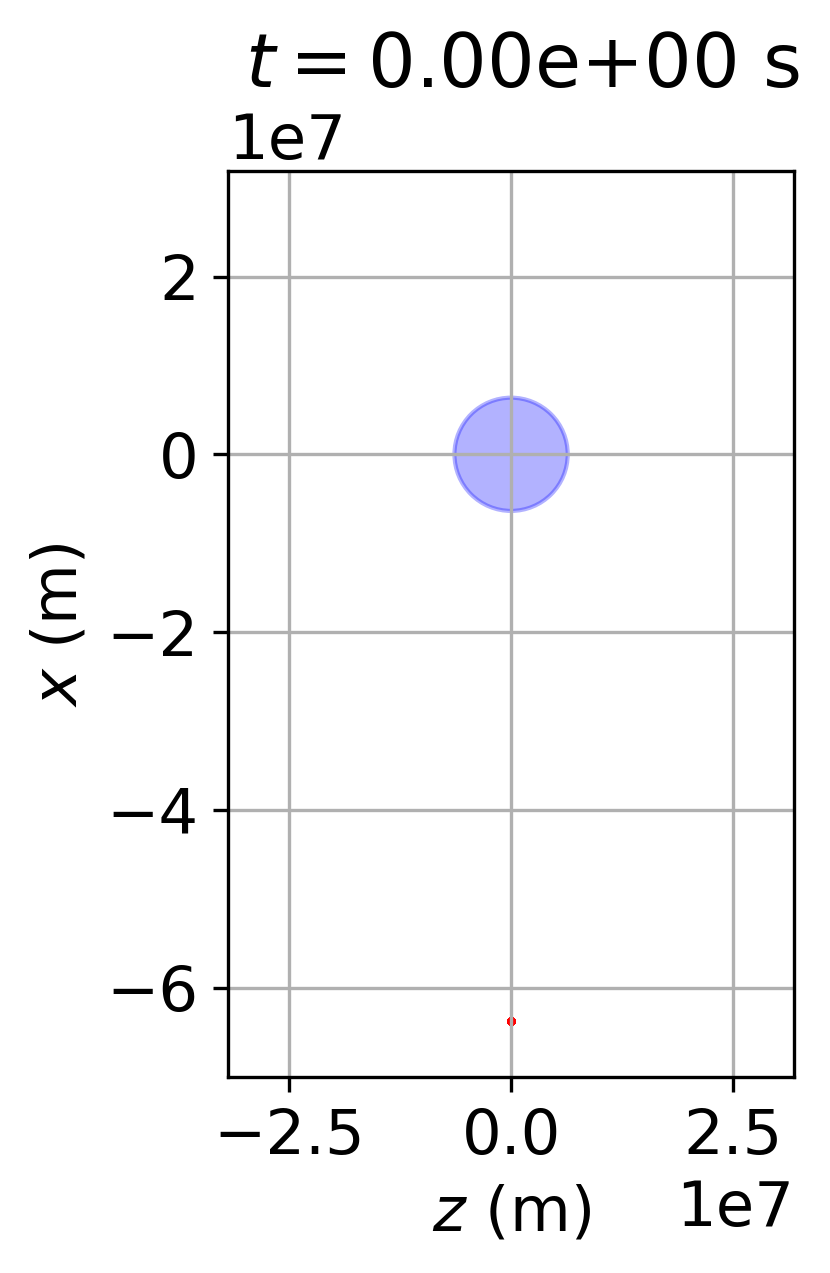}
\includegraphics[width=0.18\textwidth]{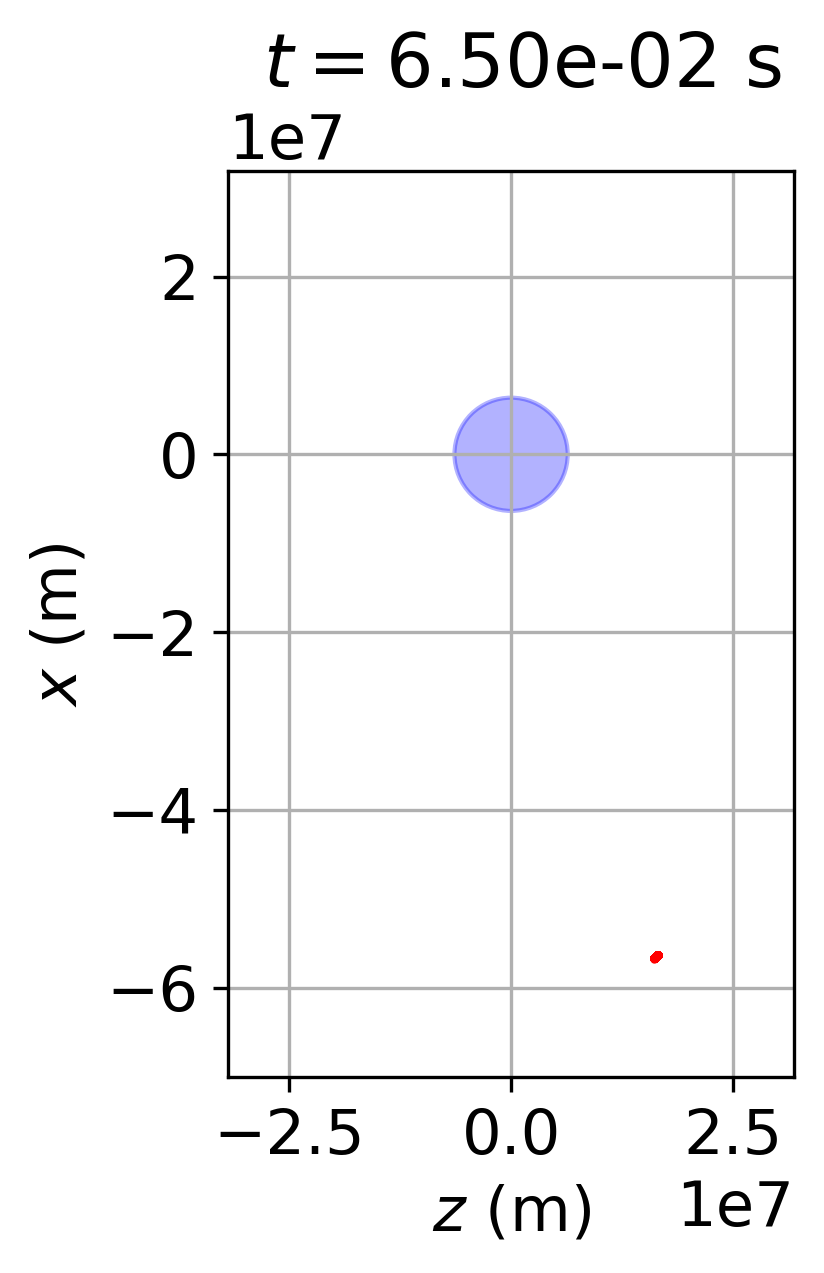}
\includegraphics[width=0.18\textwidth]{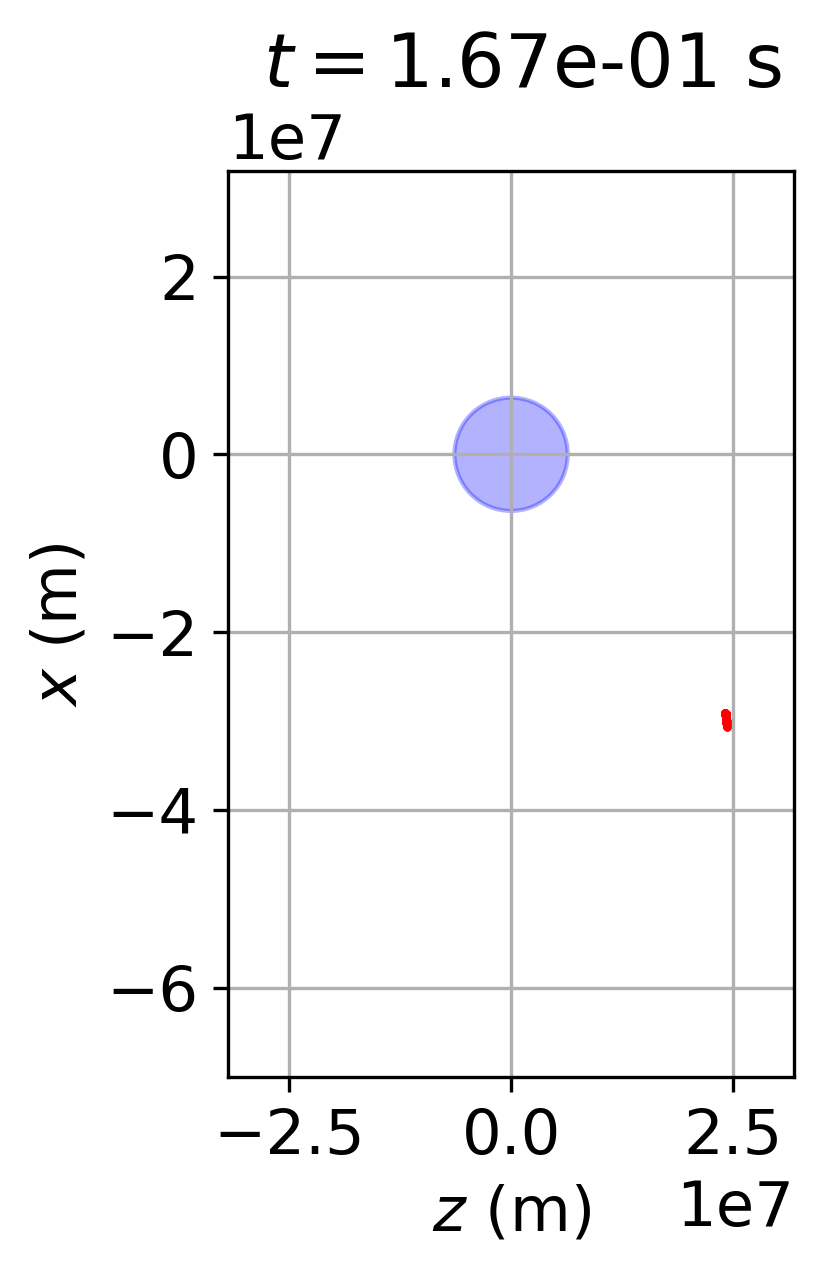}
\includegraphics[width=0.18\textwidth]{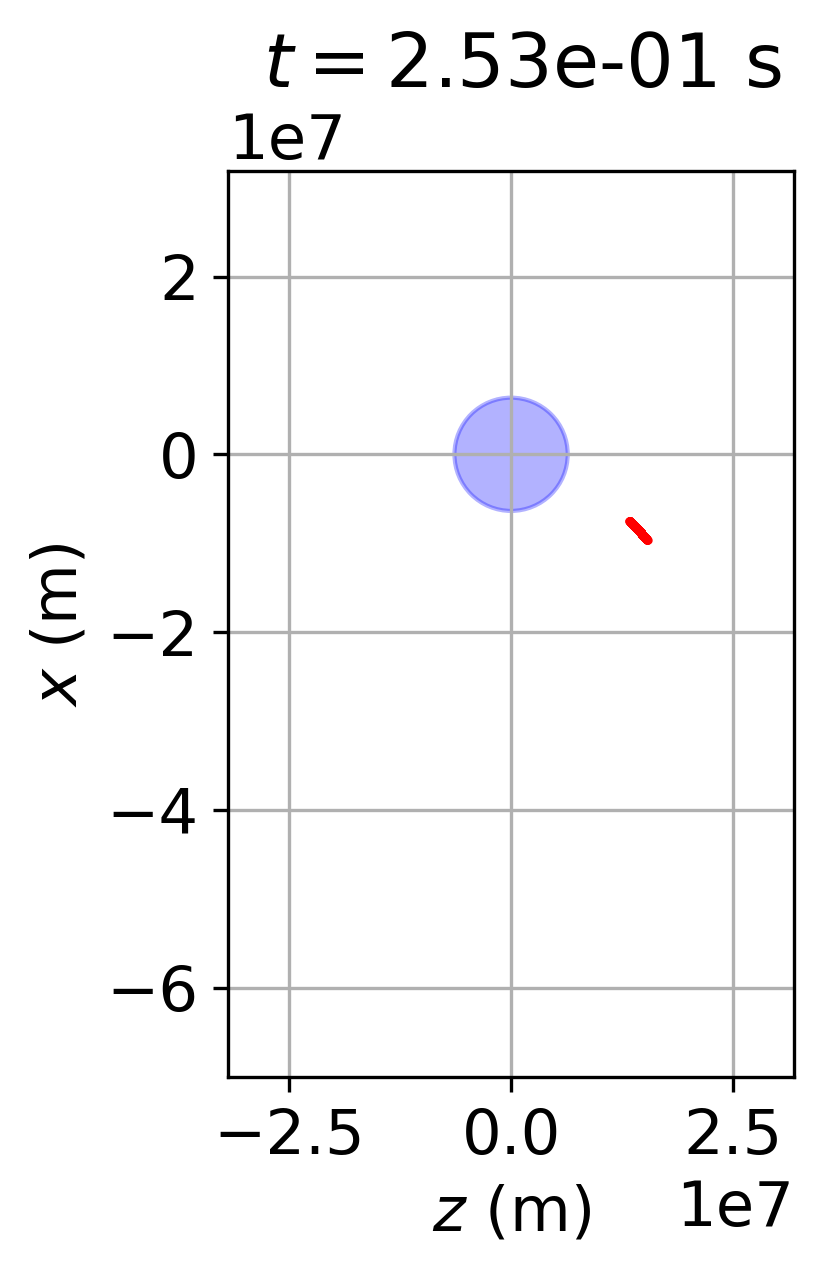}
\includegraphics[width=0.18\textwidth]{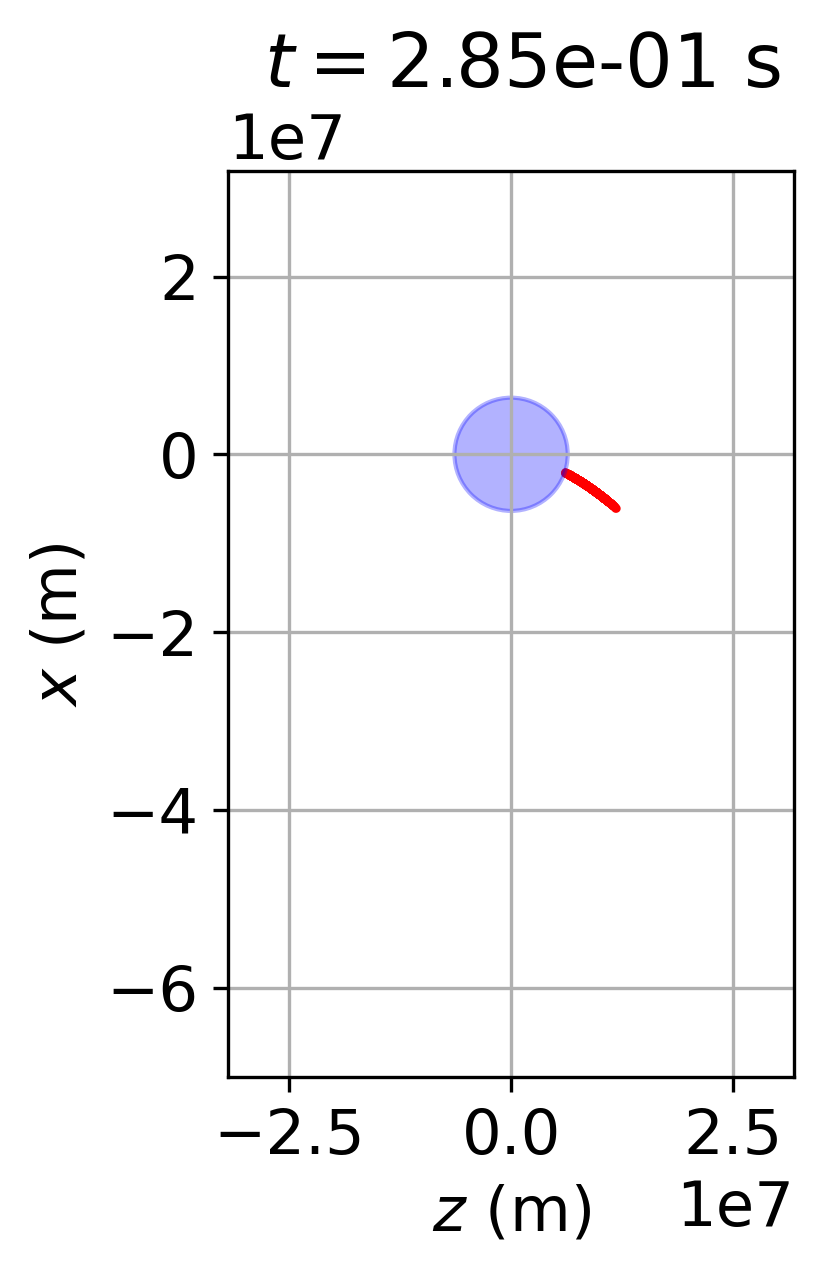}
\\
\includegraphics[width=0.25\textwidth]{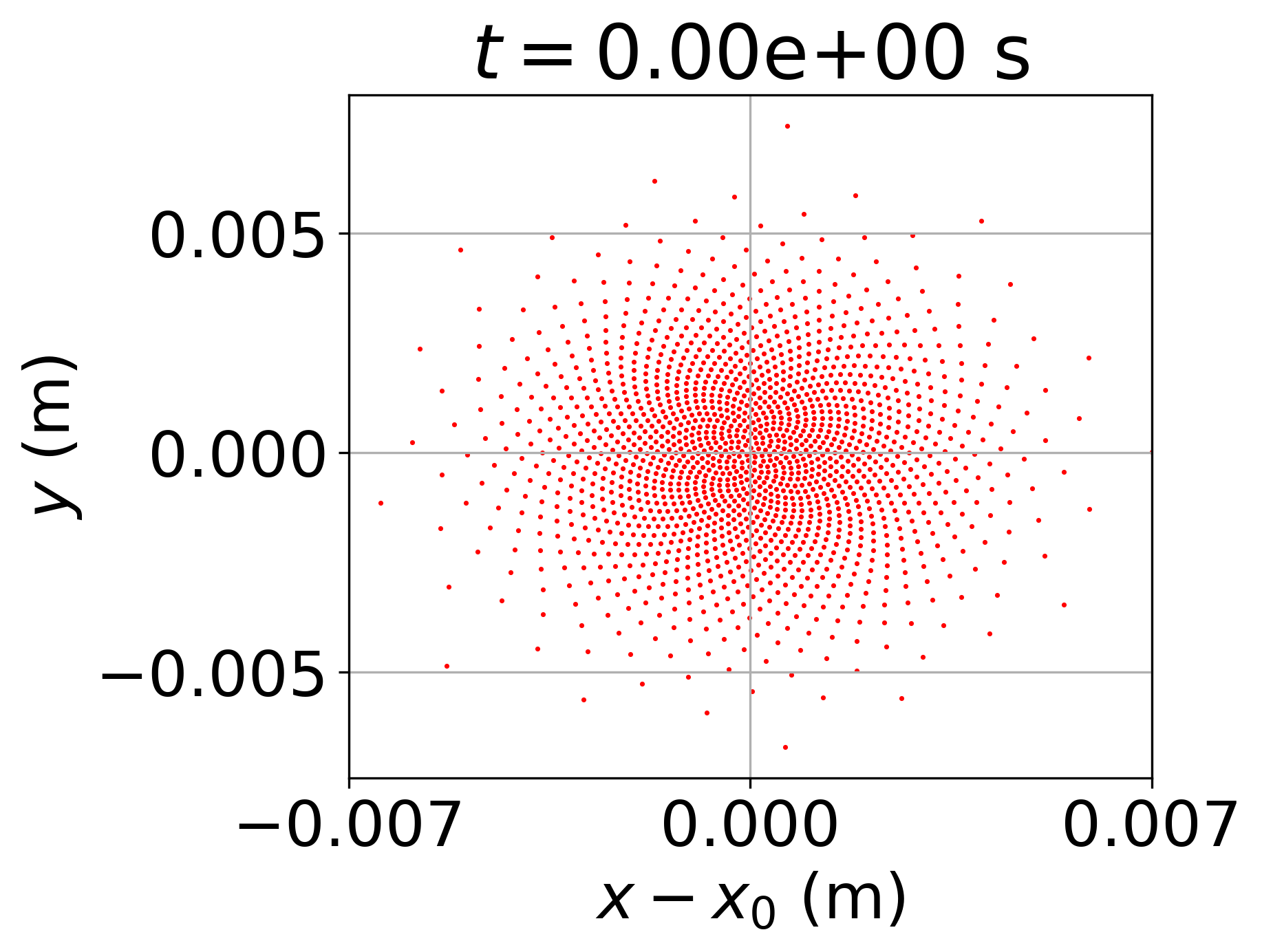}
\includegraphics[width=0.25\textwidth]{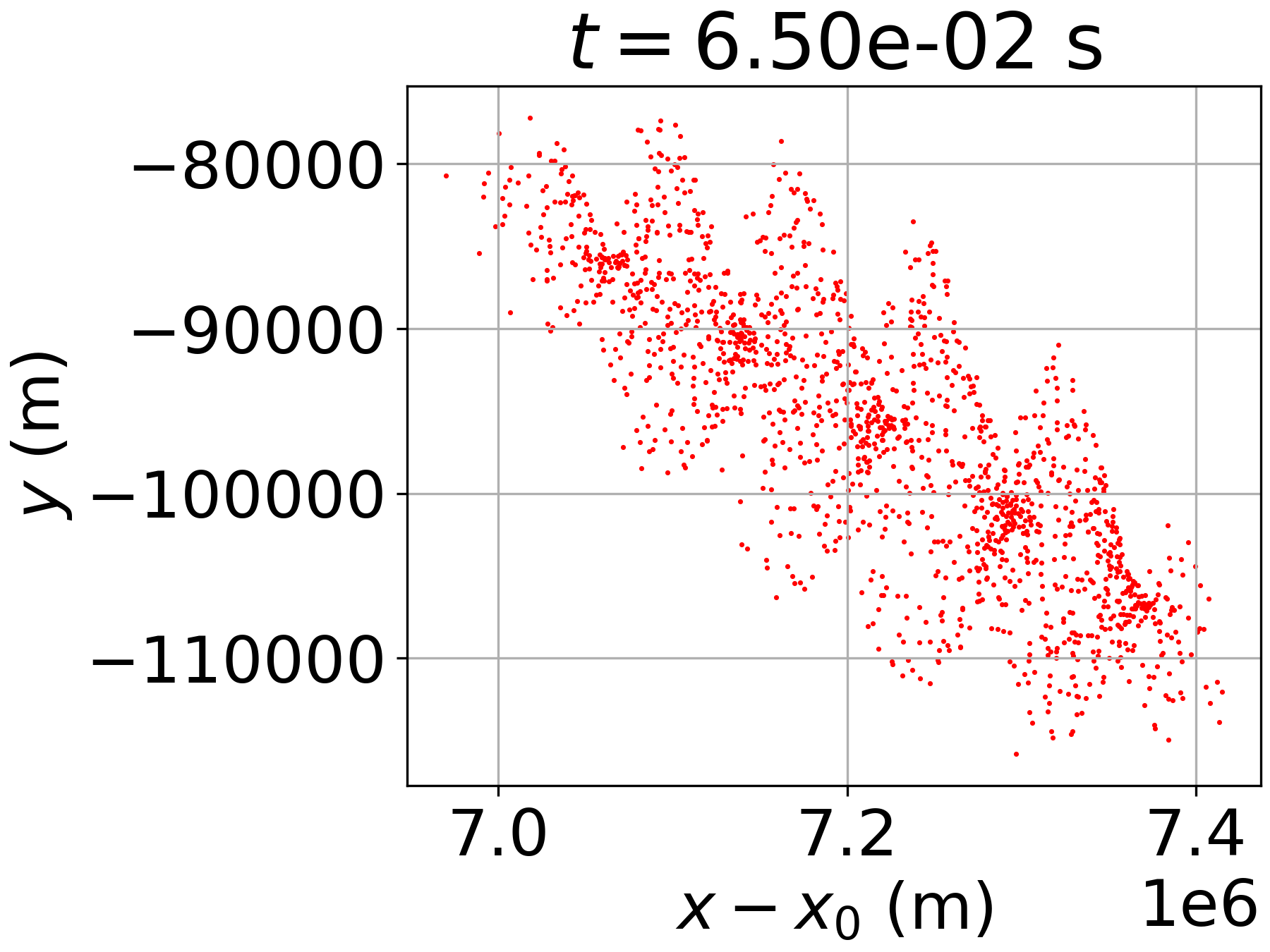}
\includegraphics[width=0.25\textwidth]{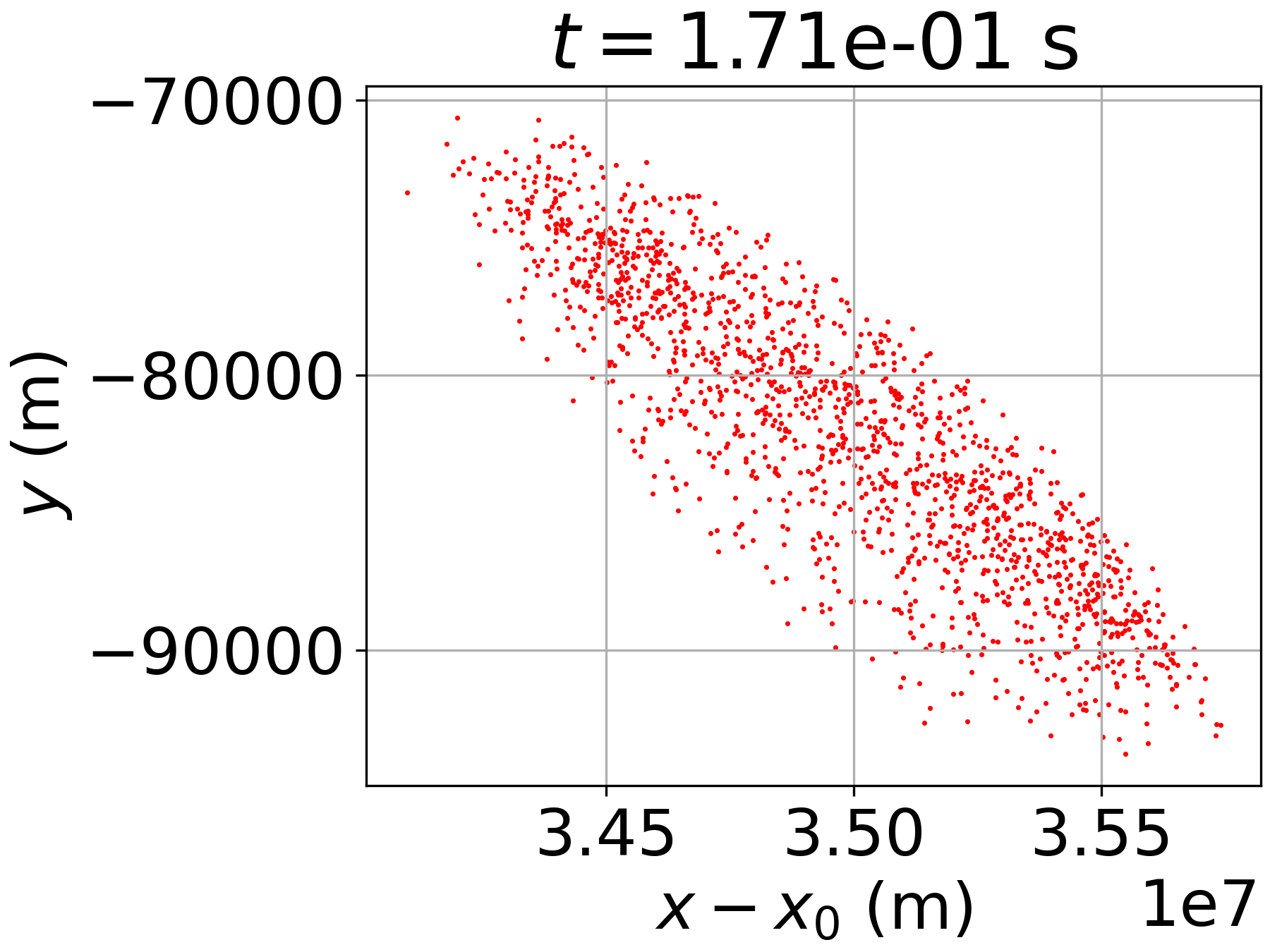}
\includegraphics[width=0.25\textwidth]{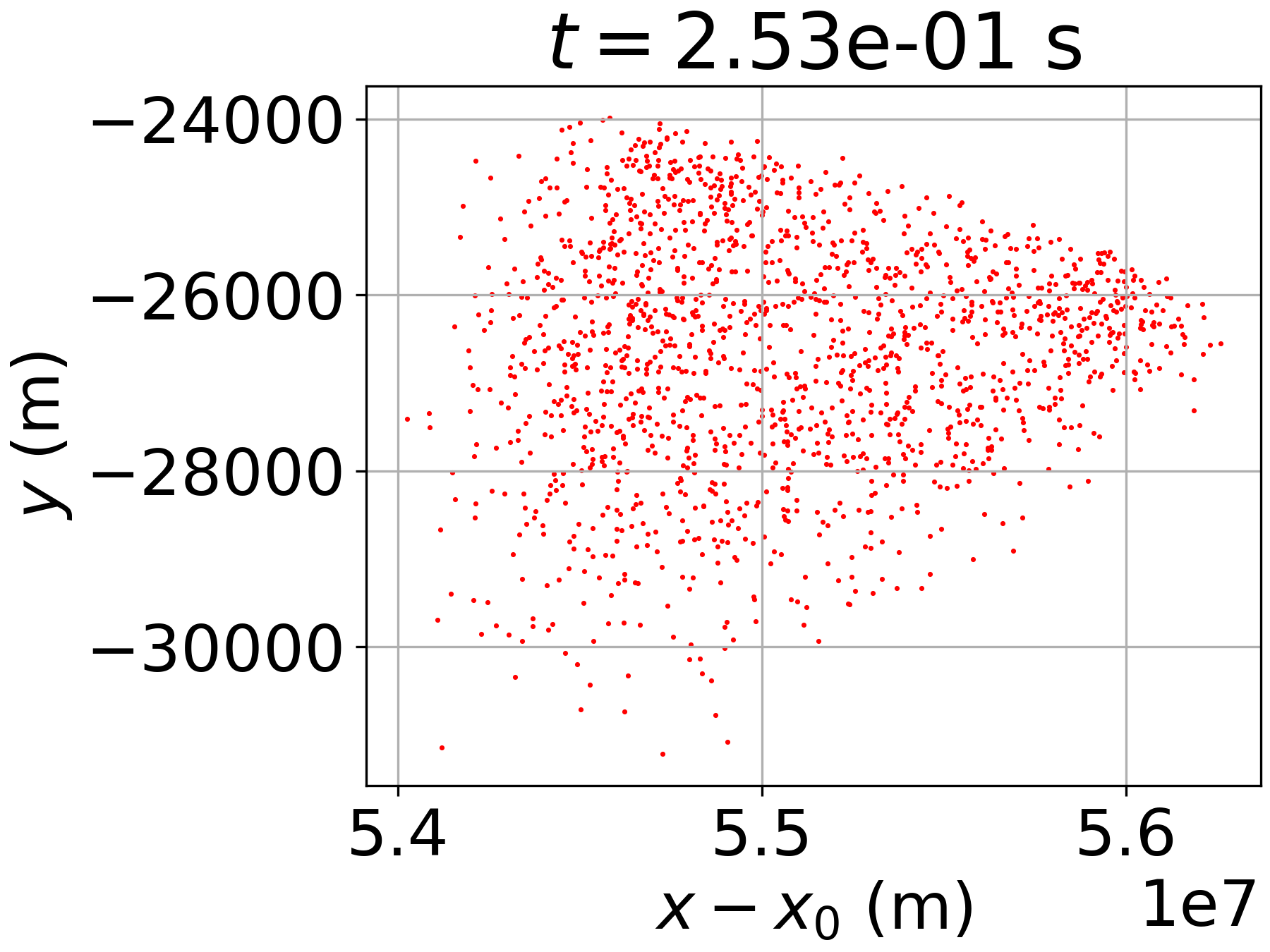}
\includegraphics[width=0.25\textwidth]{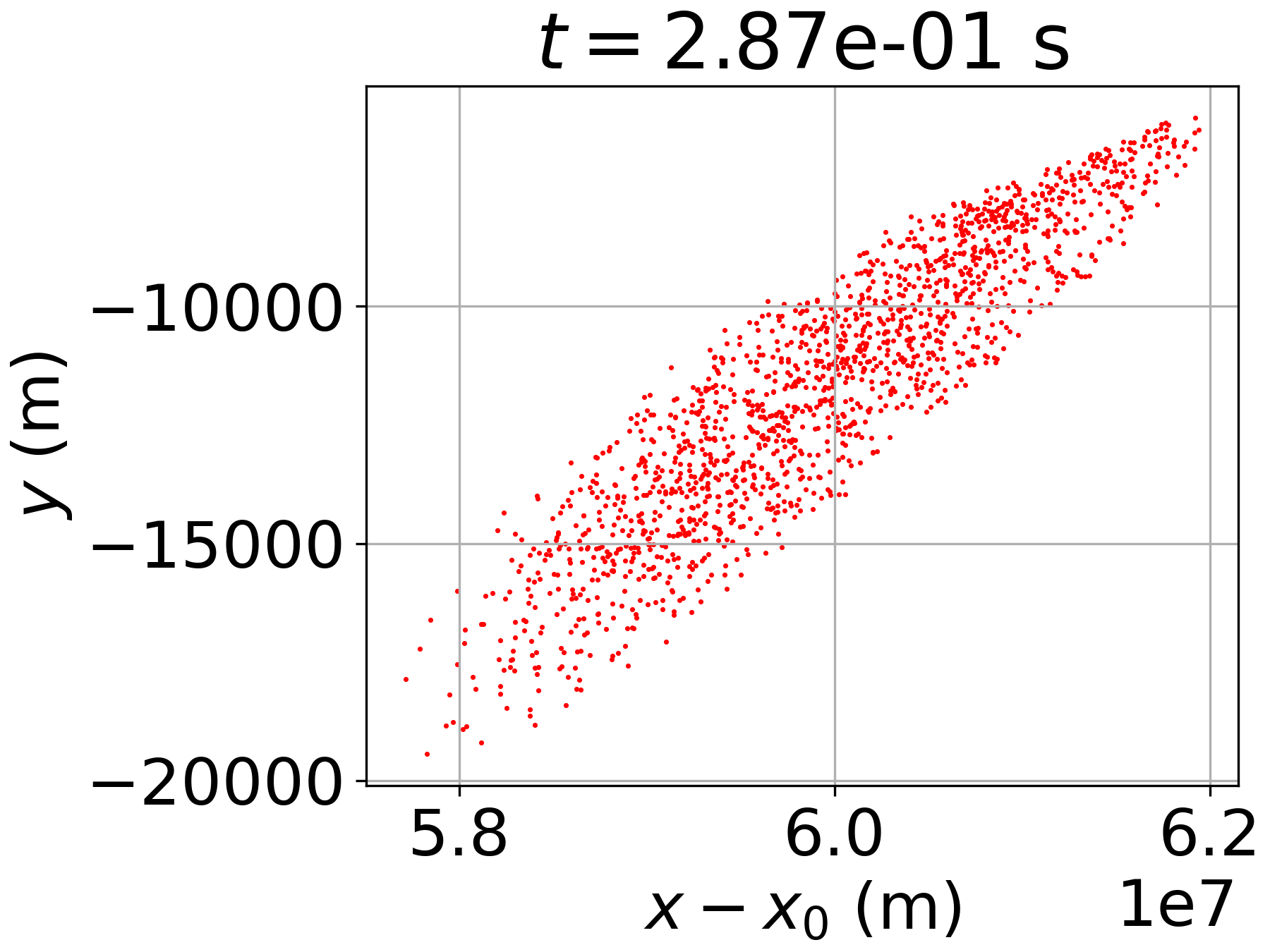}
\\
\includegraphics[width=0.25\textwidth]{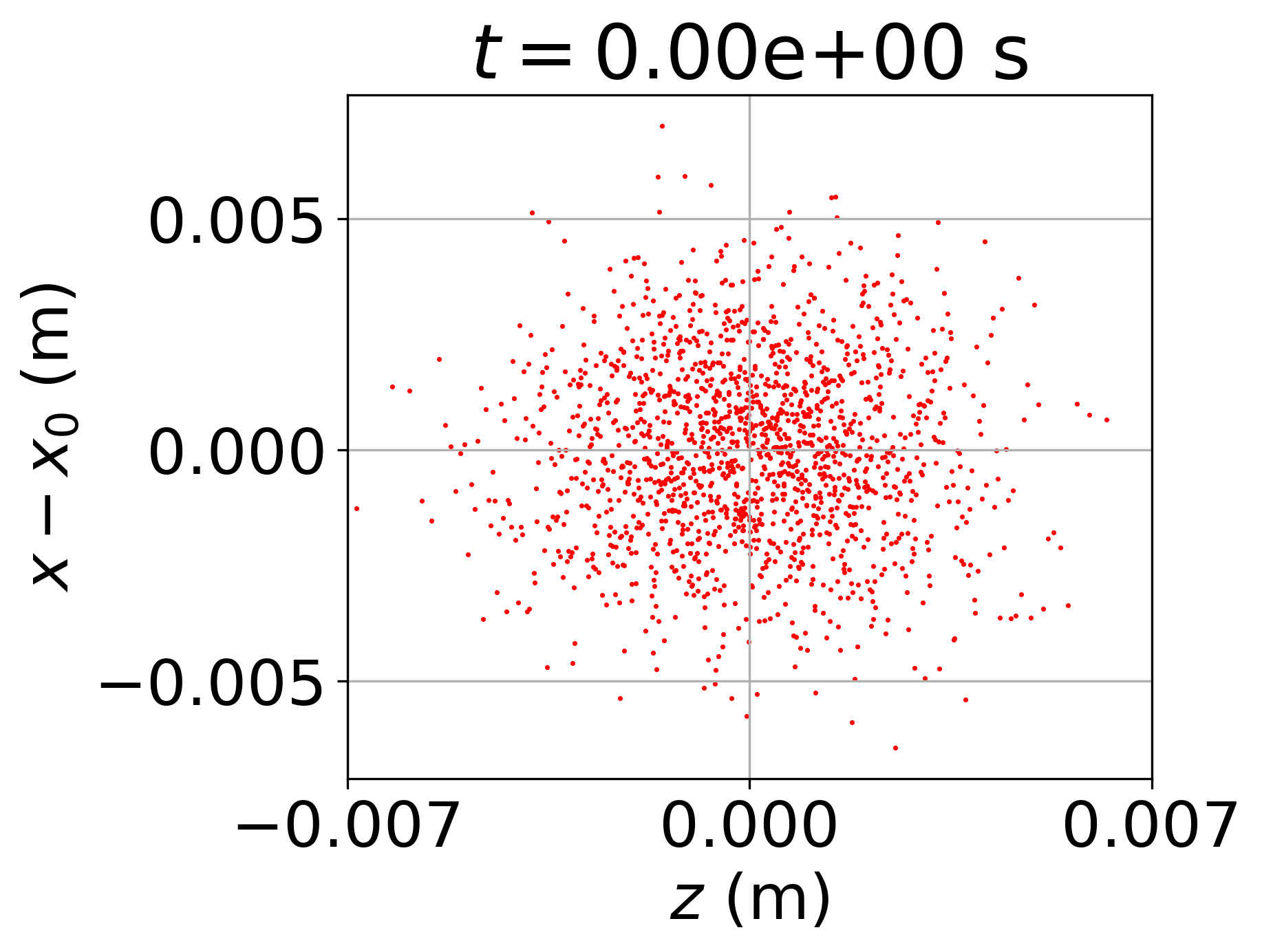}
\includegraphics[width=0.25\textwidth]{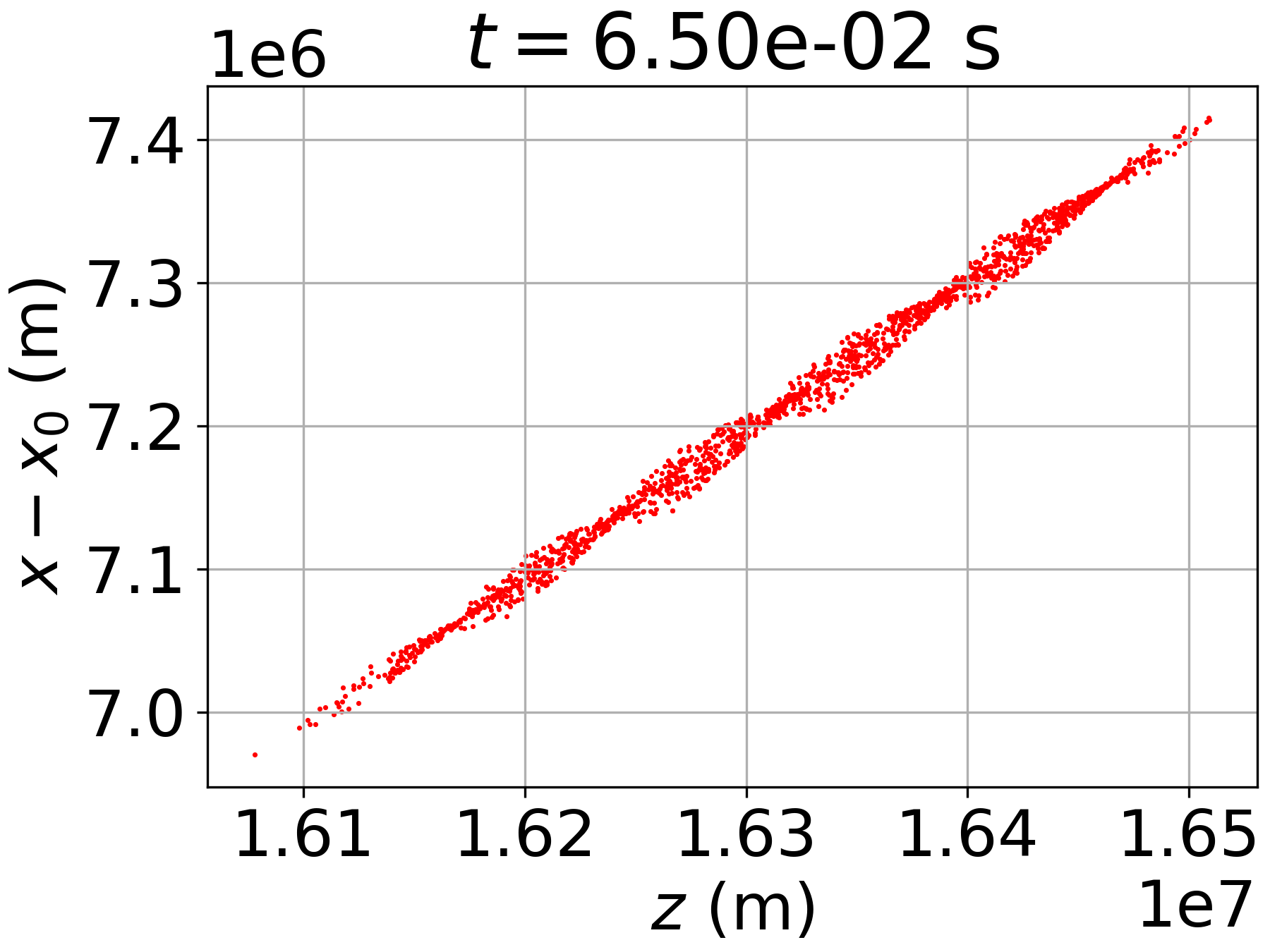}
\includegraphics[width=0.25\textwidth]{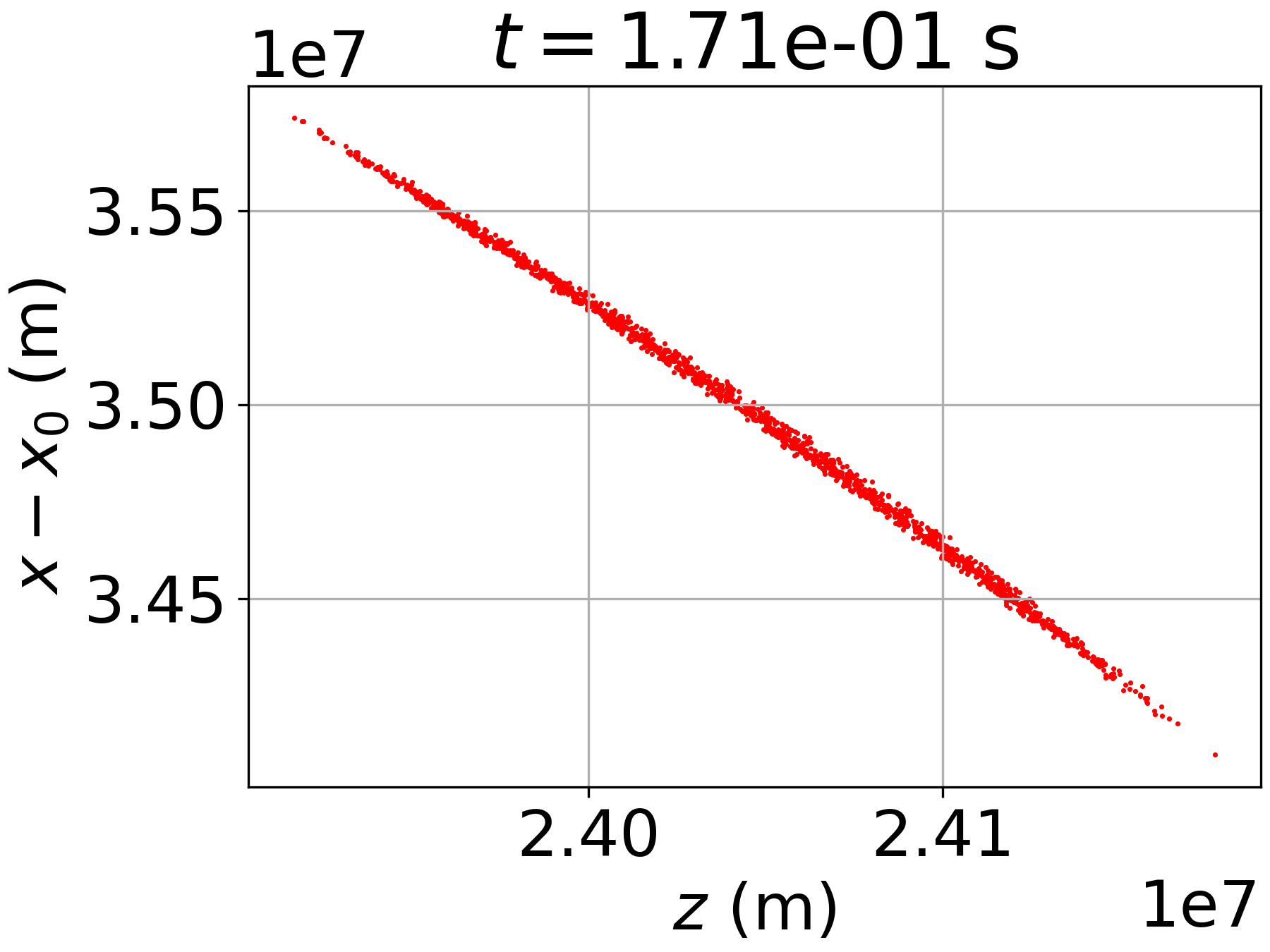}
\includegraphics[width=0.25\textwidth]{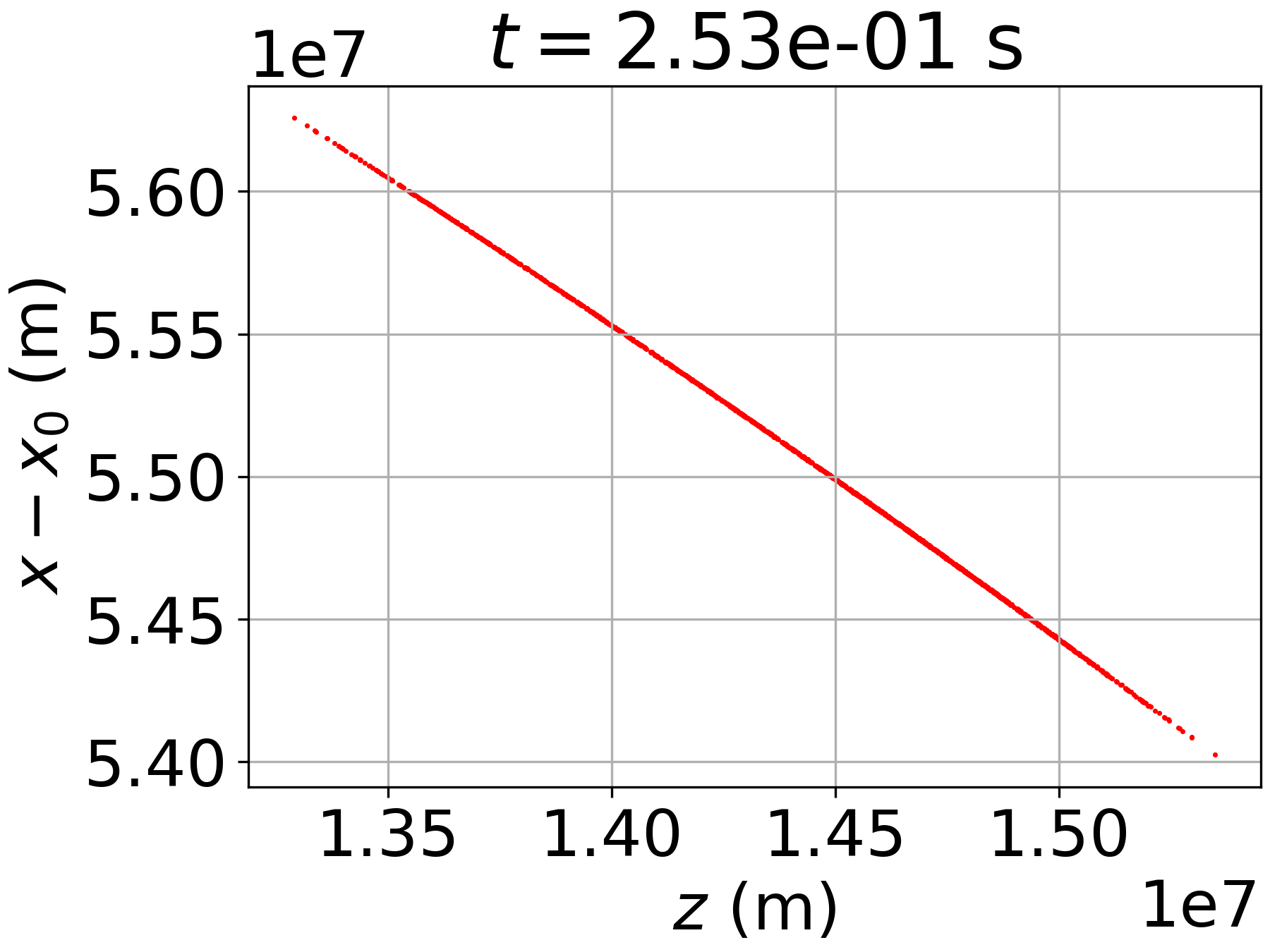}
\includegraphics[width=0.25\textwidth]{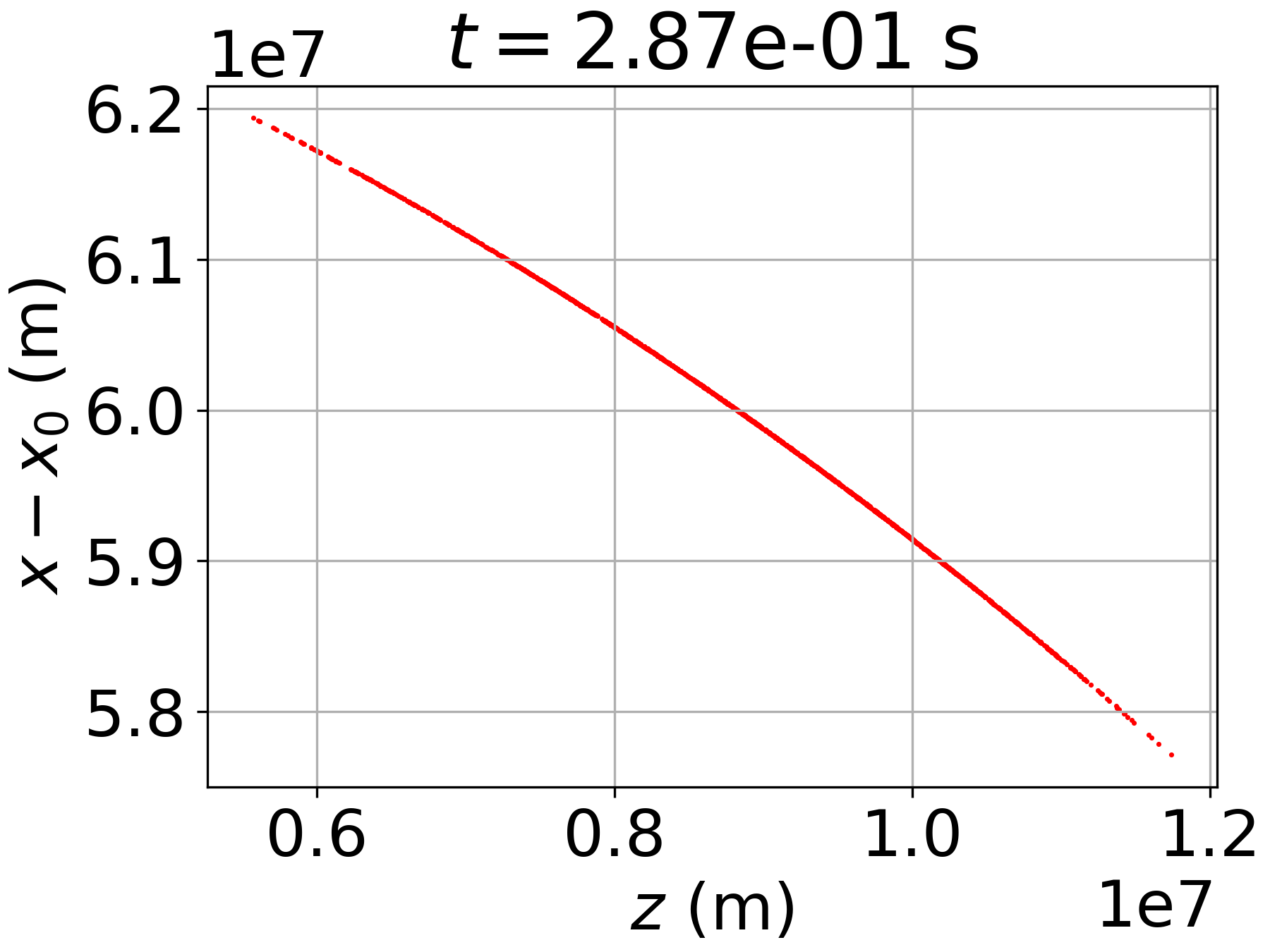}
\caption{
Long-range evolution of the relativistic electron bunch for the case
$N_p=1600$ and $E_0=1~\mathrm{MeV}$.
The top row shows the bunch trajectory in the $z$--$x$ plane
at several representative times during its propagation from
$(-10R_{\mathrm{E}},0,0)$ toward the Earth
(the blue circle denotes the Earth).
The middle two rows give the corresponding transverse particle distributions
in the $x$--$y$ plane,
and the bottom two rows show the longitudinal distributions
in the $z$--$x$ plane relative to the initial injection position.}
\label{fig:1600_1}
\end{figure*}

\begin{figure*}[!ht]
\centering
\includegraphics[width=0.18\textwidth]{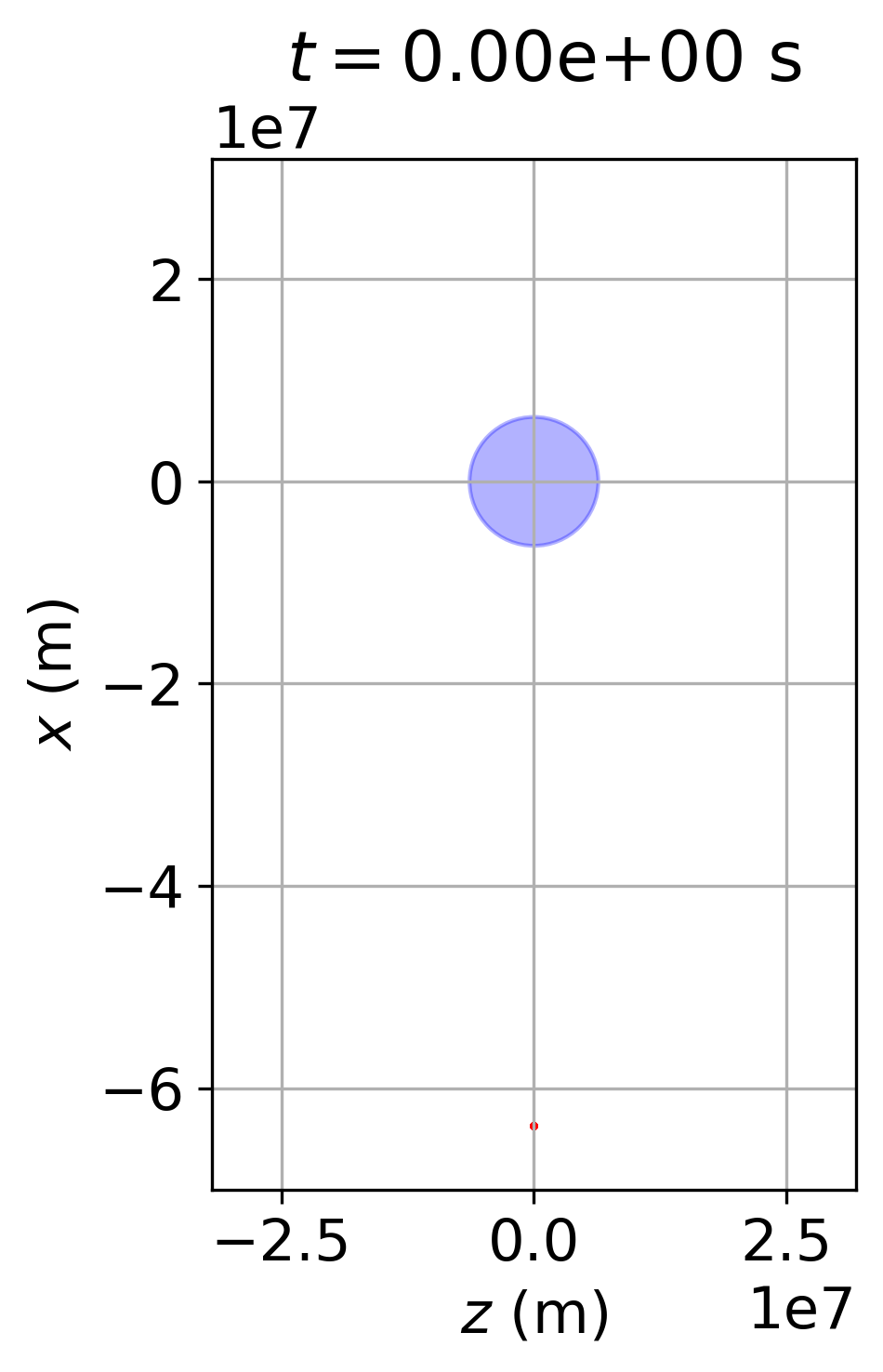}
\includegraphics[width=0.18\textwidth]{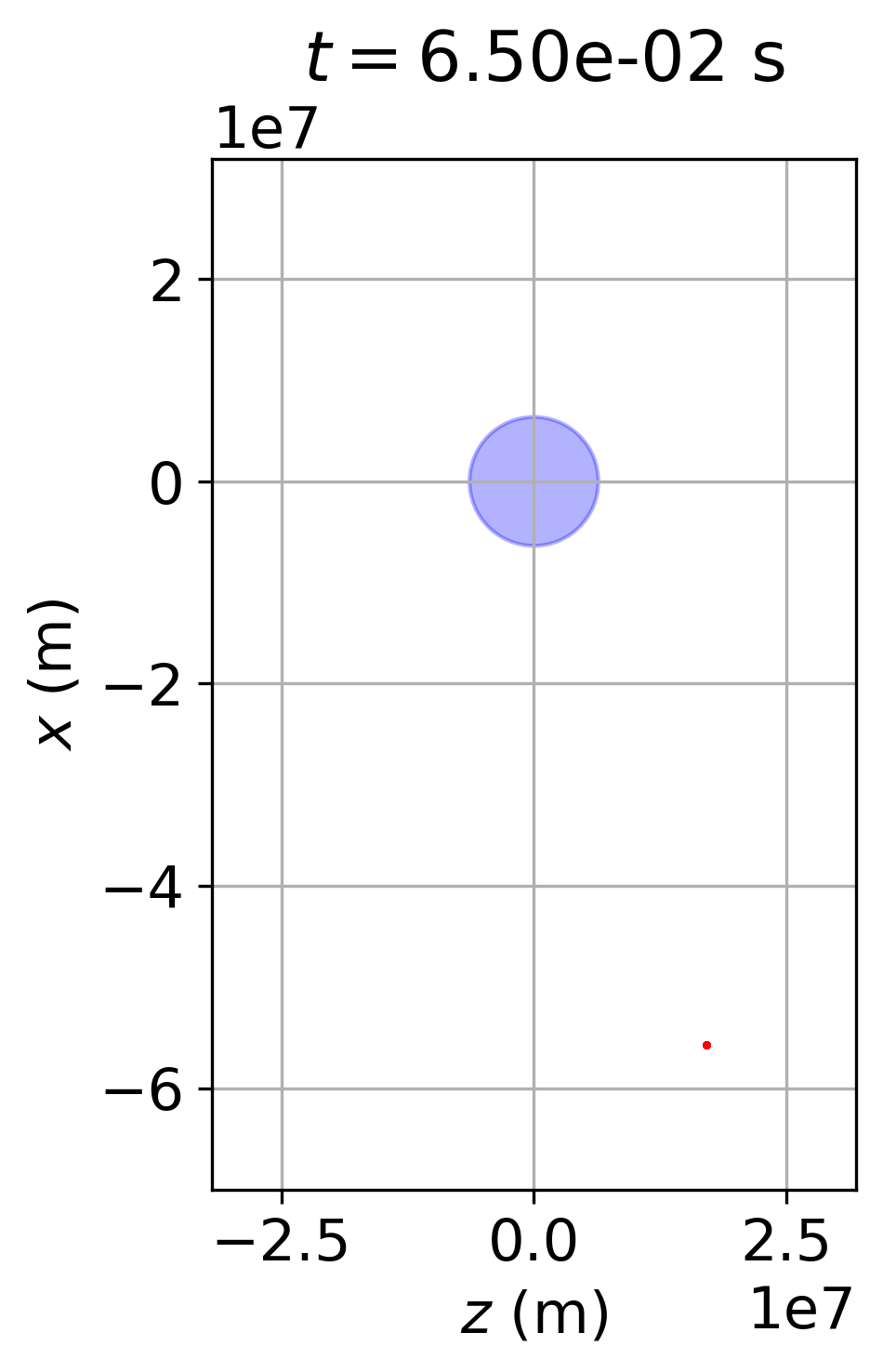}
\includegraphics[width=0.18\textwidth]{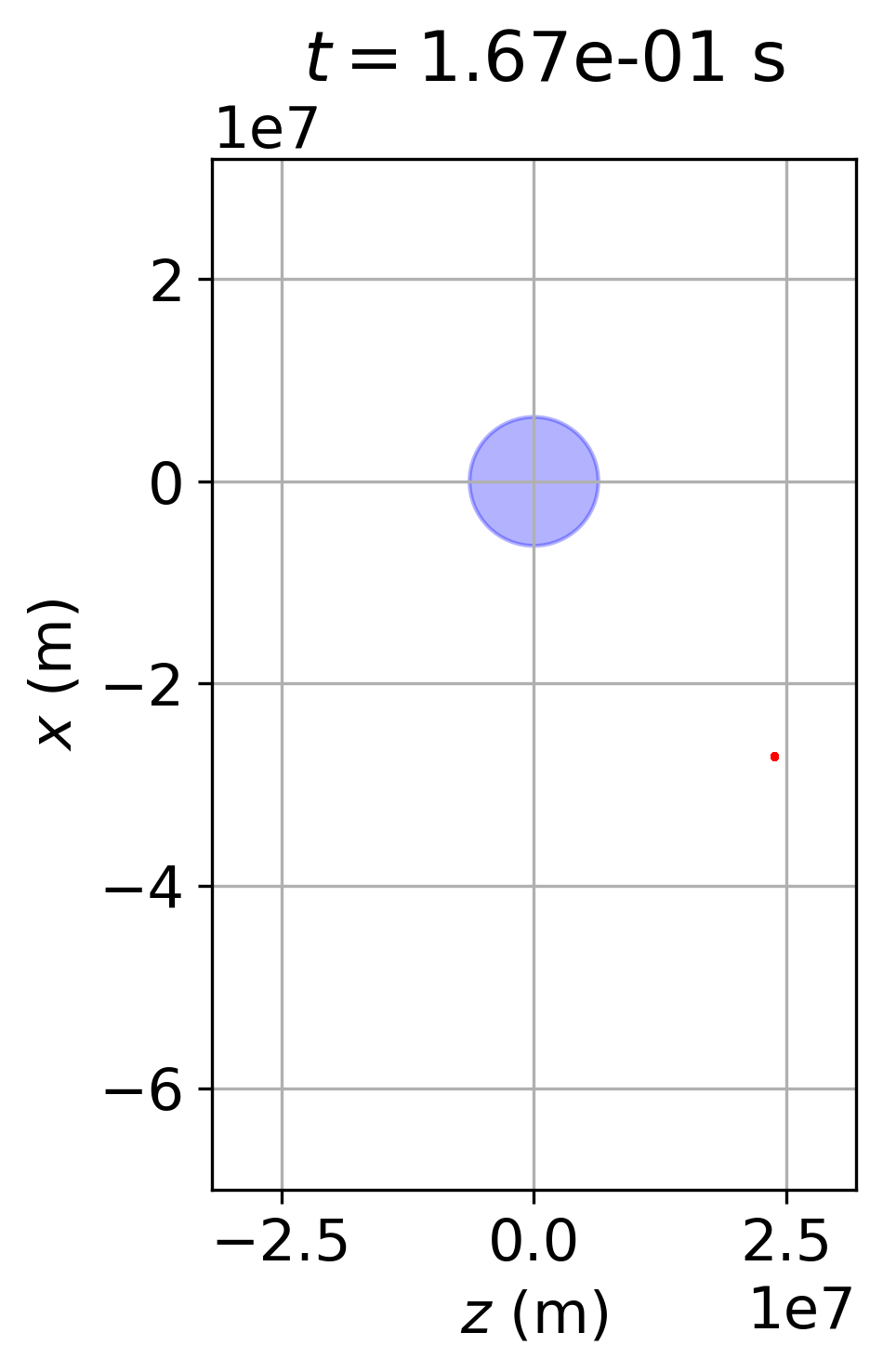}
\includegraphics[width=0.18\textwidth]{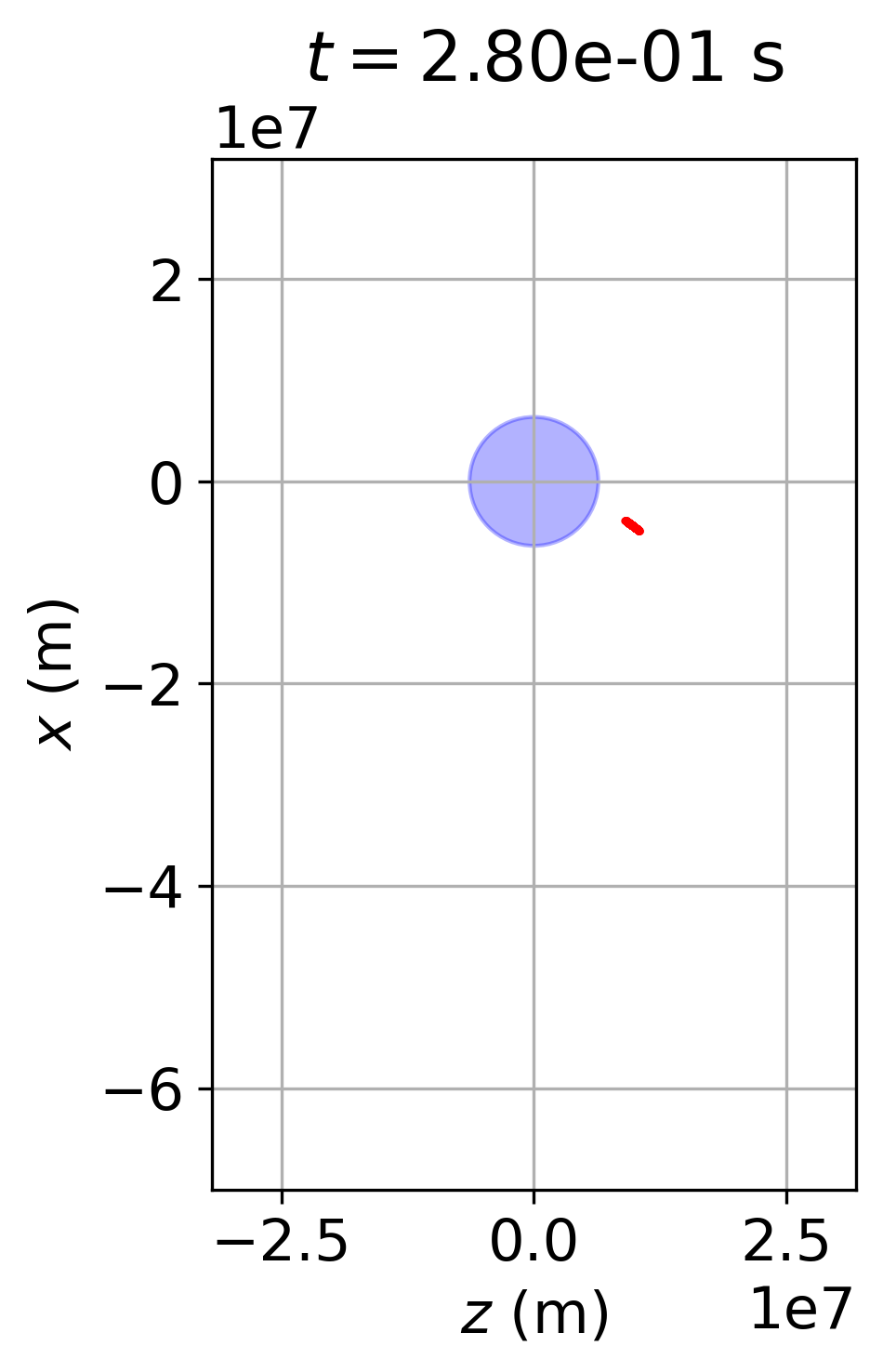}
\includegraphics[width=0.18\textwidth]{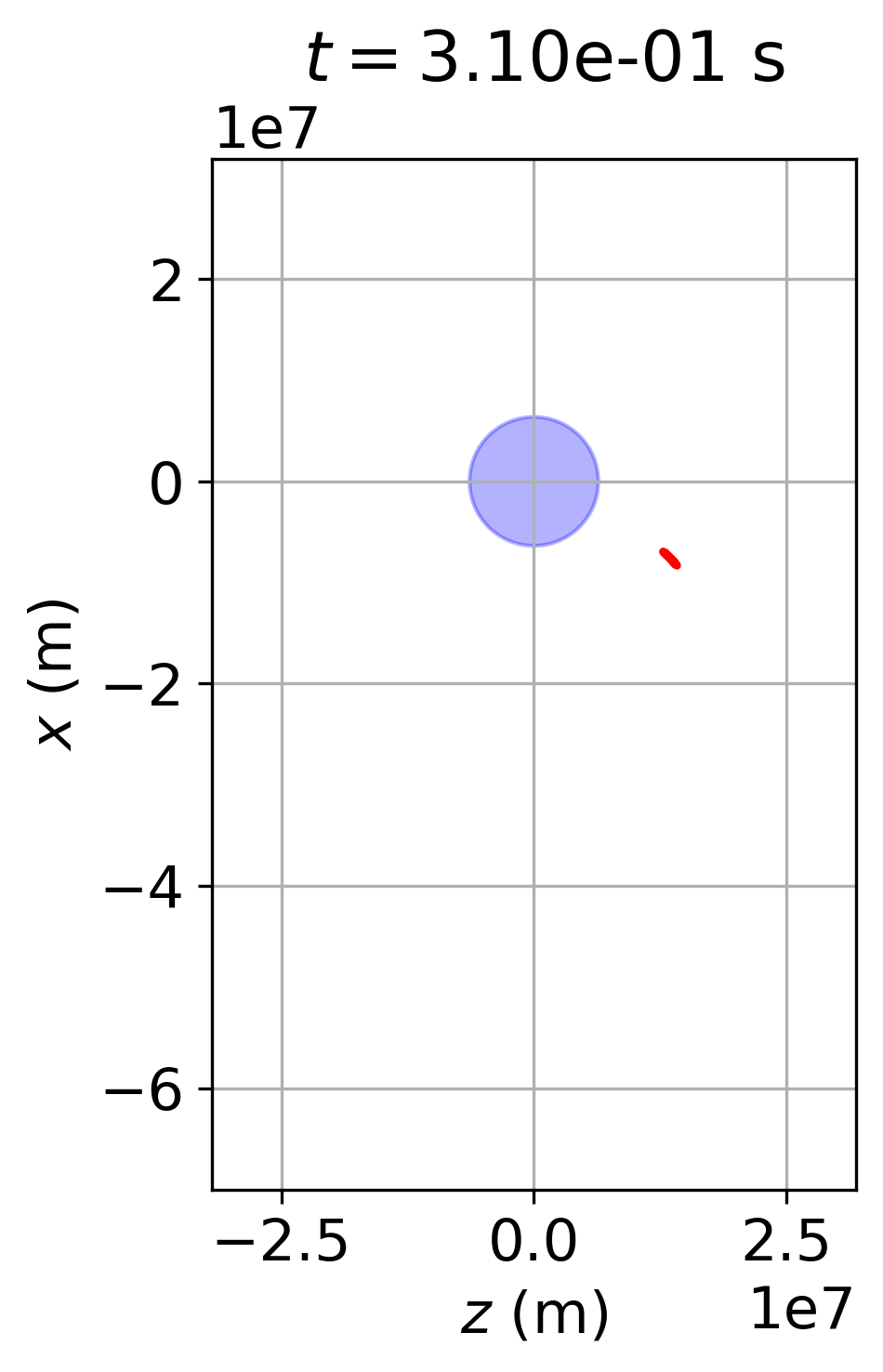}

\includegraphics[width=0.25\textwidth]{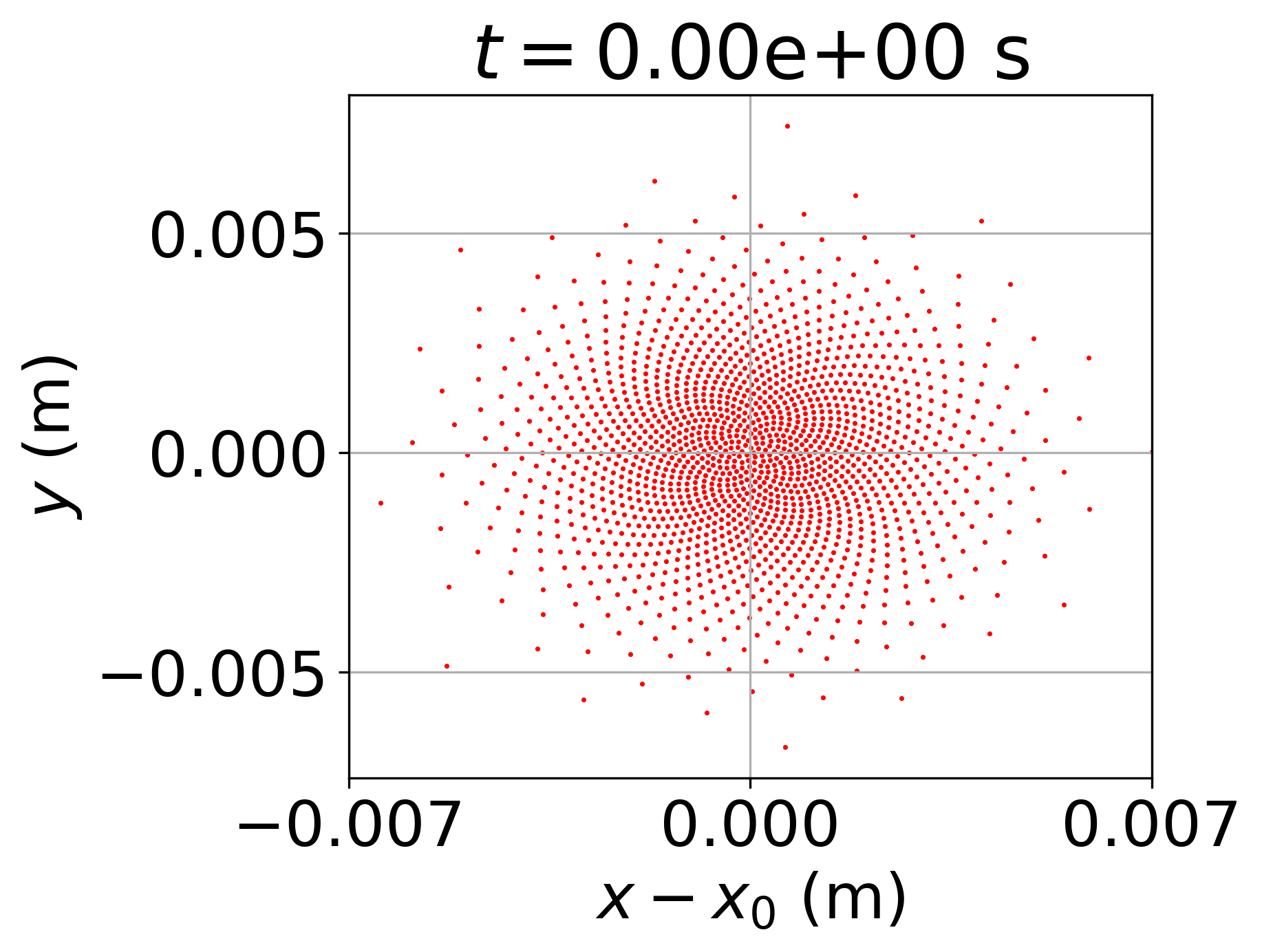}
\includegraphics[width=0.25\textwidth]{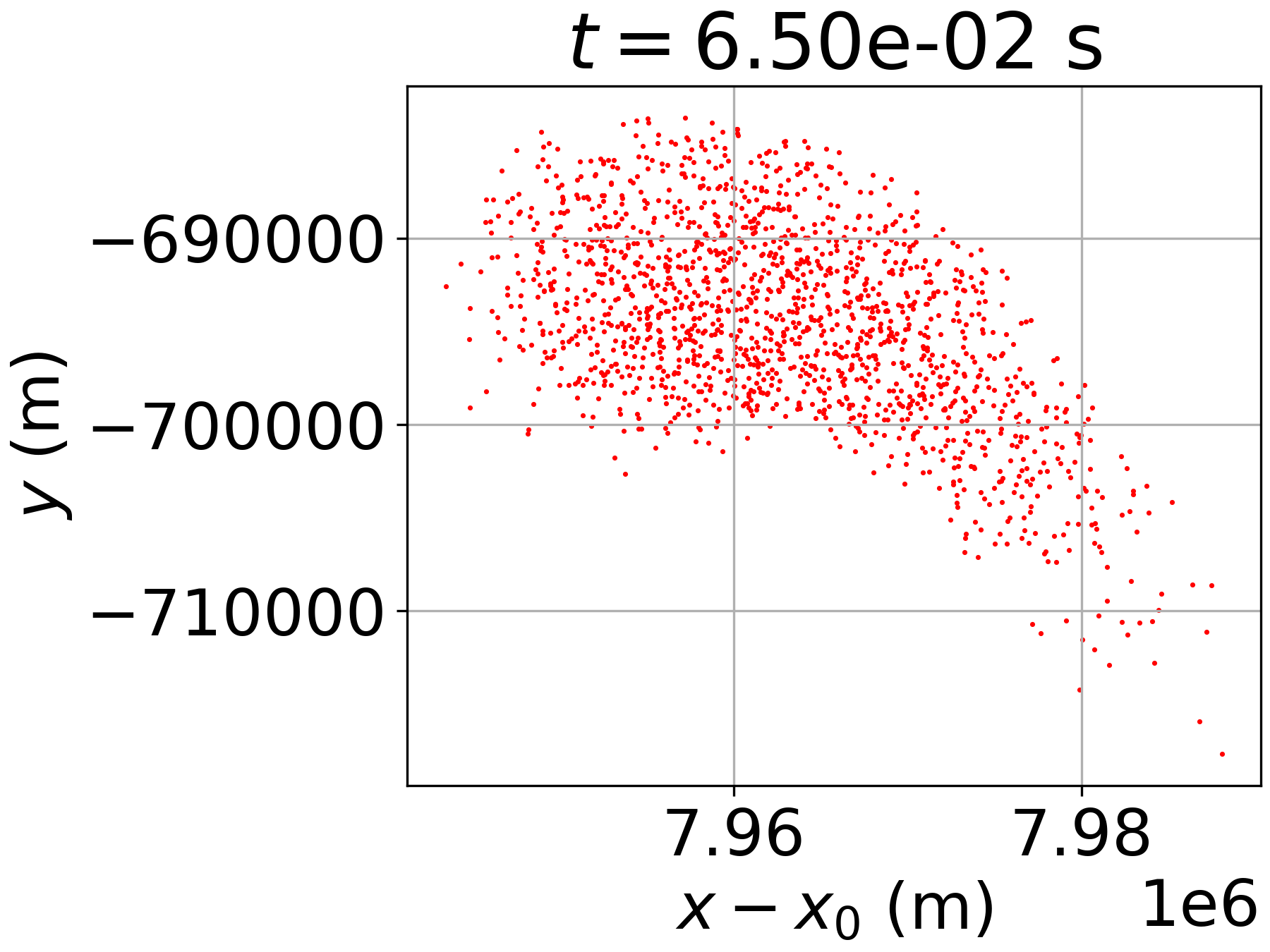}
\includegraphics[width=0.25\textwidth]{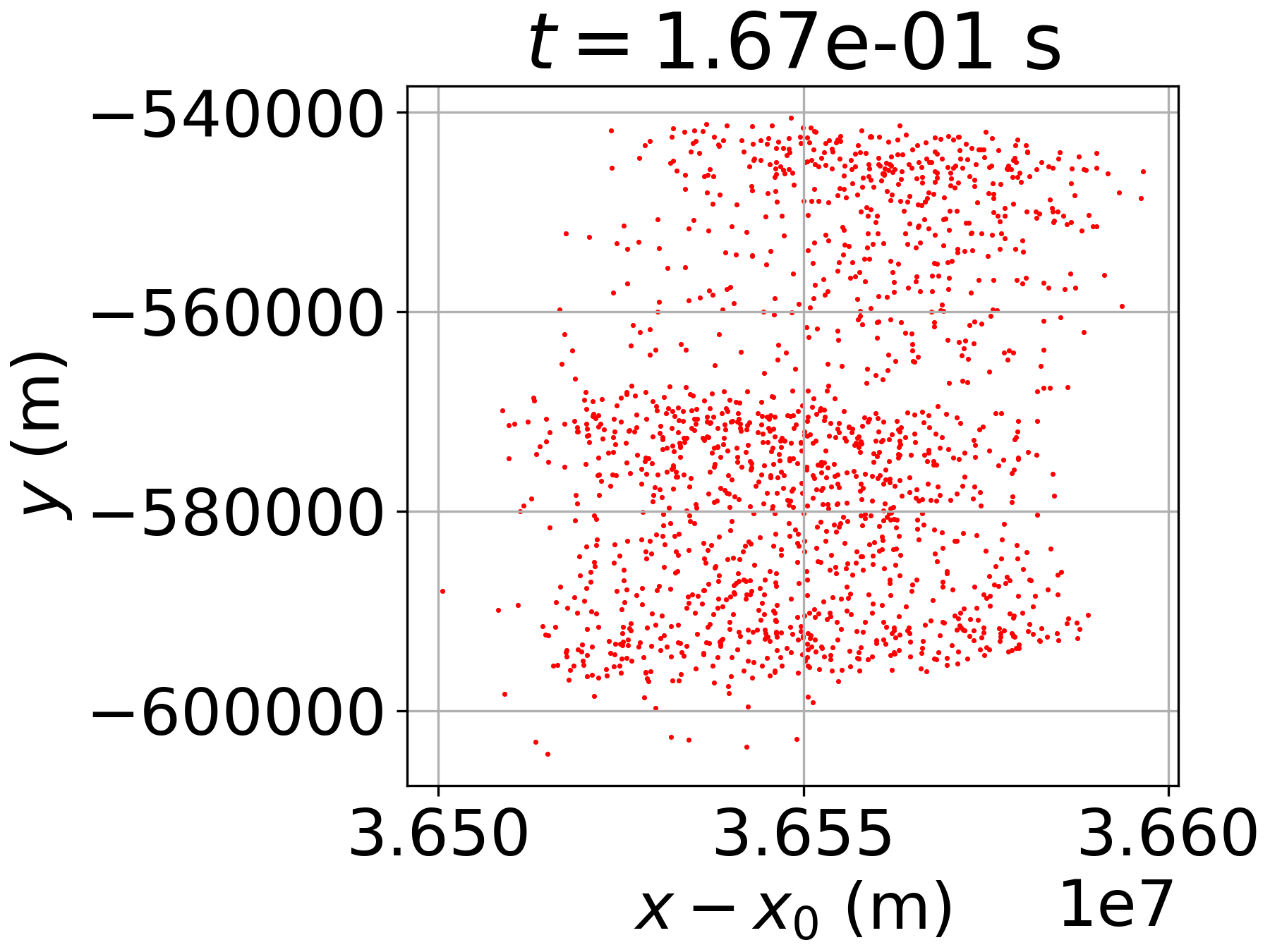}
\includegraphics[width=0.25\textwidth]{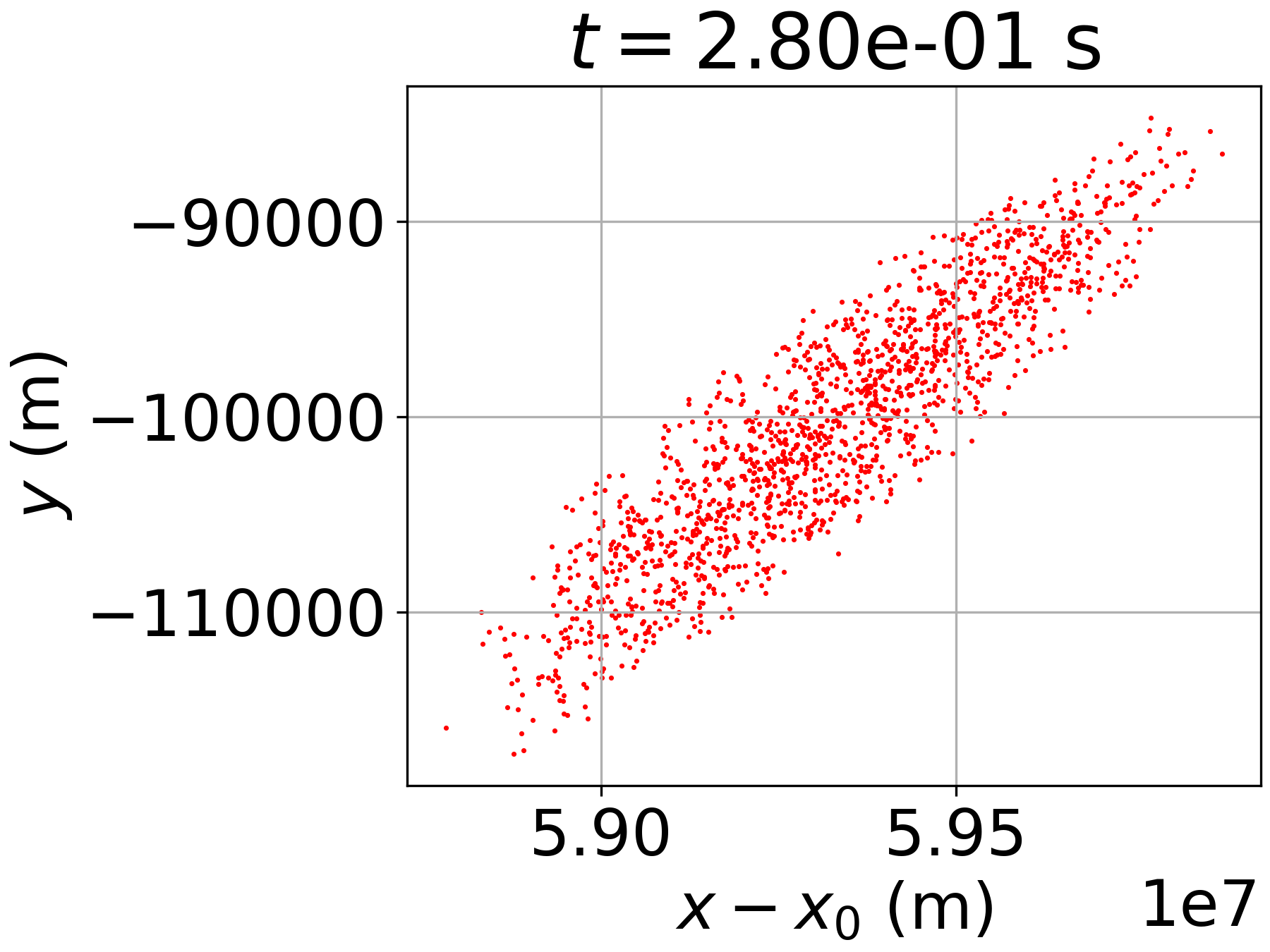}
\includegraphics[width=0.25\textwidth]{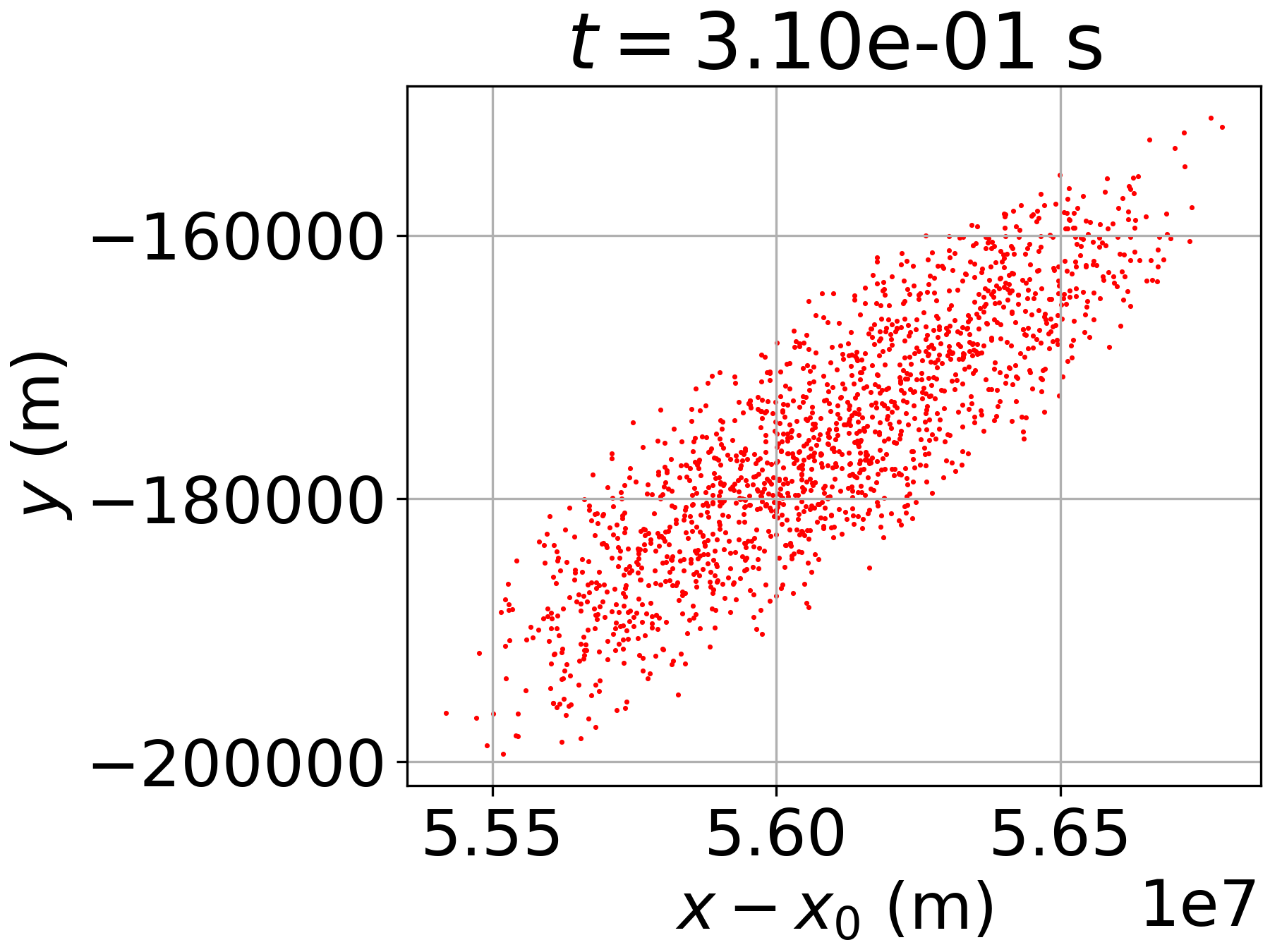}

\includegraphics[width=0.25\textwidth]{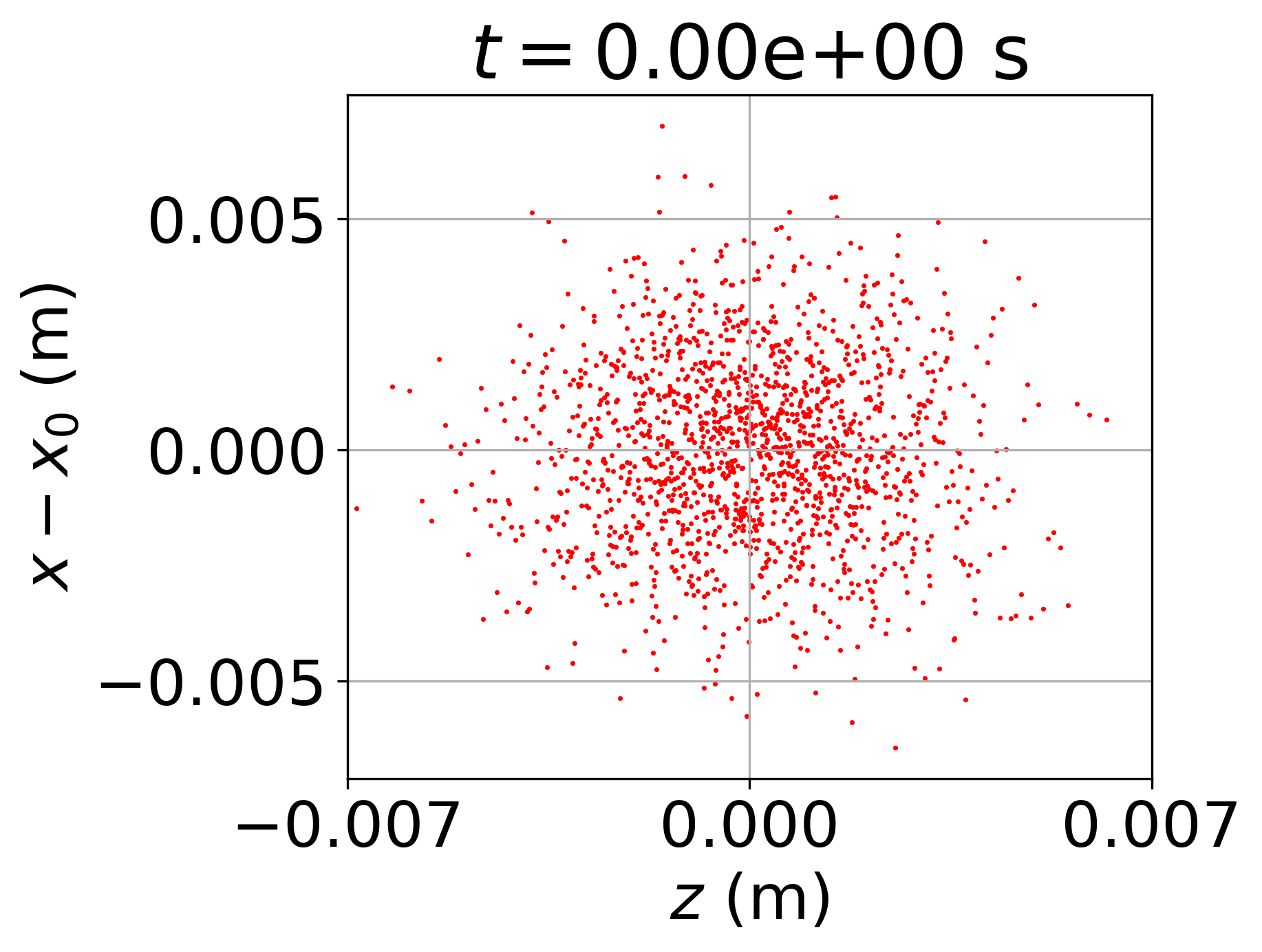}
\includegraphics[width=0.25\textwidth]{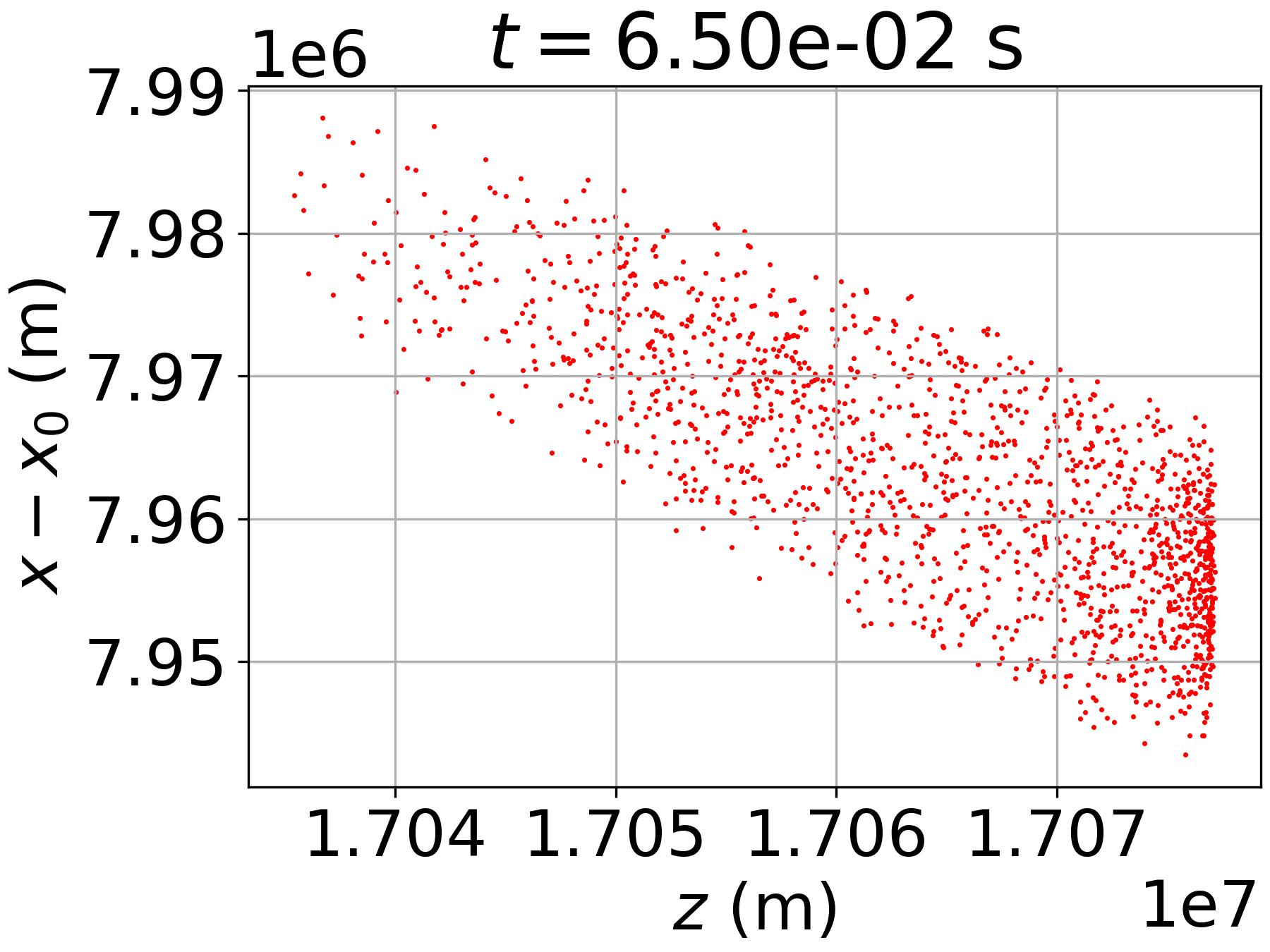}
\includegraphics[width=0.25\textwidth]{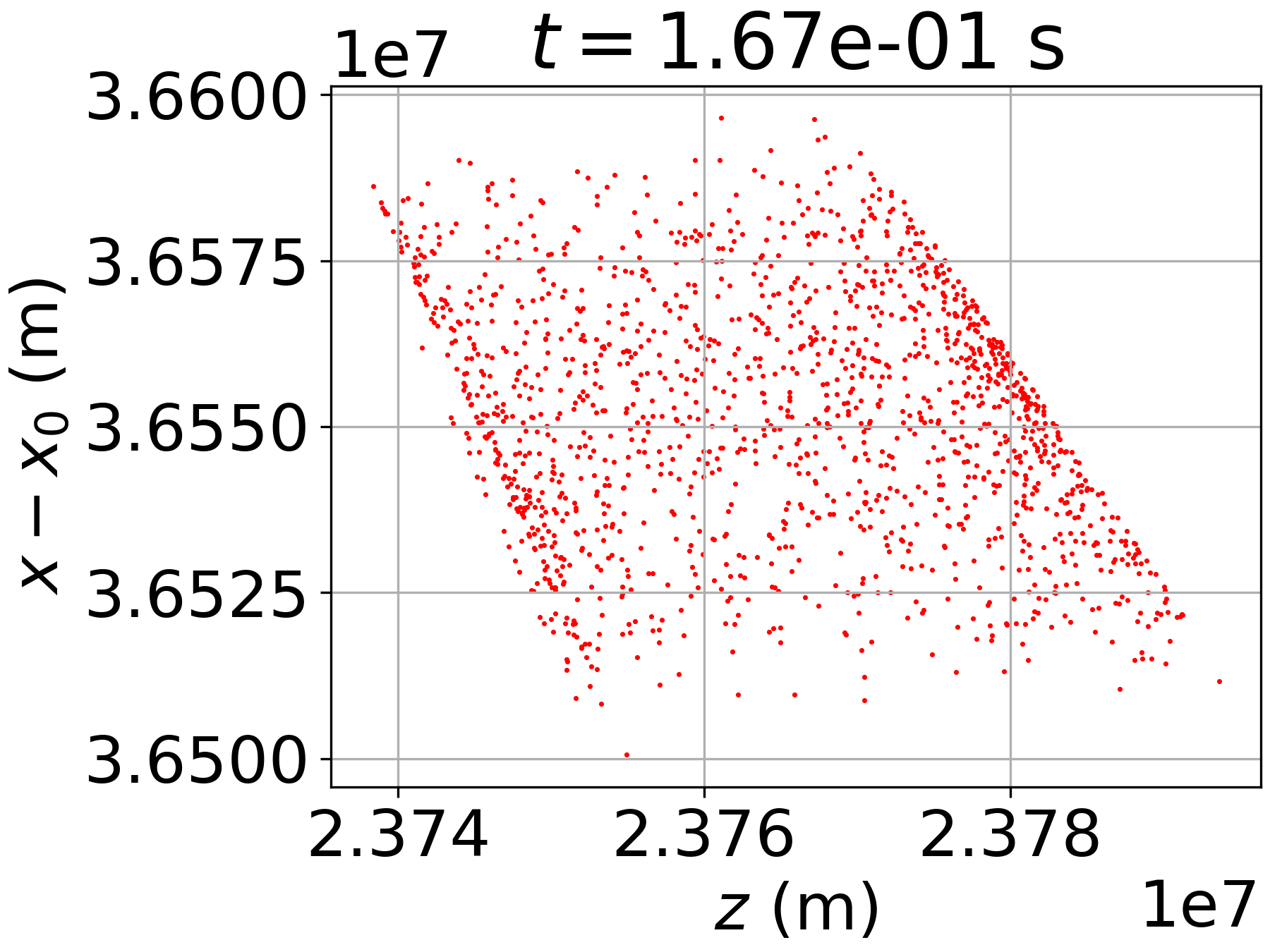}
\includegraphics[width=0.25\textwidth]{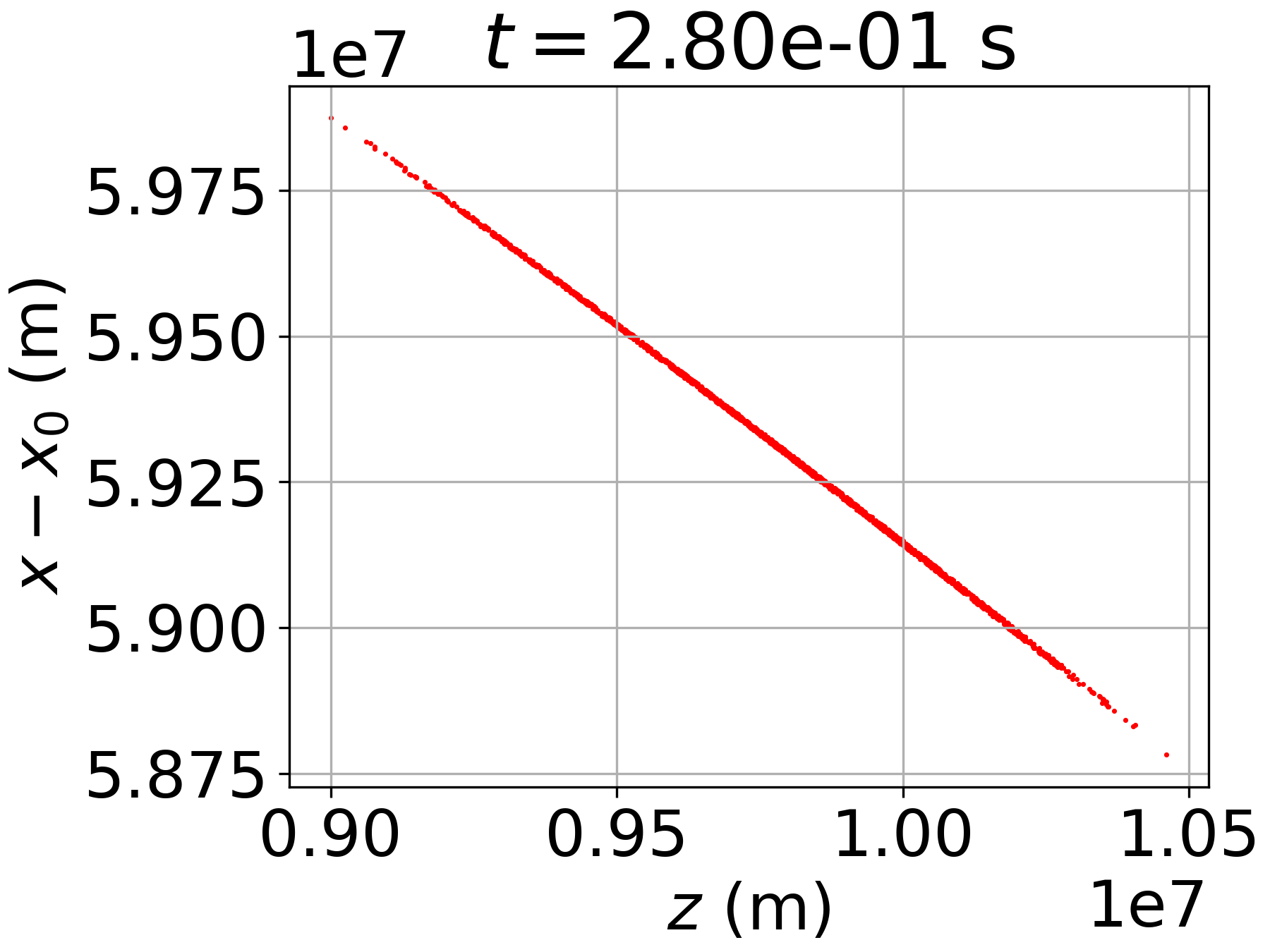}
\includegraphics[width=0.25\textwidth]{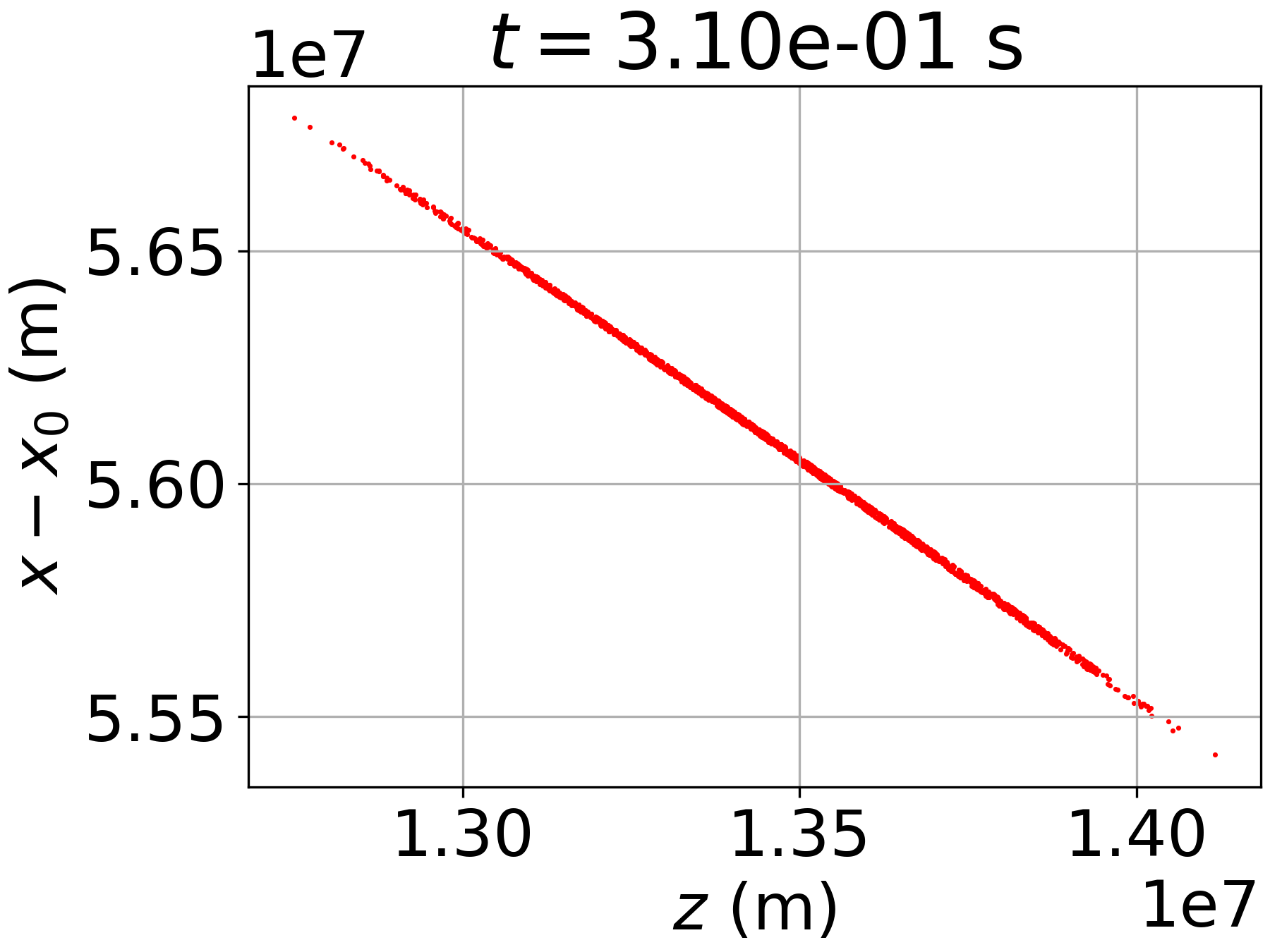}
\caption{
Long-range evolution of the relativistic electron bunch for the case
$N_p=1600$ and $E_0=10~\mathrm{MeV}$.
The top row shows the bunch trajectory in the $z$--$x$ plane
at several representative times during its propagation from
$(-10R_{\mathrm{E}},0,0)$ in the prescribed dipole geomagnetic field
(the blue circle denotes the Earth).
The middle row presents the corresponding transverse particle distributions
in the $(x-x_0)$--$y$ plane,
and the bottom row shows the longitudinal distributions
in the $z$--$(x-x_0)$ plane.
}
\label{fig:1600_10}
\end{figure*}

\subsection{Long-Range Propagation of the 1 MeV Bunch}\label{sec:1MeV}

After the initial self-field-driven expansion,
the bunch enters a long-range transport stage
in which the large-scale geomagnetic field
becomes the dominant factor controlling the particle motion.
The corresponding evolution of the bunch with
$N_p=1600$ and $E_0=1~\mathrm{MeV}$
is shown in Fig.~\ref{fig:1600_1},
where both the centroid trajectory
and the particle distributions at several representative times
are presented.

At the initial instant,
the bunch is highly compact
and remains close to the injection position at
$(-10R_{\mathrm{E}},0,0)$.
As discussed in Sec.~\ref{sec:early},
the bunch first experiences a rapid self-field-driven expansion,
which mainly increases its transverse size
and weakly stretches the bunch in the longitudinal direction.
Once this early dense stage has passed,
the particles become sufficiently separated
that the inter-particle electromagnetic interaction
is no longer the primary driver of the motion.
The later evolution is then governed mainly by the prescribed dipole field,
with the bunch following the large-scale geomagnetic guiding geometry.

The trajectory projections in Fig.~\ref{fig:1600_1}
show that the 1 MeV bunch continues to propagate
from the outer magnetosphere toward the Earth
without exhibiting strong magnetic mirroring
before reaching the near-Earth region.
Compared with the single-particle benchmark,
the overall path of the bunch centroid remains similar,
which indicates that the large-scale transport is still controlled
primarily by the geomagnetic field.
However, unlike the ideal single-particle case,
the finite-size bunch now carries a broadened spatial envelope
as a consequence of the early collective expansion.

The particle distributions at intermediate times
demonstrate this transition clearly.
At early times,
the bunch cross section expands rapidly,
reflecting the strong release of self-field energy.
At later times,
although the bunch continues to broaden gradually,
its overall morphology evolves more smoothly,
and the particles move along similar large-scale trajectories.
This behavior indicates that,
for the 1 MeV case,
the principal role of the self-field interaction
is to determine the initial relaxation and expansion of the bunch,
whereas the subsequent transport from the outer magnetosphere
toward the Earth is predominantly controlled
by the background geomagnetic field.

Another notable feature of Fig.~\ref{fig:1600_1}
is that the bunch does not remain rigid during transport.
Instead, the particle cloud develops a finite transverse extent
and a non-negligible axial spread,
which may affect the final precipitation footprint
in the near-Earth region.
Therefore, even when the later propagation is largely field-dominated,
the early self-field interaction still leaves a persistent imprint
on the beam envelope.
This result justifies the use of the two-stage strategy:
the full particle-particle interaction is essential
for accurately capturing the early dense stage,
while the later smoother transport can be computed efficiently
once the bunch has entered the weakly coupled regime.

Overall, the 1 MeV case demonstrates a physically consistent
two-stage evolution.
The bunch first undergoes a rapid self-field-driven expansion
immediately after injection,
and then propagates over a long distance
under the dominant influence of the geomagnetic dipole field
until it approaches the Earth.
This case therefore serves as the baseline reference
for comparison with the higher-energy bunches
considered in the next subsection.

\begin{figure*}[!ht]
\centering
\includegraphics[width=0.18\textwidth]{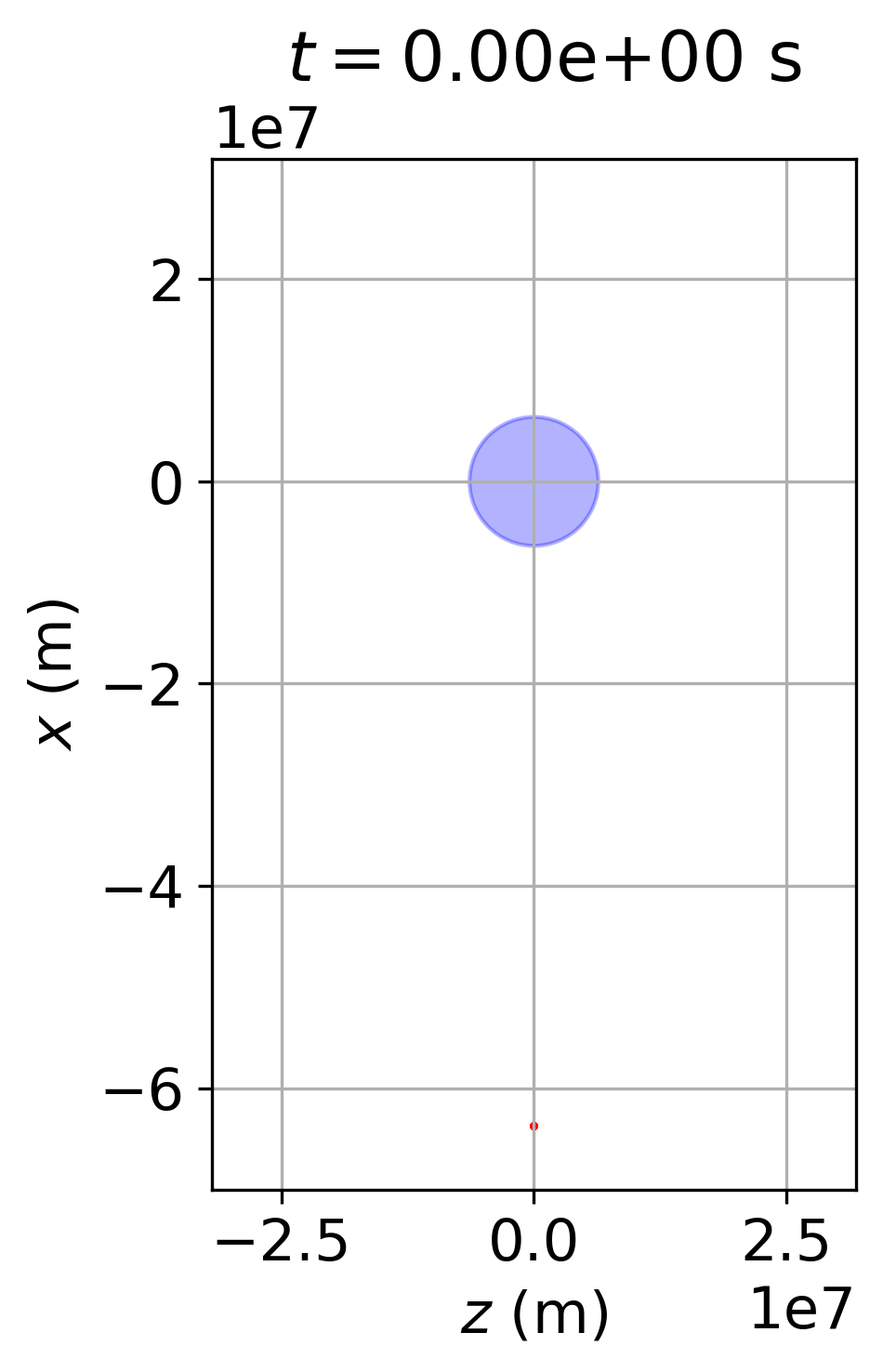}
\includegraphics[width=0.18\textwidth]{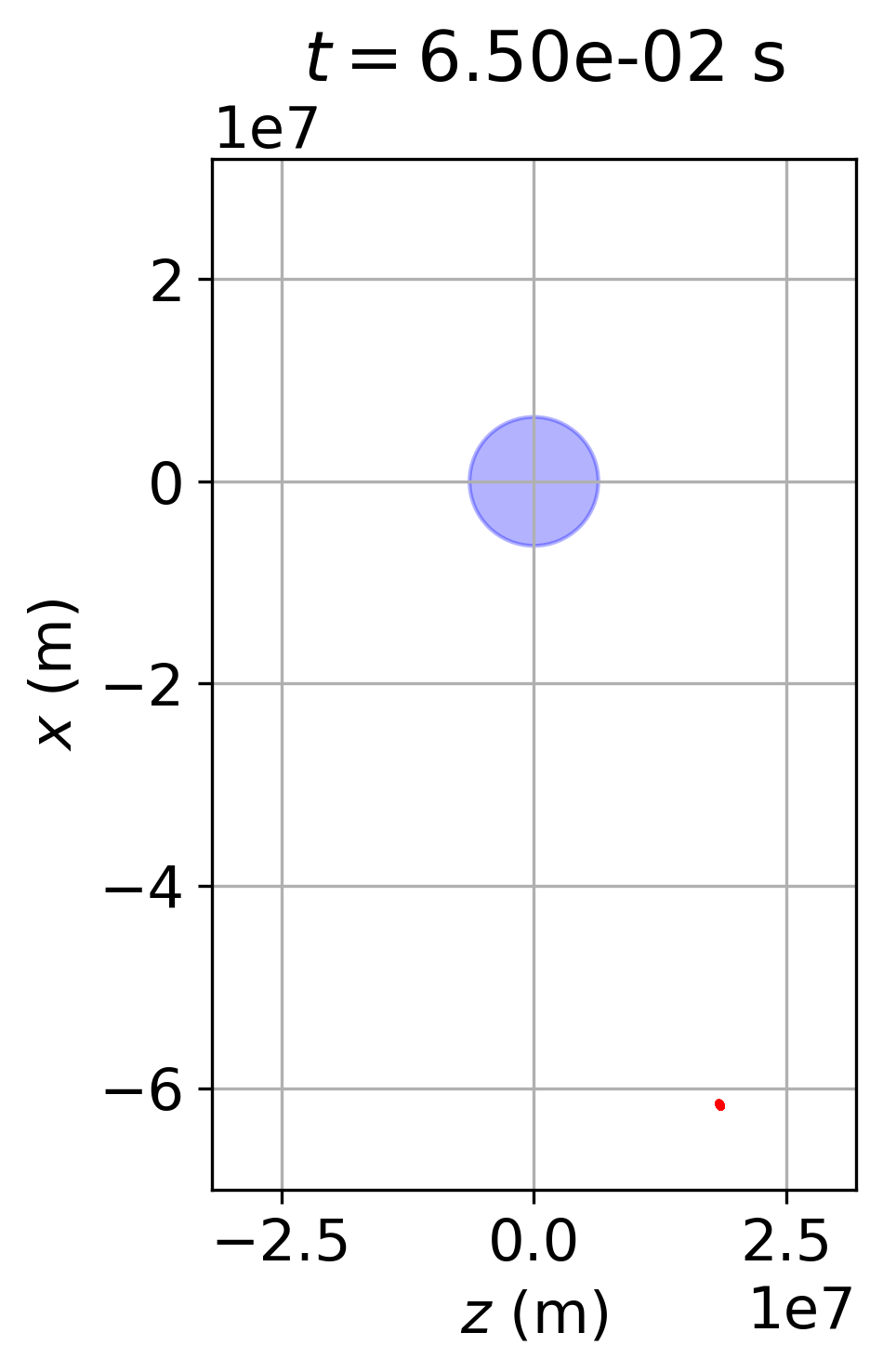}
\includegraphics[width=0.18\textwidth]{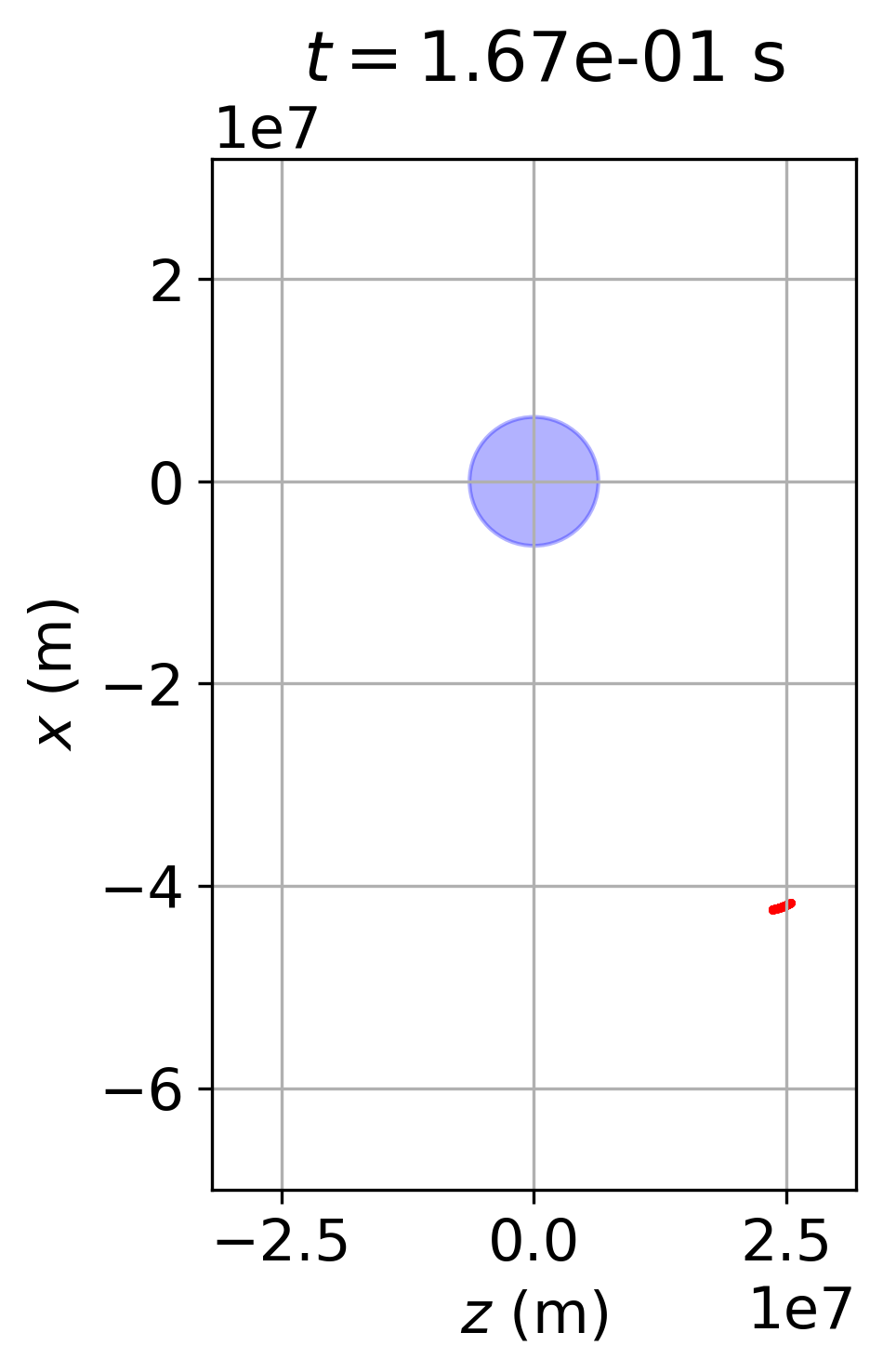}
\includegraphics[width=0.18\textwidth]{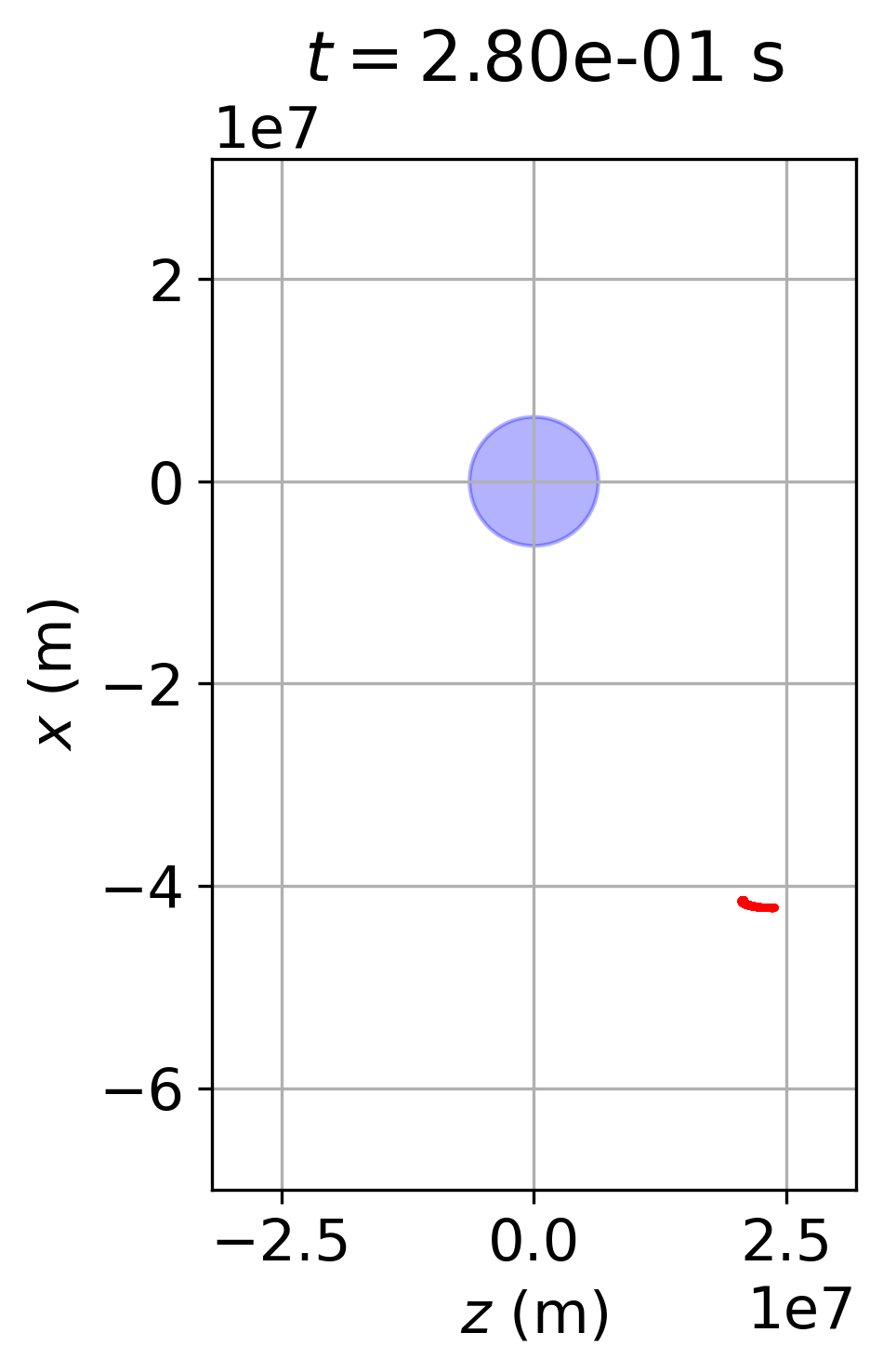}
\includegraphics[width=0.18\textwidth]{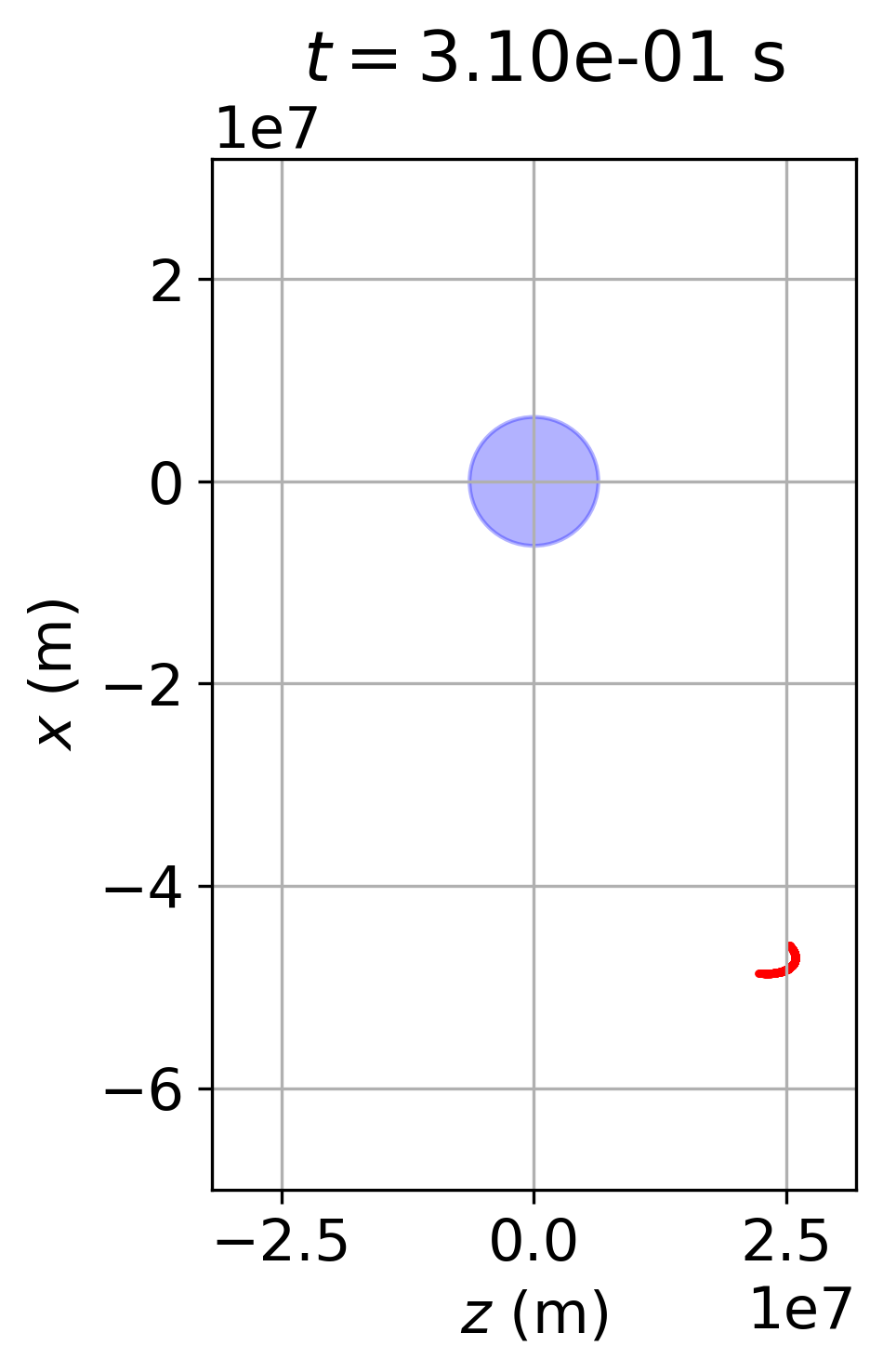}

\includegraphics[width=0.25\textwidth]{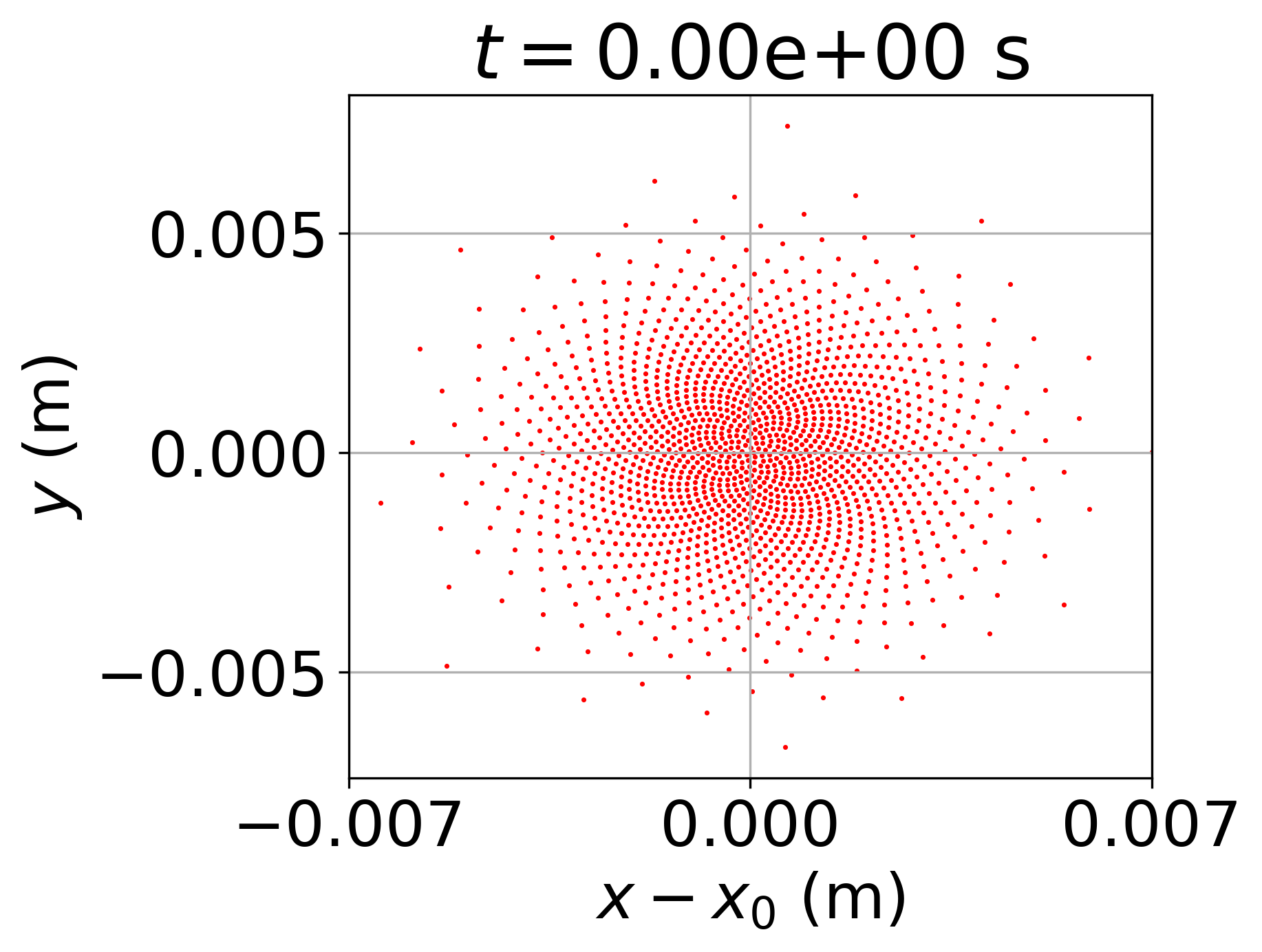}
\includegraphics[width=0.25\textwidth]{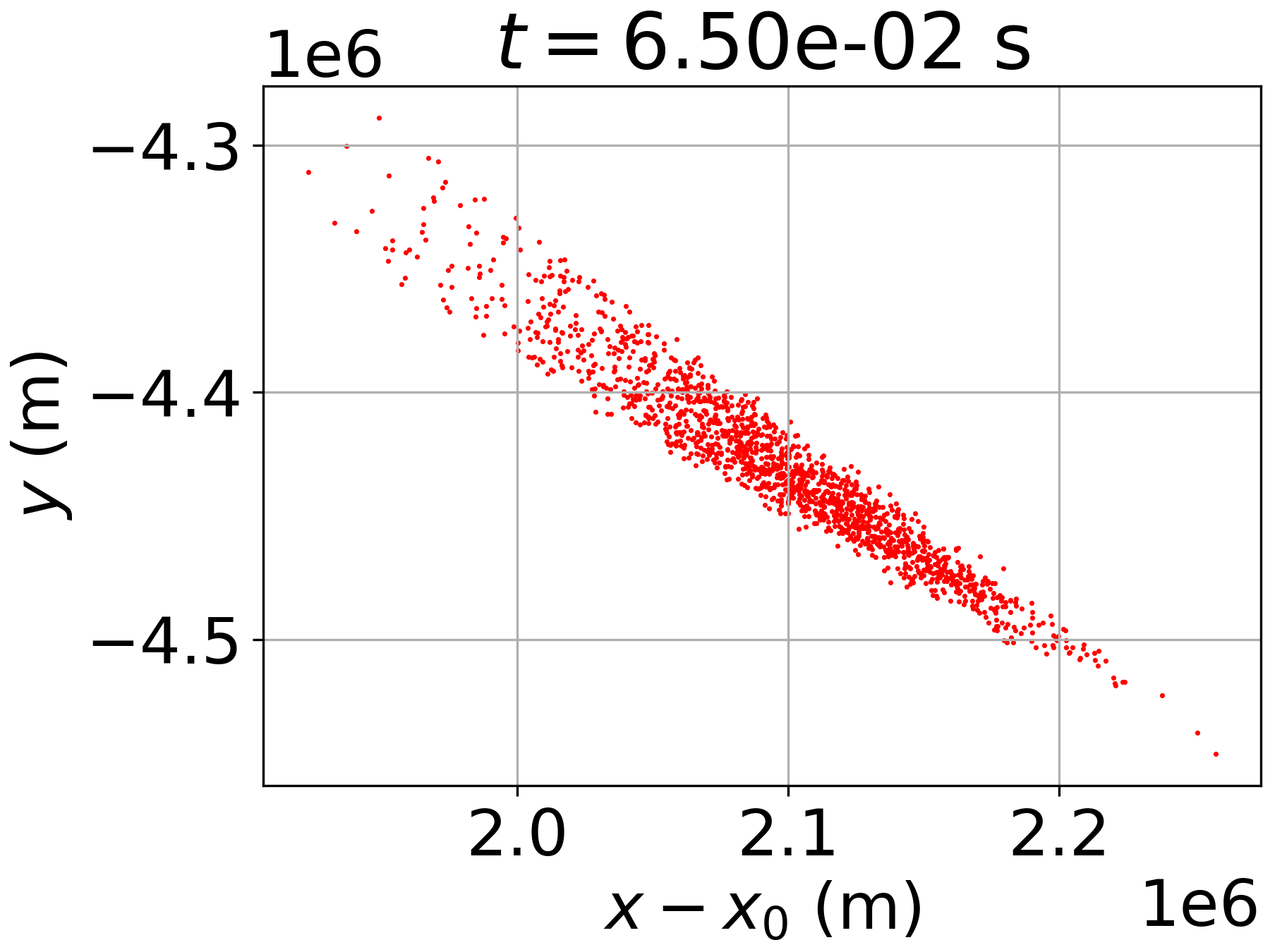}
\includegraphics[width=0.25\textwidth]{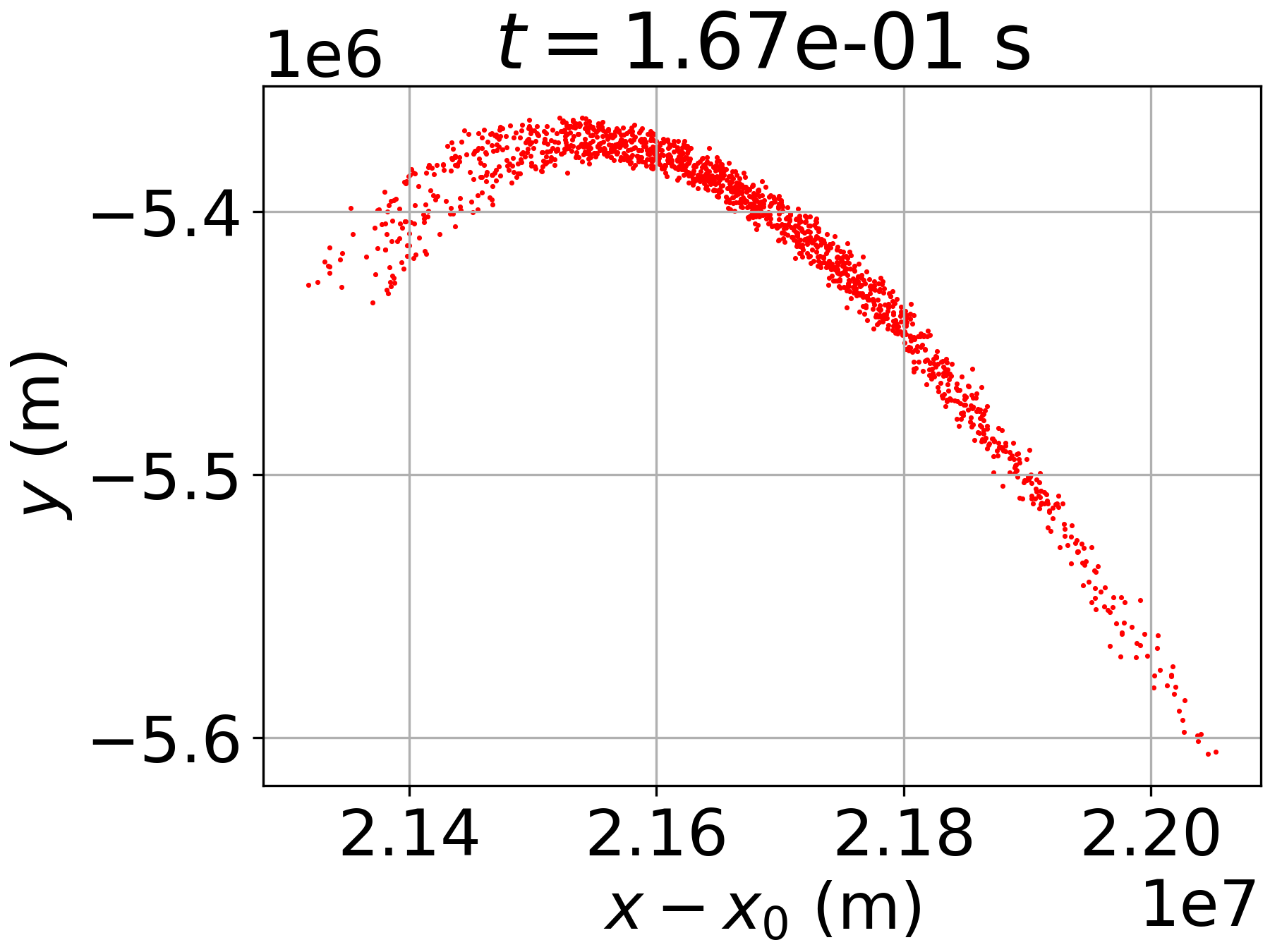}
\includegraphics[width=0.25\textwidth]{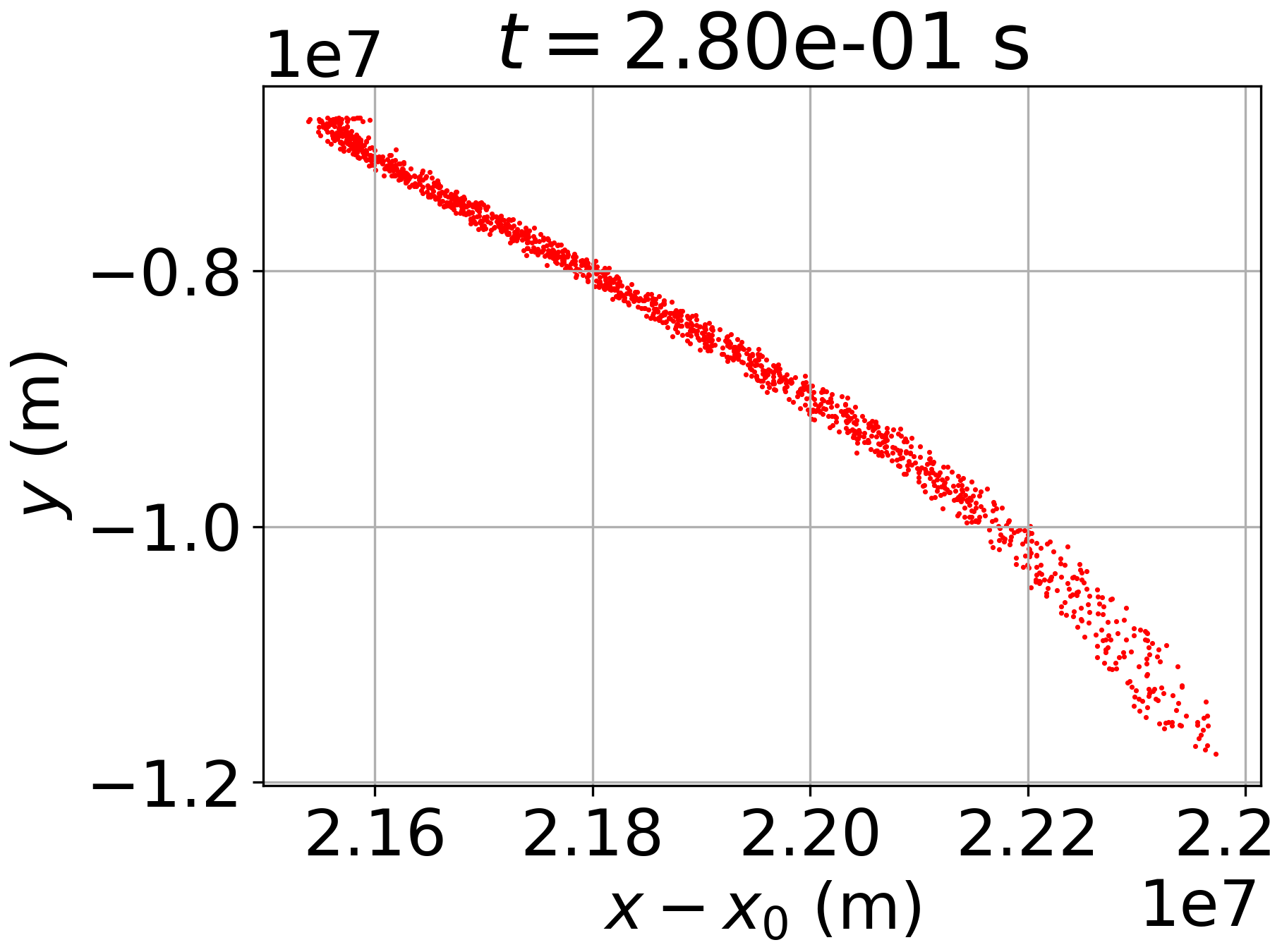}
\includegraphics[width=0.25\textwidth]{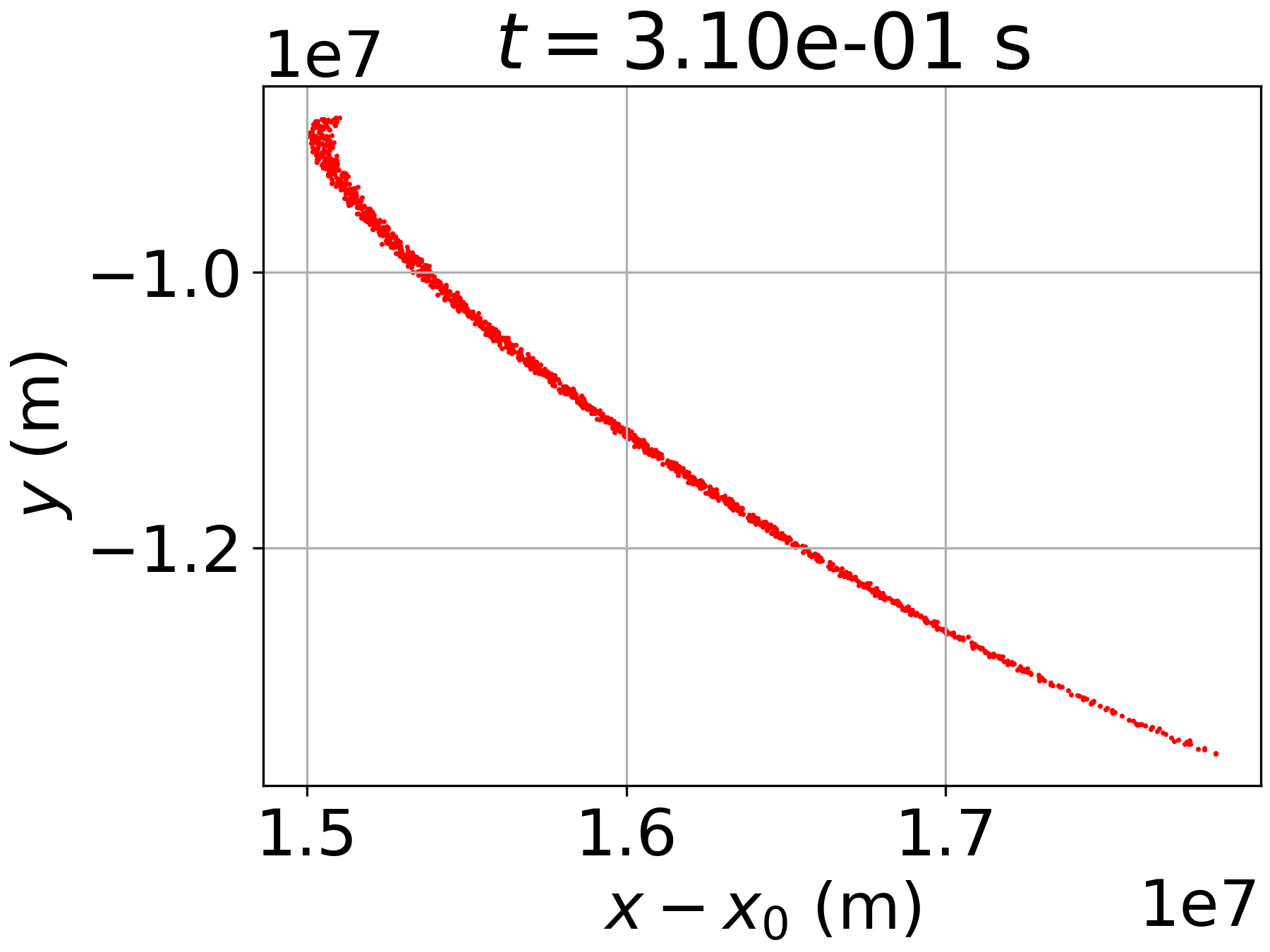}

\includegraphics[width=0.25\textwidth]{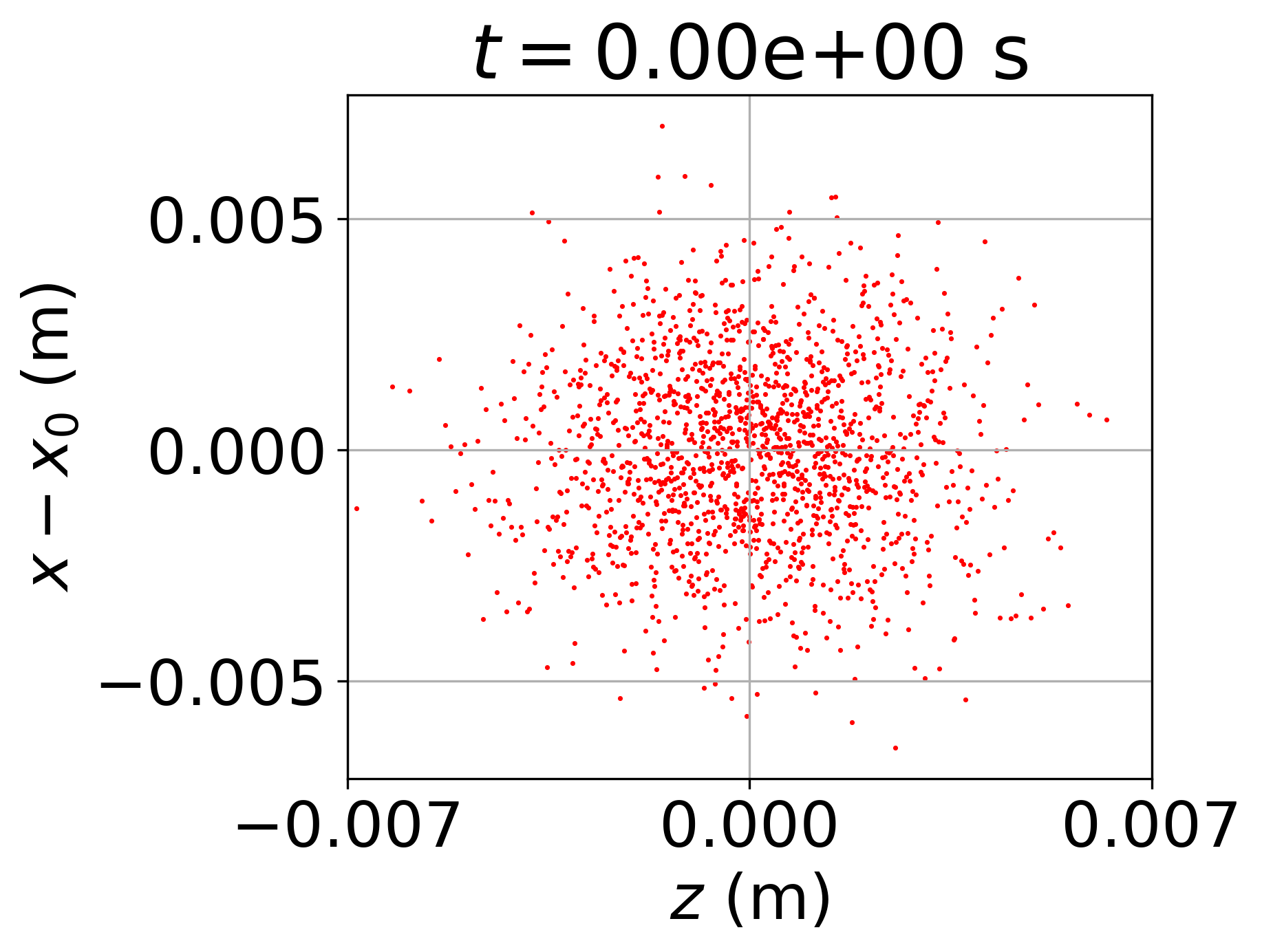}
\includegraphics[width=0.25\textwidth]{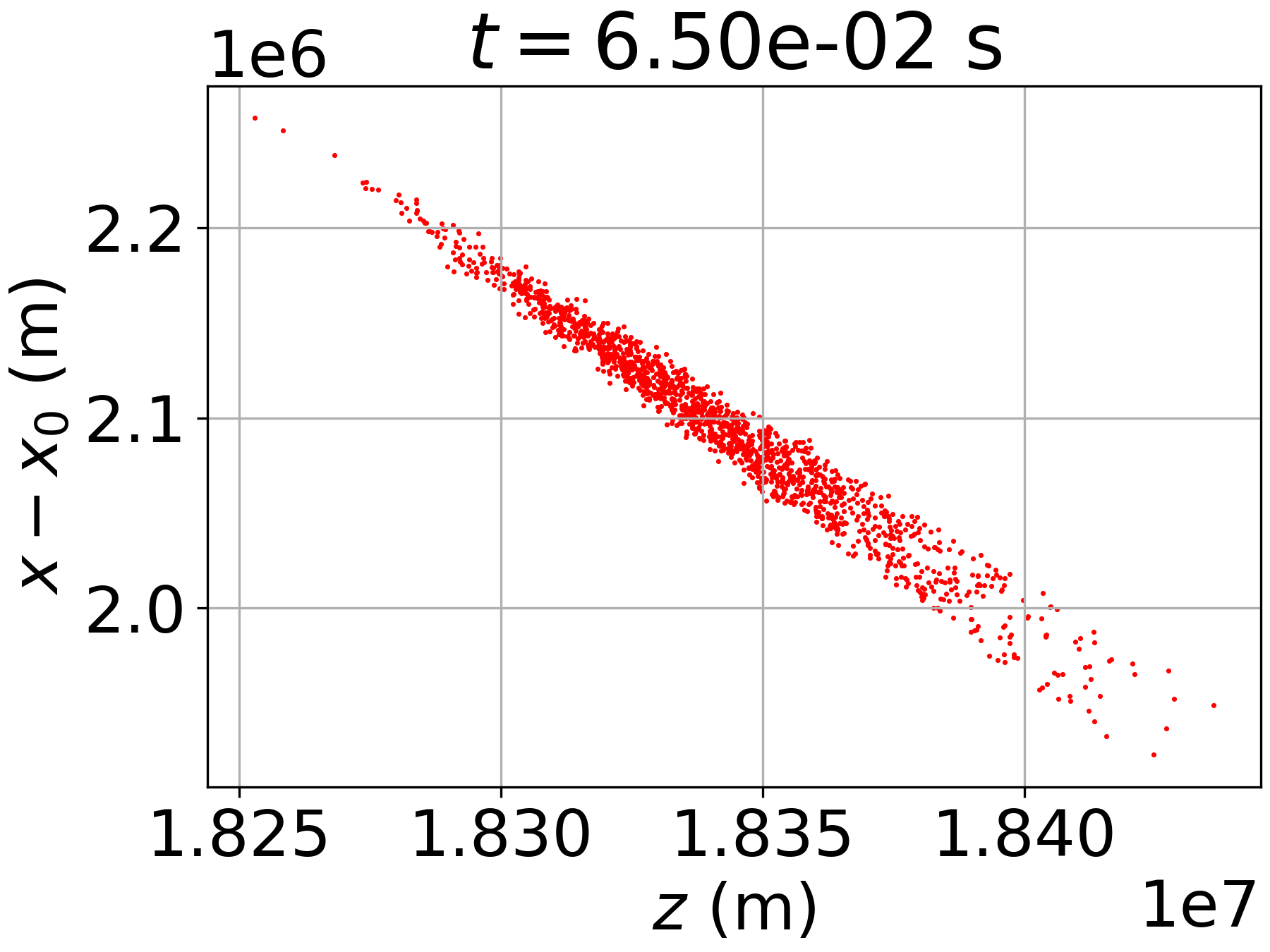}
\includegraphics[width=0.25\textwidth]{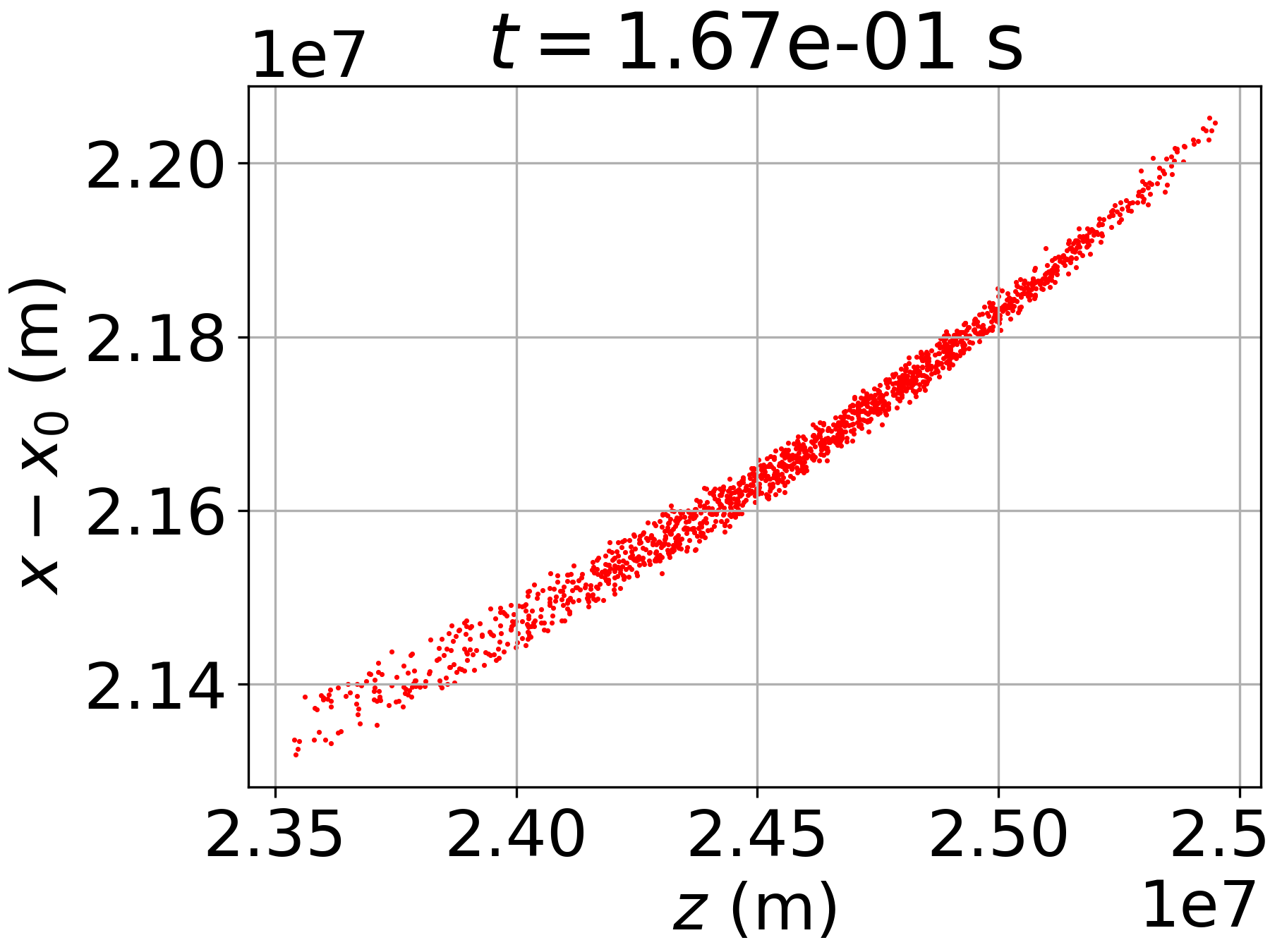}
\includegraphics[width=0.25\textwidth]{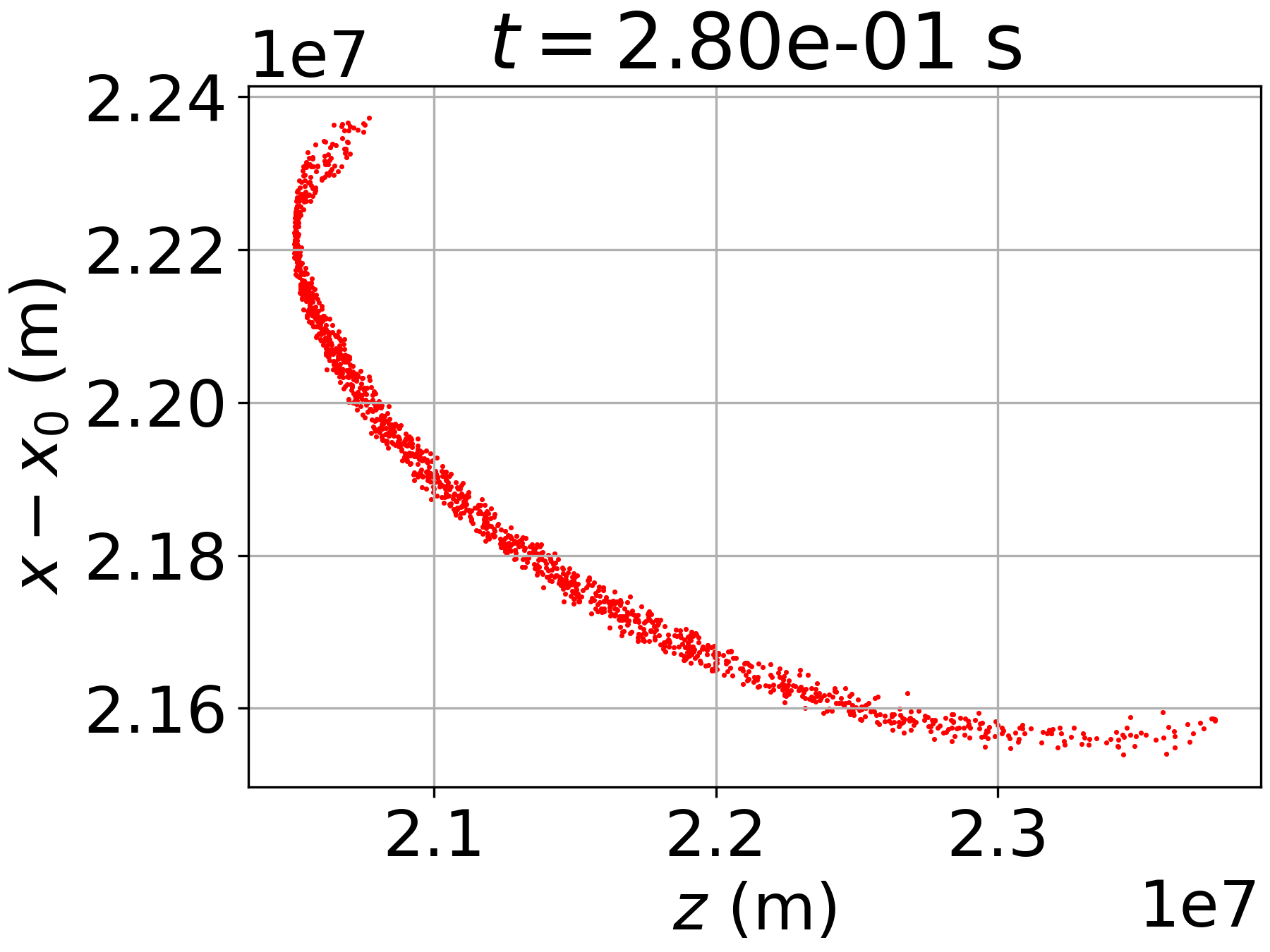}
\includegraphics[width=0.25\textwidth]{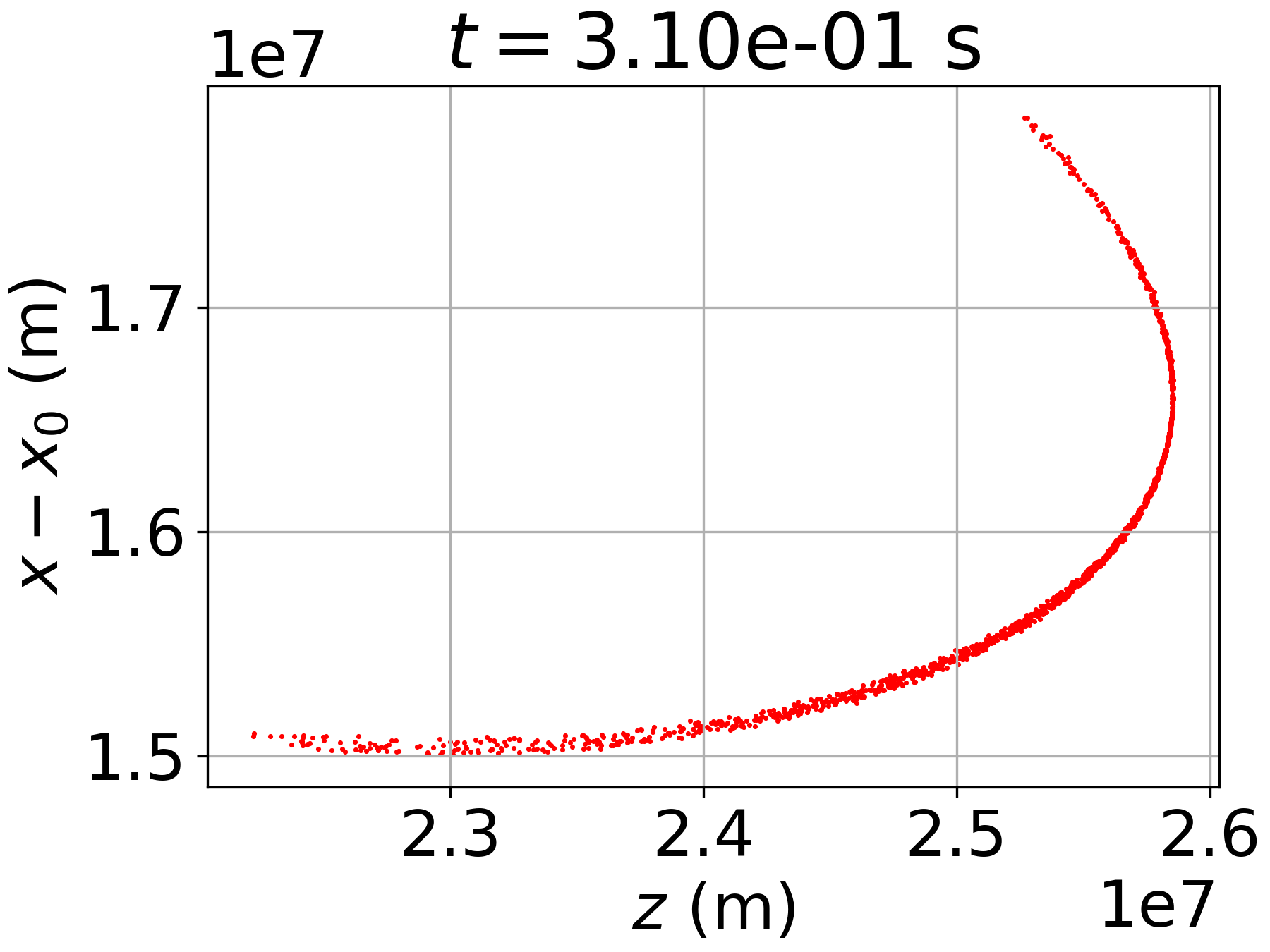}
\caption{
Long-range evolution of the relativistic electron bunch for the case
$N_p=1600$ and $E_0=100~\mathrm{MeV}$.
The top row shows the bunch trajectory in the $z$--$x$ plane
at several representative times during its propagation from
$(-10R_{\mathrm{E}},0,0)$ in the prescribed dipole geomagnetic field
(the blue circle denotes the Earth).
The middle row presents the corresponding transverse particle distributions
in the $(x-x_0)$--$y$ plane,
and the bottom row shows the longitudinal distributions
in the $z$--$(x-x_0)$ plane.
}
\label{fig:1600_100}
\end{figure*}

\subsection{Energy Dependence: 10 MeV and 100 MeV Bunches}\label{sec:highE}

To examine the influence of beam energy on the long-range transport, the baseline calculation is extended from
$E_0=1~\mathrm{MeV}$ to the higher-energy cases $E_0=10~\mathrm{MeV}$ and $E_0=100~\mathrm{MeV}$.
The corresponding bunch trajectories and representative particle distributions
are shown in Fig.~\ref{fig:1600_10} and Fig.~\ref{fig:1600_100}.
For these higher-energy cases, the switching times used in the framework
are determined from the energy-based criterion described in Sec.~\ref{sec:timestep},
so that the early strongly coupled stage and the later weakly coupled transport stage are treated consistently for each beam energy.

As in the 1 MeV case, the bunches with 10 MeV and 100 MeV still undergo an initial self-field-driven relaxation stage immediately after injection. However, as the beam energy increases,
the relative dynamical importance of this early collective relaxation becomes weaker compared with the much larger directed kinetic energy of the bunch.
Consequently, the later large-scale motion of the high-energy bunches
is controlled more strongly by the prescribed dipole geomagnetic field
than by the residual inter-particle interaction.
In this sense, increasing beam energy shifts the dynamics
from a regime in which early self-field relaxation
has a visible influence on the later transport
to a regime in which the large-scale geomagnetic geometry
plays the dominant role.

For the 10 MeV case shown in Fig.~\ref{fig:1600_10},
the bunch does not continue to precipitate toward the Earth
under the present injection conditions.
Instead, the trajectory indicates a non-precipitating motion
that is consistent with magnetic mirroring
in the dipole geomagnetic field.
The overall bunch path remains smooth and highly coherent,
which suggests that the later transport is governed mainly
by the large-scale magnetic configuration.
The early self-field interaction still contributes
to the initial bunch expansion and envelope evolution,
but it does not appear to dominate the final large-scale trajectory.
A similar behavior is found for the 100 MeV case in Fig.~\ref{fig:1600_100},
for which the bunch also fails to reach the Earth under the present injection configuration.

Under the present injection geometry, both the 10 MeV and 100 MeV cases mirror before reaching the Earth, as shown Fig. by\ref{fig:1600_10} and Fig. \ref{fig:1600_100}, consistent with the findings of Borovsky et al.~\cite{borovsky2022modification}.
For relativistic electrons propagating in the geomagnetic field, whether the bunch eventually precipitates to the Earth depends on the matching among the initial injection position, the injection direction, and the particle energy.
Therefore, when the beam energy is increased from the 1 MeV case to 10 MeV or 100 MeV, the injection direction should in general be adjusted accordingly
if precipitation to the Earth is to be maintained.
In this sense, the failure of the high-energy bunches
to reach the Earth in the present calculations
should be regarded as a consequence of the current injection geometry rather than an unexpected result.

\begin{figure}[!ht]
\centering
\includegraphics[width=0.4\textwidth]{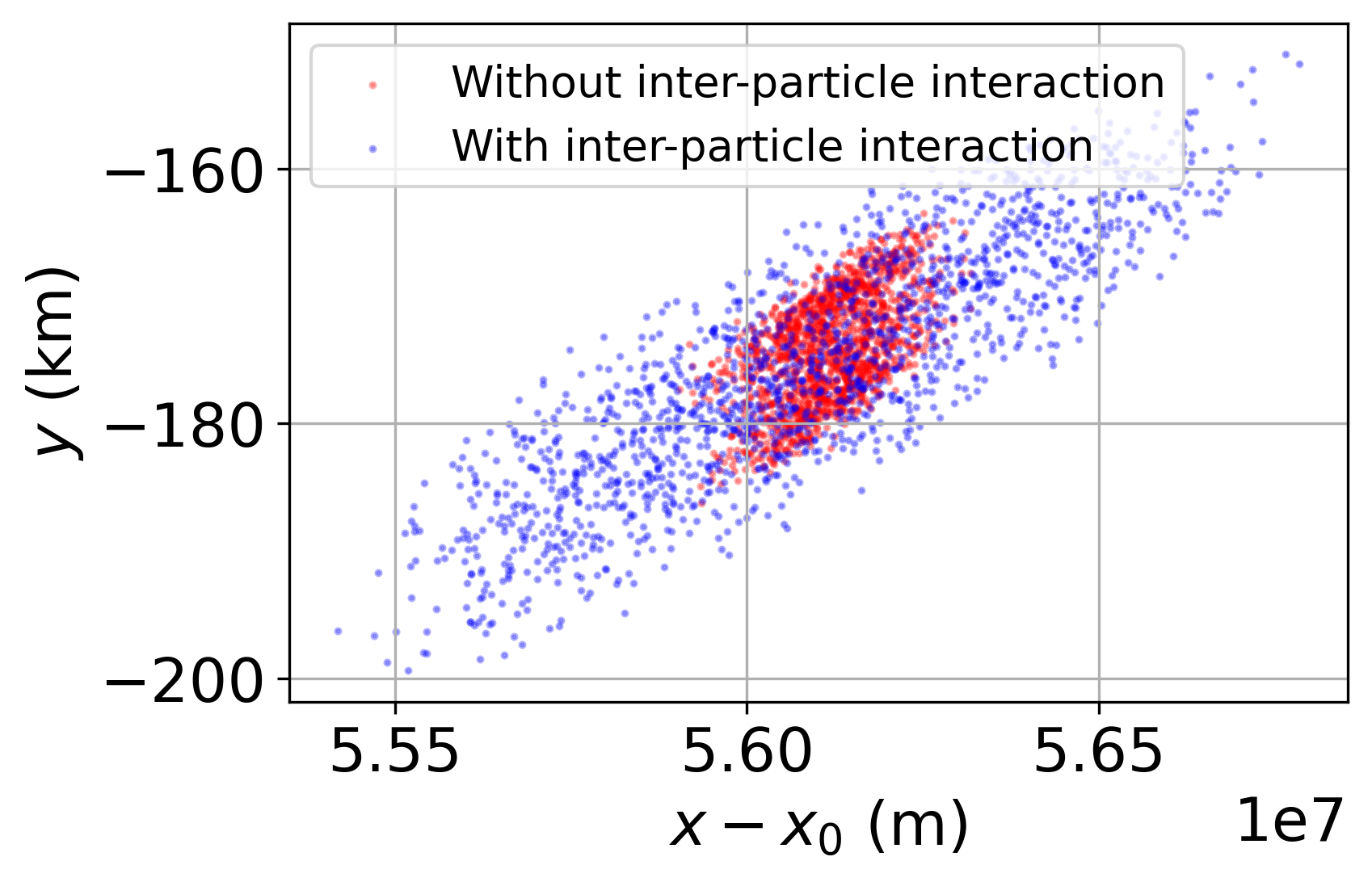}
\includegraphics[width=0.4\textwidth]{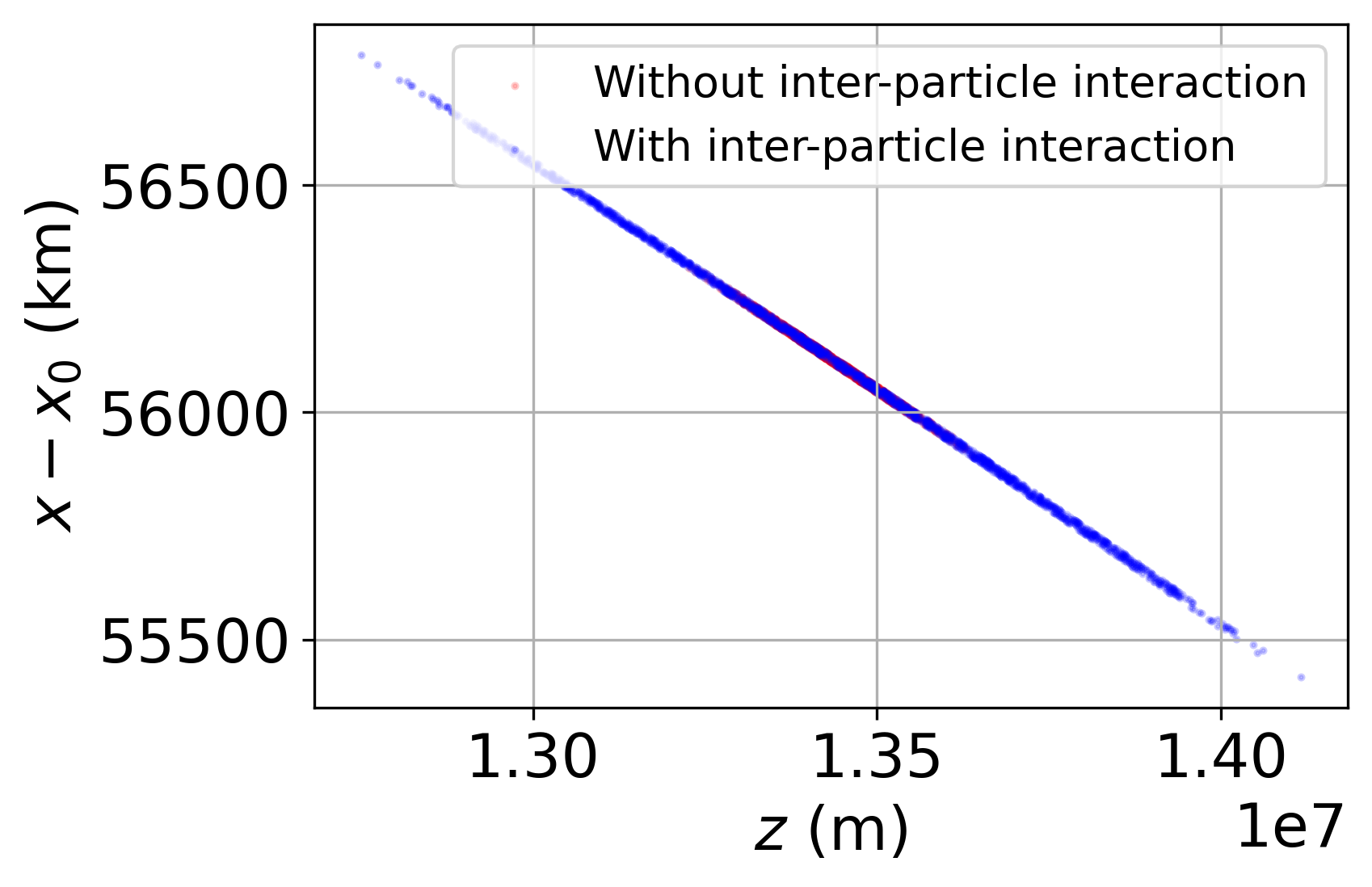}
\includegraphics[width=0.4\textwidth]{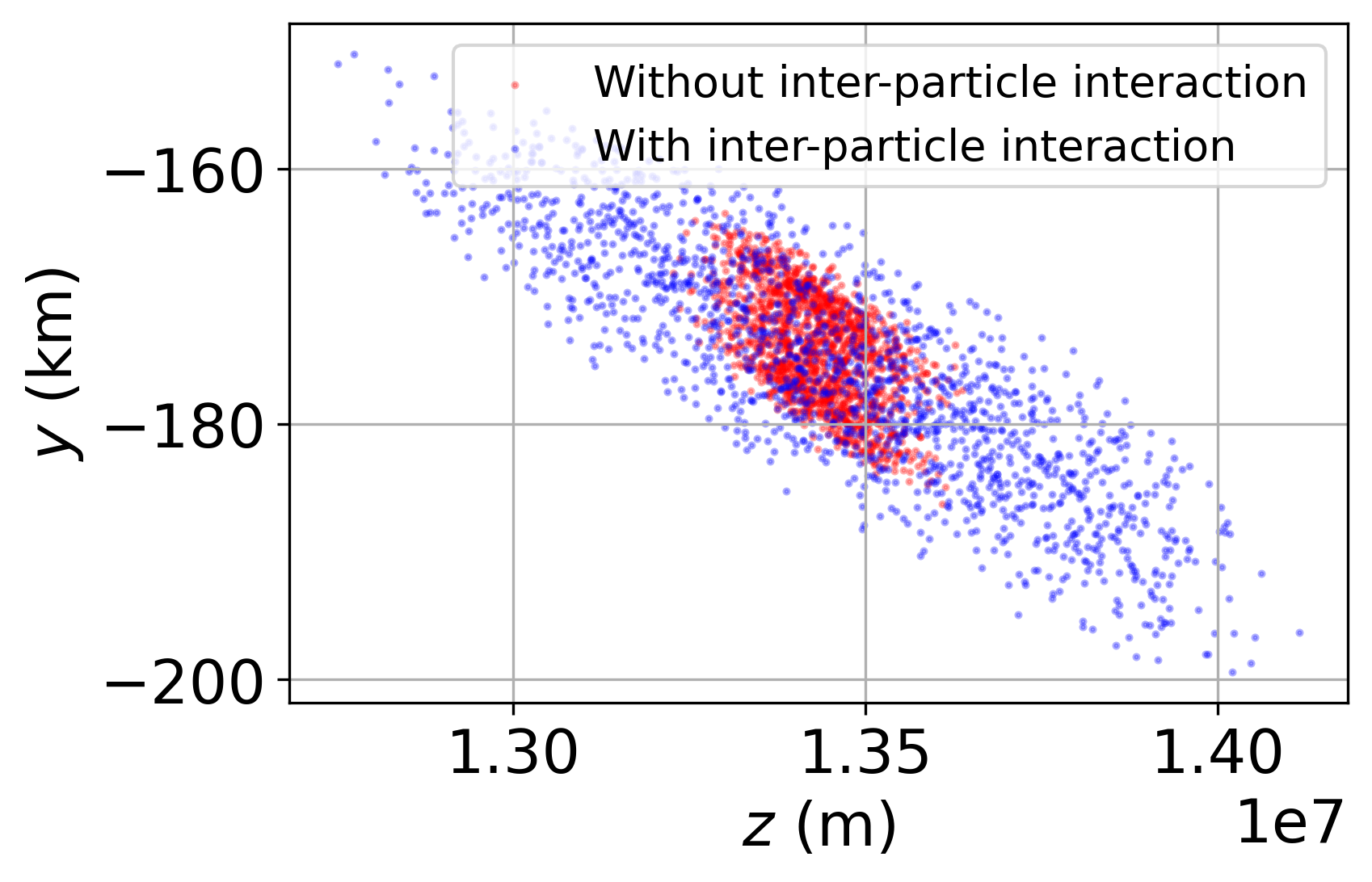}
\caption{
Comparison of the final particle distributions for the
$E_0=10~\mathrm{MeV}$ case with and without inter-particle interaction.
The red markers denote the calculation without inter-particle interaction,
whereas the blue markers denote the calculation with inter-particle interaction.
The three panels show the final distributions in the
$z$--$y$, $z$--$(x-x_0)$, and $(x-x_0)$--$y$ planes, respectively.
}
\label{fig:1600_10_force}
\end{figure}

\begin{figure}[!ht]
\centering
\includegraphics[width=0.4\textwidth]{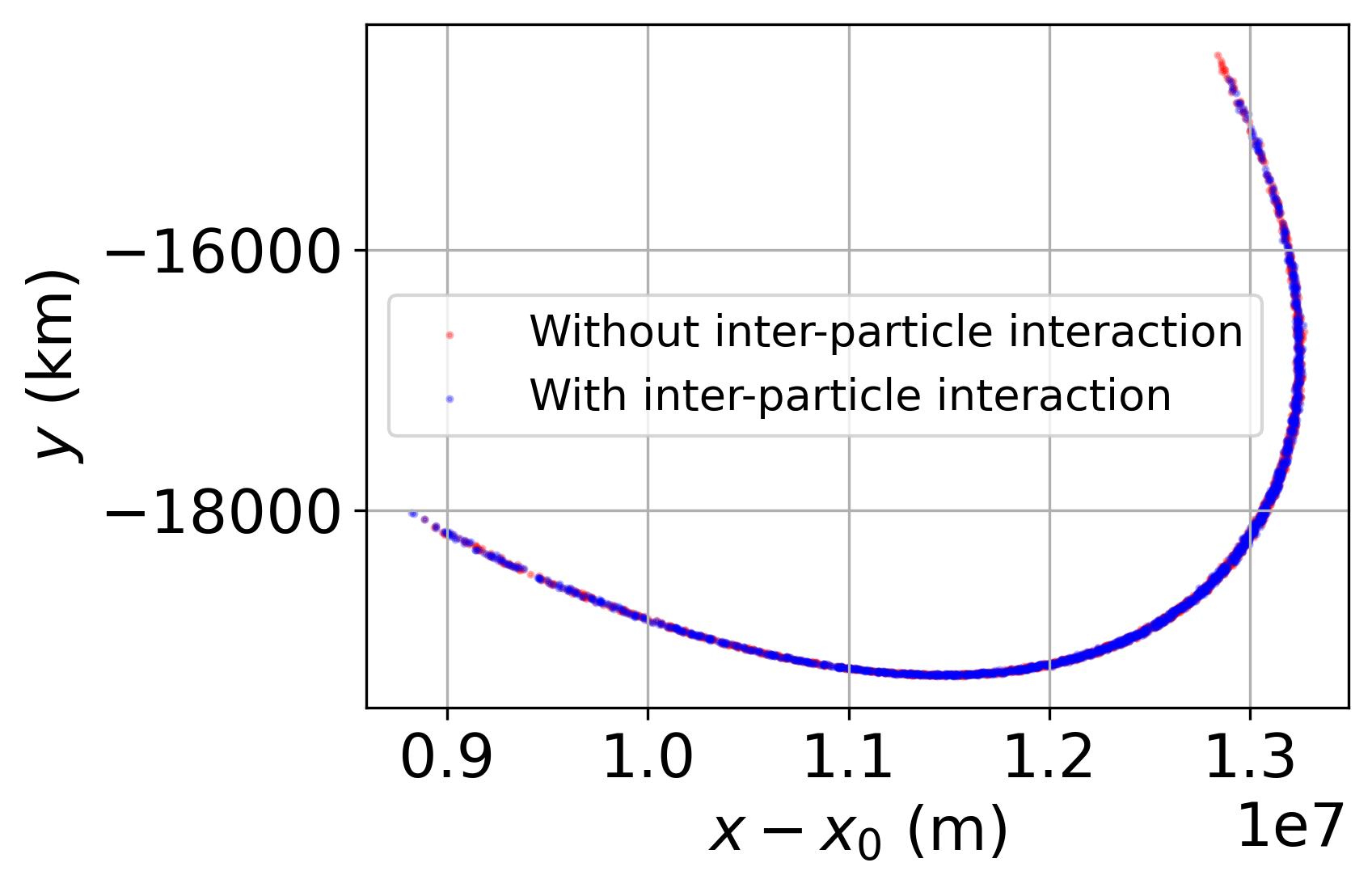}
\includegraphics[width=0.4\textwidth]{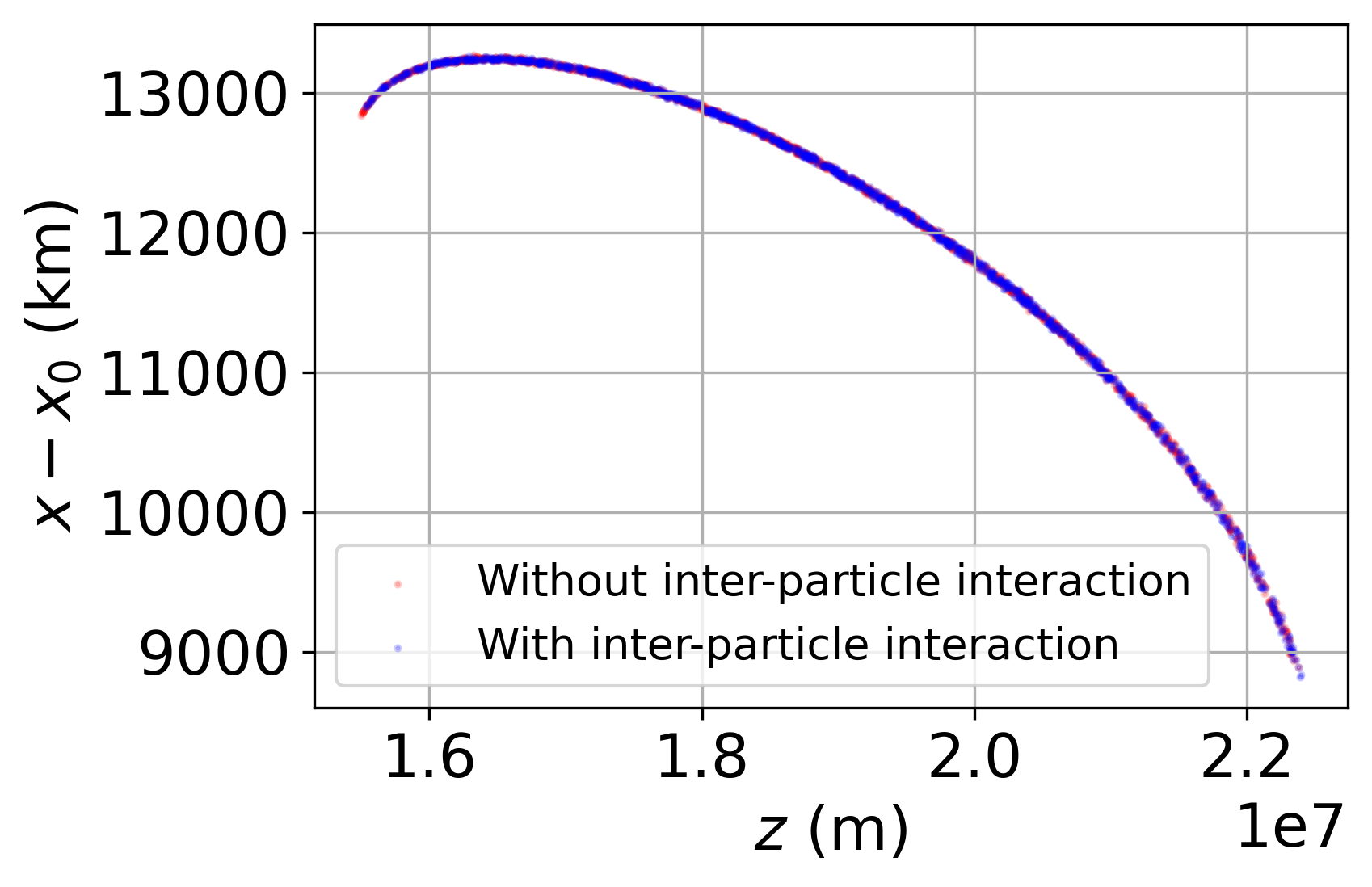}
\includegraphics[width=0.4\textwidth]{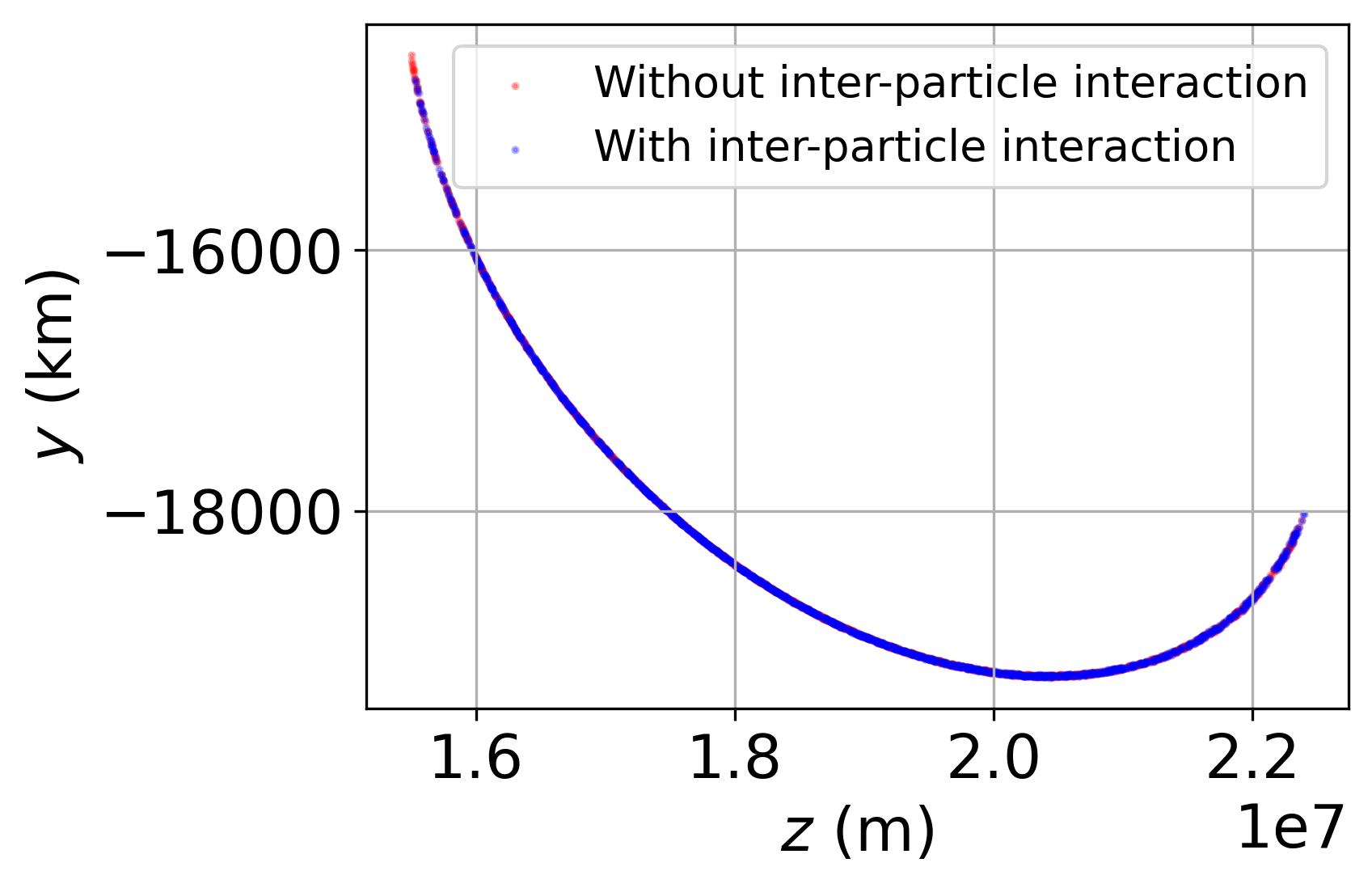}
\caption{
Comparison of the final particle distributions for the
$E_0=100~\mathrm{MeV}$ case with and without inter-particle interaction.
The red markers denote the calculation without inter-particle interaction,
whereas the blue markers denote the calculation with inter-particle interaction.
}
\label{fig:1600_100_force}
\end{figure}

To further examine the role of inter-particle interaction
in this high-energy regime,
additional control calculations were performed
for the $E_0=10~\mathrm{MeV}$ and $100~\mathrm{MeV}$ cases,
in which the inter-particle electromagnetic interaction
was neglected throughout the entire simulation.
The final particle distributions obtained with and without
inter-particle interaction are compared in
Fig.~\ref{fig:1600_10_force} and Fig.~\ref{fig:1600_100_force}.
In both cases, the mean transport path and the average final position remain nearly unchanged. This indicates that the macroscopic trajectory of the bunch is governed mainly by the large-scale geomagnetic field and is only weakly affected by the inclusion of inter-particle interaction.

The main effect of inter-particle interaction
is instead reflected in the bunch envelope.
When inter-particle interaction is included,
the bunch exhibits a visibly broader final distribution,
that is, a stronger spatial divergence,
while its average transported position remains essentially the same
as that obtained without inter-particle interaction.
Therefore, for the present 10 MeV and 100 MeV cases,
inter-particle interaction mainly modifies
the early-stage expansion and the detailed beam spreading,
but does not fundamentally alter the overall transport trajectory.

Accordingly, the question of whether a bunch
ultimately reaches the Earth
is treated here as a geometry- and energy-matching problem
under the prescribed injection setup,
rather than the primary focus of this work.
The main objective of the present study
is to accurately resolve the early dense-stage
particle-particle electromagnetic interaction
and to couple it efficiently to the subsequent long-range transport.
From this perspective,
the high-energy cases demonstrate that the present method
can capture both the self-field-driven bunch divergence
and the later field-dominated propagation
within a unified computational framework.

\section{Conclusion}\label{sec:conclusion}

In this work, the electromagnetic particle-particle (EM-PP) method,
previously developed for relativistic two-particle interaction,
has been extended to the long-range transport of relativistic electron bunches
in the Earth's magnetosphere. A two-stage strategy has been adopted
to resolve the dense early stage with full particle-particle electromagnetic interaction and to advance the later weakly coupled transport efficiently under the prescribed geomagnetic field. In this way, the method retains the mesh-free accuracy required for strong short-range interaction while remaining computationally feasible for long-distance propagation.

The numerical results show that the injected bunch undergoes a rapid early self-field-driven expansion, characterized by strong transverse broadening and weaker longitudinal stretching. Comparisons among different macroparticle numbers indicate that the main collective dynamics are only weakly sensitive to $N_p$, and that $N_p=400$ already provides a reasonable compromise
between accuracy and computational cost. The single-particle benchmark further confirms
that the present geomagnetic-field model and relativistic particle pusher correctly reproduce the large-scale reference trajectory.

For the long-range transport,
the 1 MeV bunch continues toward the near-Earth region
under the present injection conditions,
whereas the 10 MeV and 100 MeV bunches do not.
Additional control calculations show that,
for these high-energy cases,
inter-particle interaction mainly affects the bunch divergence
and envelope evolution,
but does not significantly change the macroscopic transport trajectory.
Therefore, whether a bunch ultimately reaches the Earth
should be regarded here as a geometry- and energy-matching problem
rather than the primary focus of this work.
The main contribution of the present study
is to provide a physically consistent and efficient framework
for coupling early dense-stage particle-particle interaction
to subsequent long-range transport of relativistic electron bunches.

\section*{Acknowledgment}

The authors acknowledge the support from
National Natural Science Foundation of China
(Grant No. 52472403).




\section*{Data Availability}
The data that support the findings of this study are available from the corresponding author upon reasonable request.

\section*{Conflict of Interest}

On behalf of all authors,
the corresponding author states that there is no conflict of interest.

\section*{References}
\printbibliography[heading=none]

\end{document}